\documentclass[%
prd,
twocolumn,
superscriptaddress,
nofootinbib,
 amsmath,amssymb,
 aps,
longbibliography
]{revtex4-2}
\usepackage{graphicx}
\usepackage{dcolumn}
\usepackage{bm}
\usepackage{mathtools} 
\usepackage{amsfonts}
\usepackage{booktabs}
\usepackage{siunitx}
\usepackage{verbatim}
\usepackage[normalem]{ulem} 
\usepackage[colorlinks=true,citecolor=blue,urlcolor=blue]{hyperref}
\usepackage[usenames,dvipsnames]{xcolor}
\usepackage{enumitem}

\newcommand{\Austin}{\affiliation{Center for Gravitational Physics, University of Texas at Austin, Austin, TX 78712, USA}}

\newcommand{\Cornell}{\affiliation{Cornell Center for Astrophysics and Planetary Science, Cornell University, Ithaca, New York 14853, USA}}

\newcommand{\Caltech}{\affiliation{Theoretical Astrophysics, Walter Burke Institute for Theoretical Physics,
California Institute of Technology, Pasadena, California 91125, USA}}

\begin{document}

\title{Redshift factor and the small mass-ratio limit in binary black hole simulations}

\author{Sergi Navarro Albalat}
\Austin
\author{Aaron Zimmerman}
\Austin
\author{Matthew Giesler}
\Cornell
\author{Mark A. Scheel}
\Caltech

\date{\today}

\begin{abstract}

We present a calculation of the Detweiler redshift factor in binary black hole simulations based on its relation to the surface gravity.
The redshift factor has far-reaching applications in analytic approximations, gravitational self-force calculations, and conservative two-body dynamics.
By specializing to non-spinning, quasi-circular binaries with mass ratios ranging from $m_A/m_B = 1$ to $m_A/m_B = 9.5$ we are able to recover the leading small-mass-ratio (SMR) prediction with relative differences of order $10^{-5}$ from simulations alone. 
The next-to-leading order term that we extract agrees with the SMR prediction arising from self-force calculations, with differences of a few percent. 
These deviations from the first-order conservative prediction are consistent with non-adiabatic effects that can be accommodated in an SMR expansion. 
This fact is also supported by a comparison to the conservative post-Newtonian prediction of the redshifts. 
For the individual redshifts, a re-expansion in terms of the symmetric mass ratio $\nu$ does not improve the convergence of the series. 
However we find that when looking at the sum of the redshift factors of both back holes, $z_A + z_B$, which is symmetric under the exchange of the masses, a re-expansion in $\nu$ accelerates its convergence.
Our work provides further evidence of the surprising effectiveness of SMR approximations in modeling even comparable mass binary black holes.
\end{abstract}

\maketitle

\section{Introduction}
\label{sec:level1} 

With the planned launch of the space-based gravitational wave detector LISA~\cite{Audley:2017drz} in the next decade, there is a pressing need to improve the modeling of sources of milliHertz gravitational waves.
A promising source of such waves are extreme mass-ratio inspirals (EMRIs), which are the inspiral of stellar-mass compact objects into supermassive black holes.
Such systems are expected to have mass ratios $\epsilon \sim 10^{-4}$--$10^{-6}$, providing a natural small parameter for approximating their evolution.
Meanwhile, intermediate-mass ratio inspirals (IMRIs) with $\epsilon~\sim 10^{-2}$--$10^{-4}$ may exist, sourcing gravitational waves which are detectable by LISA when both components are supermassive black holes, or by ground-based detectors when the binary is composed of a stellar mass object and an intermediate-mass black hole.
Accurate modeling of the gravitational waves produced by these IMRIs is crucial for their detection and interpretation by current generation detectors like LIGO~\cite{LIGOScientific:2014pky}, Virgo~\cite{VIRGO:2014yos} and KAGRA~\cite{KAGRA:2020tym}, and by future third-generation detectors such as Einstein Telescope~\cite{Amaro-Seoane:2018gbb} and Cosmic Explorer~\cite{Dwyer:2014fpa,Evans:2021gyd}. 

Both EMRIs and IMRIs are of special interest as probes of the strong gravity (e.g.~\cite{Ryan:1997hg,Barausse:2020rsu}).
Gravitational waves from such systems are in the detectable regime only while the binary orbit is highly relativistic, and numerical relativity (NR) simulations of such systems are challenged by the large ratio of scales that must be resolved and the large inspiral timescales required for practical signals (e.g.~\cite{Fernando:2018mov,2006.04818,Dhesi:2021yje}).
As such, the small mass-ratio (SMR) approximation, based on the gravitational self-force (GSF) expansion of the metric perturbation in integer powers of the mass ratio $\epsilon$, is currently the most practical method to solve Einstein's equation for these systems~\cite{Poisson:2011nh,Barack:2018yvs}.
Recently, gravitational wave fluxes~\cite{Warburton:2021kwk} and waveforms~\cite{Wardell:2021fyy} from non-spinning EMRIs have been computed to second order in the SMR expansion, a landmark result for LISA source modeling.

As the mass ratio becomes less extreme, as in the case of IMRIs, one expects the SMR approximation to require higher orders in the expansion to maintain the accuracy of the approximation. 
However, little is known about the convergence properties of the SMR expansion. 
It is possible that at some intermediate mass ratio the SMR series fails to converge to the exact solution at finite mass ratio.
The numerical relevance of higher order terms in the series and its convergence properties can be assessed by comparing to the exact solution provided by NR. 

The first task in such comparisons is to identify pairs of coordinate-invariant quantities whose invariant functional relation can be used as a benchmark.
Starting with \cite{LeTiec:2011bk}, a great body of work comparing different quantities to NR (see \cite{Tiec:2014lba} for a review) suggests that the SMR approximation is applicable all the way down to comparable mass binaries 
when re-expanding the SMR series in terms of the symmetric mass ratio $\nu$.
In particular, an analysis of the gravitational wave phase in \cite{vandeMeent:2020xgc} showed that, at least for most of the inspiral, $\mathcal{O}(\nu^2)$ corrections to the phase evolution seem to be small. 

In this work we use as the basis of comparison a local measure of the binary dynamics which has been studied extensively in GSF calculations, the {\it redshift factor} $z$~\cite{Detweiler:2008ft}.
In an EMRI, $z$ corresponds intuitively to the ratio between the rate of passage of proper time at infinity to proper time on the smaller body's worldline. 
For a point particle moving under conservative dynamics, this quantity is a pseudo-invariant~\cite{Detweiler:2008ft,Barack:2011ed}, meaning that is invariant under perturbative coordinate transformations that respect the symmetries (or averaged symmetries) of the orbit. 
This property makes it an ideal reference quantity to compare between different perturbative approaches and between different gauge choices. 

In addition to providing an essential benchmark for approximation methods, the redshift is closely related to the interaction Hamiltonian for the two bodies in the conservative GSF approximation~\cite{Fujita:2016igj}.
It also plays a central role in the first law of binary black hole mechanics, which relates the local redshift to the energy and angular momentum measured at infinity \cite{LeTiec:2011ab}. 
This connection has been used to compute the $\mathcal O(\nu)$ corrections to the binding energy in \cite{LeTiec:2011dp}, showing agreement with NR simulations at moderate mass ratios, and to compute the conservative ISCO shift in agreement with a previous GSF calculation~\cite{Barack:2009ey}.
As a more practical application, it has been used to inform the conservative sector of effective-one-body models, e.g.~\cite{Barausse:2011dq,Akcay:2012ea,Antonelli:2019fmq},
that can ultimately be used to generate waveforms.

The redshift factor was first calculated for circular, non-spinning binaries, to first order in the SMR approximation using the GSF expansion of the metric, and to second order in the post-Newtonian (PN) approximation for generic mass ratios~\cite{Detweiler:2008ft}.
The formalism for computing the redshift factor in the SMR limit was later extended to eccentric orbits~\cite{Barack:2011ed} and then to fully generic orbits in a Kerr background~\cite{Fujita:2016igj}. 
It has been used to compare GSF calculations in different gauges~\cite{Shah:2010bi,Thompson:2018lgb}, and has been computed for eccentric equatorial orbits in Kerr~\cite{vandeMeent:2016pee}.
Meanwhile, analytical PN predictions for the redshift factor were extended to 3PN, directly from the 3.5PN metric~\cite{Blanchet:2009sd}; and 4PN, using the 4PN binding energy in combination with the first law~\cite{LeTiec:2011ab,Bini:2013zaa,LeTiec:2015kgg,Blanchet:2017rcn,Bini:2019lcd}; and comparisons made between GSF and PN predictions for both circular and eccentric orbits, e.g.~\cite{Akcay:2015pza}.
First-order GSF results, which are valid at all orders in the PN approximation, have been used to numerically generate high-order PN predictions e.g.~\cite{Detweiler:2008ft,Akcay:2015pza}. 
Using a PN expansion of the GSF perturbation (a double expansion), accurate analytical formulas for the first order redshift have been provided. 
These include the redshift to 21.5PN order for circular orbits in a Schwarzschild background~\cite{Kavanagh:2015lva}
and lower orders for eccentric orbits and a spinning secondary. 
Similar expansions exist for a Kerr background (e.g.~\cite{Bini:2016qtx,Kavanagh:2016idg,Bini:2019lcd}), and 
we refer the reader to the Black Hole Perturbation Toolkit repository~\cite{BHPToolkit} for the latest results.

In this work we make the connection between these different approximations for the redshift factor and NR simulations.
This was first explored in~\cite{Zimmerman:2016ajr}, which provided a prescription to calculate the redshift in NR simulations using the extracted surface gravity on each black hole, and used it to test the validity of the first law of binary dynamics in a fully dynamical binary spacetime. 
In addition, in~\cite{LeTiec:2017ebm} the redshift was extracted from quasi-equilibrium initial data solutions using this surface gravity prescription, to test the zeroth and first law of binary mechanics, and found to agree remarkably well with PN and first order GSF predictions. 
Our work continues to explore the intersection between the full nonlinear NR solutions, which in our case include non-adiabatic effects from inspiral, and the SMR approximation. 
Our strategy will be similar to that in \cite{vandeMeent:2020xgc}, starting from a set of non-spinning and quasi-circular NR simulations at different mass ratios, we perform fits across mass ratios at a fixed orbital frequency to test whether NR data alone can recover the SMR prediction. 
We are also able to estimate the value of higher-order coefficients in an SMR expansion, and assess the validity of the adiabatic approximation by comparing to the conservative PN and SMR predictions for $z$.

We find that a polynomial fit of order $N=5$ to the NR redshift captures all variability in the data.
The fit recovers the leading SMR prediction (geodesic) to one part in $10^{-5}$, without requiring any input from geodesic or perturbation theory.
The next-to-leading order coefficient agrees well with the existing conservative SMR prediction to a few percent. 
We argue that this deviation is consistent with the size of non-adiabatic effects, which we estimate from the quantity $\dot{\Omega}/\Omega^2$. 
Moreover our result shows that these non-adiabatic effects can be accommodated in the SMR expansion. 
A comparison to the conservative PN prediction also supports this conclusion, with no improvement in agreement with successive PN orders past 2PN, when radiation-reaction becomes relevant at 2.5PN. 
Finally, we confirm that a re-expansion in $\nu$ doesn't accelerate the convergence of the individual redshift series, but we propose a new symmetric combination of redshift factors on the large and small black holes, $Z \coloneqq z_A+z_B$.
Fits of this symmetric quantity converge very rapidly with $\nu$, demonstrating once again the compelling result that re-expansions of low-order SMR predictions using $\nu$ rather than $\epsilon$ can be accurate even at equal masses. 
We extract the SMR coefficients of $Z$ from our numerical simulations and compare them with the predictions from perturbation theory for both black holes, showing again a few percent deviation of order $\mathcal{O}(\nu)$ from the purely conservative prediction, consistent with the measurement of non-adiabatic effects.

{\it Conventions} -- In this study we use Greek indices for spacetime quantities and Latin indices from the middle of the alphabet for spatial quantities. 
We denote the component masses as $m_a$, using the Latin index $a=A,B$ to refer to each of the binary black holes: the primary component (larger mass) is labeled $A$ and the secondary (smaller mass) $B$. The total mass is $m = m_A+m_B$.
Since we consider the mass ratio as a small parameter, we define the small mass ratio $\epsilon \coloneqq m_B/m_A \leq 1$, and unconventionally define the inverse mass ratio as $q\coloneqq m_A/m_B = 1/\epsilon$ with $q\geq 1$.
The symmetric mass ratio is $\nu=m_A m_B/(m_A+m_B)^2$, and note $\nu \leq 1/4$. 
The mass unit of the numerical simulations is $M$, and is very nearly equal to the initial ADM mass of the simulations. 
Numerical indices and $k$ are used to indicate the order in $\epsilon$ of the SMR expansion.

\section{Redshift factor}
\label{sec:Redshift}

\subsection{Redshift factor for circular orbits}
\label{sec:GSFredshift}

The key property of pseudo-invariance of the redshift factor
is precise only when the binary spacetime has a global helical Killing vector field (HKVF).
This HKVF takes the form
\begin{align}
    K^\mu=\partial_t^\mu+\Omega\partial_\phi^\mu \,.
\end{align}
Here $\partial_t^\mu$ is a vector field that is timelike outside the history $\mathcal{T}$ of some sphere, $\Omega$ is a constant corresponding to the orbital frequency, and $\partial_\phi^\mu$ is spacelike with integral curves of length $2\pi$ \cite{Bonazzola:1997gc}. 
For an asymptotically flat spacetime, $\partial_t^\mu$ and $\partial_\phi^\mu$ limits to asymptotic timelike and rotational Killing vector fields, respectively. 
When considering a binary in a spacetime with an exact HKVF, and where the secondary is treated as a point particle, the orbit of the particle coincides
with the integral curves of the HKVF.
In this case the redshift factor is~\cite{Detweiler:2008ft}
\begin{align}
    z_B=\frac{1}{u^t} \,,
\end{align}
where $u^\mu$ is the four-velocity of the particle.

For example, consider the circular orbit of a point particle of mass $m_B$ around an Schwarzschild black hole of mass $m_A$.
We are interested in the limit of small mass ratio.
At leading order, the point particle behaves like a test mass: the worldline corresponds to an affinely parametrized geodesic of the Schwarzschild metric $g_{\mu\nu}$. 
Then the redshift of the test mass is related to the orbital frequency of the circular geodesic by
\begin{align}
    \label{z0}
    z_{B,0}^{\rm{SMR}}=\sqrt{1-3(m_A\Omega)^{2/3}} \,.
\end{align}

At linear order in the mass ratio, the metric becomes $g_{\mu\nu}=g^0_{\mu\nu}+\epsilon h_{\mu\nu}$ where $h_{\mu\nu}$ is the metric perturbation due to the presence of the small mass on the background metric $g^0_{\mu\nu}$. 
For a point particle $g_{\mu\nu}$ diverges on the worldline and so does the redshift. 
Instead, the redshift is defined from the geodesic motion in an effective metric 
\begin{align}
\tilde{g}_{\mu\nu}=g^0_{\mu\nu}+\epsilon h^{\text{R}}_{\mu\nu} \,,
\end{align}
where $h^{\text{R}}_{\mu\nu}$ is a certain regular piece of the retarded metric perturbation~\cite{Detweiler:2002mi}. 
The regular metric perturbation can be further split into dissipative (time antisymmetric) and conservative (time symmetric) pieces~\cite{Detweiler:2002mi}. 
The dynamics due to the conservative part alone retains the symmetry of the HKVF, and one can calculate from it an invariant $\mathcal{O}(\epsilon)$ contribution to the redshift given by~\cite{Detweiler:2008ft}
\begin{align}
    \label{z1tilde}
    z_{B,1}^{\rm{SMR}}=-\frac{1}{2}[1-3(m_A\Omega)^{2/3}]h^{\text{R,cons}}_{\mu\nu}u^{\mu}_0u^{\nu}_0 \,.
\end{align}
This conservative, first order SMR term has been calculated with very high precision using GSF codes in various gauges.
It can be calculated for any (stable and unstable) circular orbit labelled by its invariant $\Omega$.
To compare to our NR results we use the 4PN accurate formula for $z$ in~\cite{LeTiec:2017ebm} and the 21.5PN analytic formula for $z^{\rm{SMR}}_{B,1}(m_A\Omega)$ provided by \cite{Kavanagh:2015lva}.

\subsection{Redshift factor and surface gravity}
\label{sec:zBH}

A different but equivalent approach to calculating a redshift is to consider the surface gravity when the small particle is a black hole. 
If one insists in the existence of a global HKVF in our binary spacetime, this can only be achieved by having equal amounts of asymptotically ingoing and outgoing radiation, in which case the spacetime is not asymptotically flat \cite{Gibbons}. 
In such scenario $K^\mu$ is proportional to the Killing horizon generators and the surface gravity (uniform across each horizon \cite{cmp/1103858973}) is given by \cite{Friedman:2001pf}
\begin{align}
    \label{kappa}
    K^\mu\nabla_\mu K^\nu|_{\mathcal{H}_a}=\kappa_a K^\nu \,.
\end{align}

The problem with this construction is that, without asymptotic flatness, there is no natural normalization of $K^\mu$ and the value of $\kappa$ is free. 
One way around this is to consider the small mass-ratio limit, with no incoming radiation needed to preserve the HKVF.
In this limit, using matched asymptotic expansions, one can show the redshift in Eq.~\eqref{z1tilde} corresponds to~\cite{PoundRedshift:2015}
\begin{align}
    \label{zAdam}
    z_B=4m_B\kappa_B\,.
\end{align}
Briefly, imagine near the small black hole, using coordinates such that the geometry is described by the stationary, tidally perturbed, Schwarzschild black hole in~\cite{Poisson:2005pi}. 
In the limit of large separation, $v$ is the usual Eddington-Finkelstein advanced-time coordinate. One can show that at this level of approximation, $\partial/\partial v$ are the generators of the perturbed horizon. 
Thus, the generators of the local symmetry $k^\mu=(\partial/\partial v)^\mu$ must be proportional to the generators of the global symmetry $K^\mu$. 
If we define the redshift by the constant of proportionality between these normalizations of the HKVF,
\begin{align}
    \label{matching}
    k^\mu=\frac{1}{z}K^\mu \,,
\end{align}
where $\bar{\kappa}_B$ denotes the effective surface gravity from equation $k^\mu\nabla_\mu k^\nu\rvert_{\mathcal{H}_B}=\bar{\kappa}_Bk^\nu$ and using Eqs.~\eqref{kappa} and~\eqref{matching}, one has
\begin{align}
\label{zdef}
    z_B=\frac{\kappa_B}{\bar{\kappa}_B} \,.
\end{align}
 
More physically, we can understand this construction as comparing the rate of passage of proper time between two inertial observers. 
The first is at rest and asymptotically far from the binary.
The second is comoving with the smaller black hole, close enough to neglect the curvature scale of the larger black hole but far enough from it (in the so-called buffer region) so that the divergent piece of the metric perturbation goes to zero and the metric perturbation is dominated by $h^{\rm{R}}_{\mu\nu}$~\cite{Zimmerman:2016ajr}.
This construction is more suitable to our numerical spacetime since it makes no explicit reference to a worldline and it also allows us to define a redshift for the larger black hole. 
 
We also define the redshift of the primary from Eq.~\eqref{zdef} applied to black hole $A$.
A first order correction to the surface gravity of the larger black hole due to an orbiting ``moon" in co-rotation was found in \cite{Gralla:2012dm}. 
Using Eq.~\eqref{zdef} to translate it into a redshift and neglecting the numerically small contribution from the small spin required by co-rotation, we have
\begin{align}
\label{zA0}
    z_{A,0}^{\rm{SMR}}&=1\,,
\\
\label{zAletiec}
    z_{A,1}^{\rm{SMR}}&=-\frac{(m_A\Omega)^{2/3}}{\sqrt{1-3(m_A\Omega)^{2/3}}}\,.
\end{align}
This can be easily derived when assuming the integral version of the first law~\cite{Friedman:2001pf,LeTiec:2011ab} and our mapping between the redshift and the surface gravity, which together give $m_A z_A +m_B z_B=M_B-2\Omega J_B$. 
Expanding both sides of the equation in $\epsilon$ one has
  \begin{align}\label{zA2}
    z_{A}^{\rm{SMR}}=1&+\epsilon[E^{\rm{SMR}}_0-2(m_A\Omega) L^{\rm{SMR}}_0-z_{B,0}^{\rm{SMR}}]\nonumber\\
    &+\epsilon^2[-E^{\rm{SMR}}_0+E^{\rm{SMR}}_1-2(m_A \Omega)L^{\rm{SMR}}_1-z_{B,1}^{\rm{SMR}}]\nonumber\\
    &+\mathcal{O}(\epsilon^3) \,,
\end{align}
where $E=(M_B-m)/\mu$ and $L=J_B/(\mu m)$ are the specific binding energy and angular momentum of the binary. 
Their first order corrections are directly related to $z^{\rm{SMR}}_{B,1}$ \cite{LeTiec:2011bk}. 
Note how this assumption also gives a second order correction to the surface gravity of the larger black hole in terms of known quantities, which we state here for the first time and test using NR simulations.

\subsection{Redshift factor in NR simulations}
\label{sec:DynamicalRedshift}

The connection~\eqref{zdef} between surface gravity and the redshift factor provides our starting point for defining a redshift factor $z_a$ for the two black holes in NR simulations.
We use this relation in our simulations, although in reality the emission of gravitational waves means that there is no global HKVF; the best we can hope for is a slowly evolving, approximate HKVF.
Thus Eq.~\eqref{zdef} is only strictly true in the adiabatic limit, where the system evolves through a sequence of  conservative spacetimes labeled by $\Omega$.
This is a good approximation during the inspiral phase of our simulations, while $\Omega$ evolves ``slowly" on the orbital timescale, but it fails as the secondary approaches the innermost stable circular orbit (ISCO).
As a measure of the non-adiabaticity of the system, we track the evolution of the quantity $\dot{\Omega}/\Omega^2$, which remains $\mathcal O(10^{-2})$ through much of each simulation, as discussed in Sec.~\ref{sec:PN}.

Moreover, in a numerical simulation we do not track the event horizons of the black holes, and what we consider to be the horizon should be the dynamical horizon $H$ of the type defined in~\cite{Ashtekar:2003hk}. 
However, in the adiabatic approximation, they can be approximated by Killing horizons. 
As a first implementation of our NR definition of the redshift, we proceed from Eq.~\eqref{zdef}, deriving a practical expression for the surface gravity assuming the adiabatic approximation.
Regardless of whether the evolution is adiabatic, we can take the result as our operational definition of the redshift factor in NR.

To begin with, in our numerical spacetime it is useful to express the HKVF in terms of 3+1 quantities
\begin{align}
\label{KNR}
    K^\mu=N n^\mu+\beta^\mu+\Omega\partial_\phi^\mu \,,
\end{align}
where $N$ is the lapse, $n^\mu$ is the normal vector to the surface of constant time $\Sigma_t$, and $\beta^\mu$ is the shift.
The overall normalization of the Killing field is fixed by our choice of inertial frame at infinity. 
In that frame $K^\mu=(1,0,0,\Omega)$. On $H$ we calculate null normals $\ell^\mu$ with the following default normalization
\begin{align}
    \label{nullnormal}
    \ell^\mu=\frac{n^\mu+s^\mu}{\sqrt{2}} \,,
\end{align}
where $s^\mu$ is the unit normal to the 2-sphere $S_t$ corresponding to the intersection of $H$ with $\Sigma_t$. We fix the re-scaling freedom of the null normals $\ell^\mu \rightarrow \alpha\ell^\mu$ by having them match the Killing field~\eqref{KNR} on the horizon. 
Matching the time component one finds $\alpha=\sqrt{2}N$. The re-scaled null normals are 
\begin{align}
    \xi^\mu=N(n^\mu+s^\mu) \,.
\end{align}
Equipped with this choice of null normals we calculate the surface gravity pointwise on $H$.
Consistent with the adiabatic approximation, we neglect the term $\xi^{\mu} \nabla _{\mu}\ln N$.
The result, expressed with spatial quantities on $\Sigma_t$, is 
 \begin{align}
 \label{kappaNR}
  \kappa_\xi=  s^i \partial_i N-NK_{ij} s^is^j \,,
 \end{align}
where $K_{ij}$ is the extrinsic curvature of $\Sigma_t$. 
This is consistent with Eq.~(10.10) in \cite{Gourgoulhon:2005ng}, when the evolution of the lapse along the generators is neglected. 

\begin{figure}[tb]
\includegraphics[width = 0.98\columnwidth]{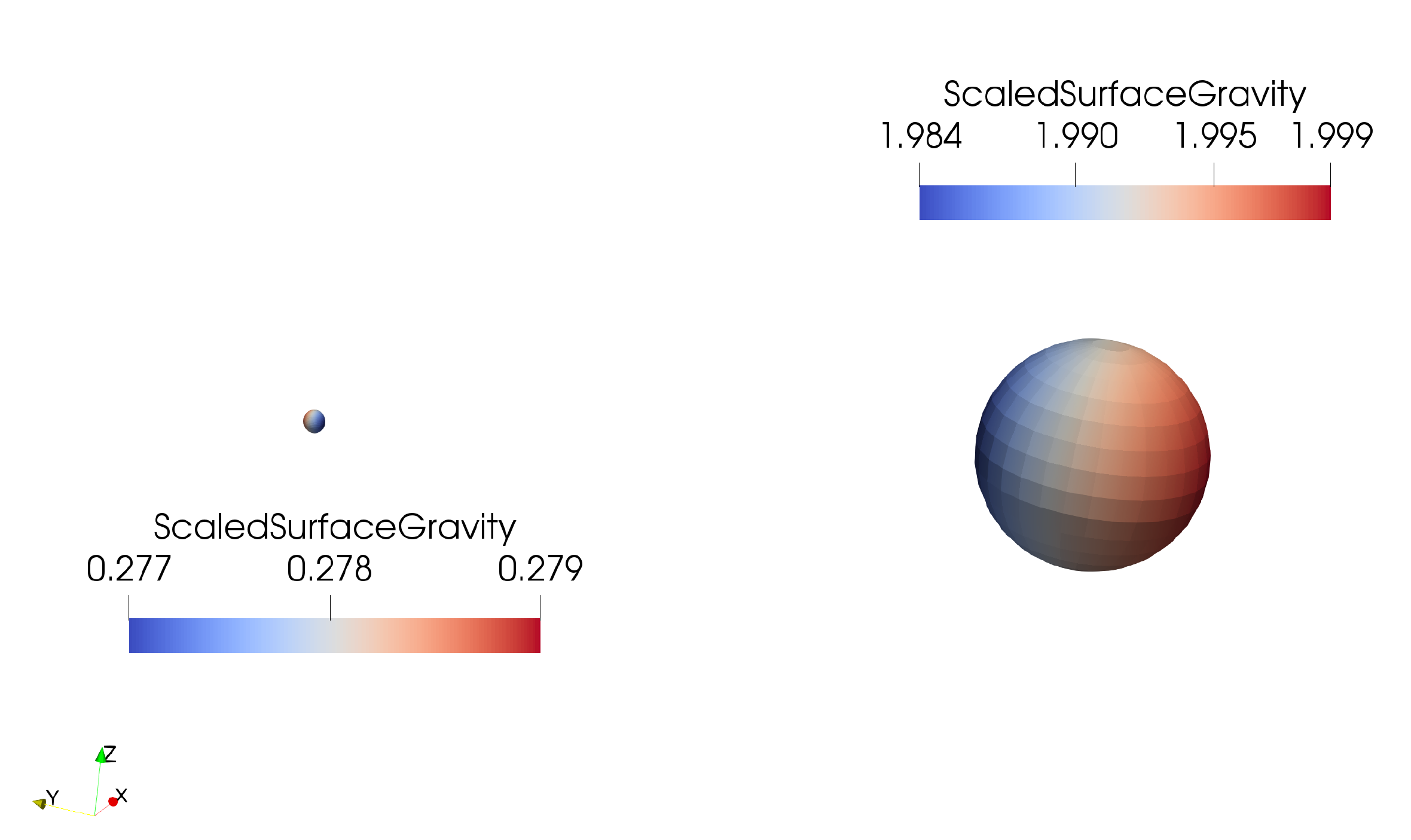}
\caption{Scaled surface gravity $\kappa_\xi$ on each of our $q=8$ black hole horizons.}
\label{fig:SurfaceGravity}
\end{figure}

Since the horizons are not precisely Killing, $\kappa_\xi$ varies across each black hole horizon.
Figure~\ref{fig:SurfaceGravity} depicts the scaled surface gravity $\kappa_\xi$ on the horizons for a quasi-circular, non-spinning binary simulation with $q = 8$, illustrating the variance of this quantity around its average.
As a final step we average $\kappa_\xi$ across each horizon at every time step,
\begin{align}
    \langle \kappa_{\xi}\rangle = \frac{1}{A}\int \kappa_{\xi} dA \,.
\end{align}
With the averaged surface gravity, we then calculate the redshift of each black hole using
 \begin{align}
 \label{zfromkappa}
 z_a=\frac{\langle\kappa_{\xi}\rangle_a}{\bar{\kappa}_a}\,.
 \end{align}
where in our simulations $\bar{\kappa}_a=1/(4m_a)$ is the Schwarzschild surface gravity calculated from the quasi-local areal mass $m_a=\sqrt{A_a/16\pi}$.

\section{Numerical simulations}
\label{sec:NR}

In this section, we present the binary black hole simulations used in this study and our extracted redshift quantities $z_a$ on each black hole. 
Equally important in our comparisons to PN and SMR approximations is the estimation of an appropriate pseudo-invariant orbital frequency $\Omega$.
Before presenting $z_a$ as a function of $\Omega$, we discuss several definitions of $\Omega$, ultimately selecting a co-rotating frame frequency $\Omega_{\rm co}$ derived directly from the extrapolated gravitational waves~\cite{Boyle:2013nka}.

\subsection{Simulations used in this study}
\label{sec:Sims}

\begin{table*}[t]
\centering
\setlength{\tabcolsep}{1em}
\begin{tabular}{cccccccc}
\toprule
    $q$ & \text{Type} & $M\Omega_0$ & $N_\text{cycles}$ & $e_0$ & $|\delta x_{\rm{CoM}}|$ & $|v_{\rm{CoM}}|$ & $\text{Highest Levs}$ \\ \midrule
    1 & SKS & 0.01233 & 27.96 & 1.355e-4 & 3.1363e-06 & 1.5145e-08 & $\text{5,6}$  \\
    1  & SHK  & 0.01453 & 20.78 & 2.4e-3 & 7.409e-07 & 1.923e-08 & $\text{4,5}$  \\
    1.5  & SKS  & 0.01250 & 28.98 & 5.77e-5 & 0.001803 & 7.129e-06 & $\text{2,3}$ \\
    2  & SHK  & 0.01554 & 20.70 & 2.408e-4 & 0.0002893 & 1.815e-06 & $\text{2,3}$  \\ 
    2  & SKS    & 0.01842 & 15.45 & 2.890e-4 & 0.001761 & 5.1151e-06 & $\text{2,3}$  \\
    3  & SKS   & 0.01707 & 20.44 & 9.64e-5 & 0.001900 & 4.287e-06  & $\text{2,3}$  \\ 
    3.5  & SKS   & 0.01477 & 27.76 & 2.665e-4 & 0.01348 & 4.126e-05 & $\text{4,5}$ \\ 
    4  & SKS   & 0.01600 & 25.67 & 8.702e-4 & 0.03156 & 1.613e-05 &  $\text{4,5}$  \\
    4  & SHK   & 0.01824 & 20.07 & 8.25e-5 & 0.001338 & 2.075e-06 & \text{4,5} \\
    4.5  & SKS   & 0.01616 & 27.37 & 8.289e-4 & 0.0165 & 3.399e-05  & $\text{4,5}$  \\
    5  & SKS  & 0.01589 & 29.13 & 2.236e-4 & 0.0233 & 3.217e-05 & $\text{4,5}$   \\ 
    5.5  & SKS  & 0.01592 & 30.81 & 4.442e-4 & 0.03242 & 4.0443e-05 & $\text{4,5}$   \\ 
    6  & SKS   & 0.01588 & 32.62 & 5.864e-4 & 0.022980 & 4.0374e-05 & $\text{4,5}$   \\ 
    6.5  & SKS   & 0.01599 & 34.43 & 7.263e-4 & 0.037534 & 3.9212e-05 & $\text{4,5}$   \\ 
    7  & SKS  & 0.01577 & 36.16 & 3.612e-4 & 0.02493 & 1.4183e-05  & $\text{4,5}$  \\ 
    7.5  & SKS   & 0.01597 & 37.89 & 5.524e-4 & 0.04963 & 3.694e-05  & $\text{4,5}$ \\ 
    8  & SKS   & 0.01584 & 39.53 & 6.688e-4 & 0.05589 & 5.876e-05  & $\text{5}$  \\ 
    8.5  & SKS   & 0.01594 & 41.31 & 8.578e-4 & 0.04370 & 3.00178e-05  & $\text{5}$  \\ 
    9  & SKS   & 0.01583 & 43.16 & 2.010e-4 & 0.02375 & 3.5280e-05 & $\text{4,5}$   \\ 
    9.5  & SKS   & 0.01585 & 44.93 & 1.584e-4 & 0.03413 & 3.8326e-05  & $\text{1,4}$ \\
    \midrule
    $14^*$  & SHK   & 0.02292 & 27.70 & 3.814e-4 & 0.0016026 & 1.747e-06  & $\text{2,3}$ \\
    $15^*$  & SHK   & 0.02317 & 27.94 & 3.692e-4 & 0.001791 & 2.815e-06  & $\text{2,3}$ \\
    \bottomrule
\end{tabular}

\caption{Properties of the SpEC simulations used in this analysis. The subscript zero denotes the reference time (time at which junk radiation has sufficiently decayed). The orbital frequency at that time is $M\Omega_0$. The approximate number of orbital cycles before merger is $N_{\rm cycles}$. The center of mass position $|\delta x_{\rm{CoM}}|$ and velocity $|v_{\rm{CoM}}|$ have been estimated using the \texttt{estimate\textunderscore avg\textunderscore com\textunderscore motion} function from \texttt{Scri} \cite{Boyle:2015nqa,Boyle_scri_2020}. 
The parameters shown correspond to the highest resolution setting (Lev). *The $q=14$ and $q=15$ simulations are used in the PN comparisons but are not included in the fits for the reason discussed in Sec.~\ref{sec:PN}}.
\label{table:1}
\end{table*}

We extract the redshift factor from a sequence of non-spinning binary black hole simulations, with mass ratios varying from $q=1$ to $q=15$.
These simulations were carried out with the SpEC code ~\cite{Boyle:2019kee,SpECwebsite} and are summarized in Table~\ref{table:1}.
SpEC is a pseudospectral code that uses the Extended Conformal Thin-Sandwich Method for initial data~\cite{York:1998hy,Pfeiffer:2002iy,Pfeiffer:2002wt}, damped harmonic gauge~\cite{Lindblom:2009tu,Choptuik:2009ww,Szilagyi:2009qz} for the evolution, and excision to remove the interior of each black hole~\cite{Hemberger:2012jz}.
Our simulations use one of two types of initial data: superimposed Kerr-Schild (SKS)~\cite{Lovelace:2008tw}
 or superposed harmonic Kerr (SHK)~\cite{Varma:2018sqd} which has the advantage of reducing the initial junk radiation at the expense of not being able to resolve high spins. 
This initial data uses the improvements described in~\cite{Ossokine:2015yla} to reduce the center of mass (CoM) motion.
However, residual CoM motion remains in our simulations, and the drift of the CoM introduces oscillations into our extracted redshifts, as discussed below in Sec.~\ref{CoMoscillations}.
These binaries are initialized in quasi-circular orbits after using iterative eccentricity reduction~\cite{Pfeiffer:2007yz,Buonanno:2010yk,
Mroue:2012kv}, so that initial the orbital eccentricity is $e_0\lesssim 10^{-3}$.

We make use of data after the relaxation time $t_0$ at which junk radiation has sufficiently decayed~\cite{Boyle:2019kee}.
For each simulation we record the coordinate centers $\vec{x}_a$ of the apparent horizons and the Christodoulou masses $m_a$ of the black holes, with $a = A,B$. 
We take as the value for each mass the time average between $t_0$ and $t_0+4000M$, and use these masses to calculate the mass ratio $q=m_A/m_B$.
We use the average of the surface gravity over the apparent horizons to compute the redshift factor through Eq.~\eqref{zfromkappa}.

We also make use of the gravitational wave emission from our simulations to construct gauge-invariant measures of the orbital frequency as described in Sec.~\ref{sec:omega}.
From each simulation, the gravitational waves are extracted
at finite radii and extrapolated to infinity, as described in~\cite{Boyle:2009vi}.
For our analysis, we choose the fourth-order polynomial in inverse areal radius $r$ for our extrapolation,
which is more accurate than lower order polynomials during the early inspiral~\cite{Boyle:2009vi}.
The retarded time is chosen according to Eq.~(12) in \cite{Boyle:2019kee}.
The gravitational wave strain is further corrected for the center of mass motion of the binary, using the method of~\cite{Boyle:2015nqa}.

Nearly all our simulations have multiple resolutions (Levs), and where appropriate we plot results from our two highest resolutions. 
We repeat our analysis with both these resolutions and incorporate the range of results in our error estimates for the results presented in Sec.~\ref{sec:Results}.

\subsection{Orbital frequency}
\label{sec:omega}

A reliable extraction of the orbital frequency, which provides a gauge-invariant parametrization of the orbit, is as important as extracting the redshift.
The reason is that even though $z(t)$ and $\Omega(t)$ are pseudo-invariant quantities in the presence of a HKVF, their functional relation to $t$ is not (because $t$ itself is gauge dependent). 
Instead, considering $z(\Omega)$ guarantees the same functional relation independent of the gauge.

In the presence of a HKVF an invariant definition of $\Omega$ is provided by the Killing condition. 
However such a binary spacetime, as discussed earlier, would be eternally rotating at a constant frequency. 
In the presence of dissipation $\Omega$ evolves in time and we cannot rely on the HKVF to define it. 
We do not attempt to define a new invariant frequency measure in this work.
Instead, we consider four different definitions of $\Omega$ that have been used in previous NR analysis and base our choice in the intuitive requirements that $\Omega$ should coincide with the HKVF $\Omega$ in the limit of a perfect circular orbit and should be insensitive to gauge ambiguities. 

We consider the following definitions of $\Omega$, based on:
\begin{enumerate}[label=(\roman*)]
\item The coordinate motion of the black hole centers,
\begin{align} 
\label{omega_coor}
\Omega_{\text{coor}}\coloneqq\frac{\rvert \vec{r} \times \dot{\vec{r}} \rvert}{r^2} \,,
\end{align}
where $\vec{r}=\vec{x}_B-\vec{x}_A$ is the relative position vector between the two black holes.
\item The time derivative of the $l=2$, $m=2$ mode of the gravitational wave phase $\Phi_{22}$ (sometimes denoted by $\varpi$),
\begin{align} 
\Omega_{22}\coloneqq\frac{1}{2}\frac{d \Phi_{22}}{dt} \,.
\end{align}
\item The definition of the co-rotating frame, for which the time-dependence of the waveform is minimized, and using the angular frequency of this frame,
\begin{align} 
\label{omega_cor}
\Omega_{\rm co}\coloneqq\omega_z \,,
\end{align}
where $\bf \omega$ is calculated according to formula (7c) in~\cite{Boyle:2013nka} (see reference for details on the calculation). We use the built in function for it in \texttt{Scri}~\cite{Boyle_scri_2020}.
\item The flux relation for circular orbits:
\begin{align} 
\Omega_{\rm circ}\coloneqq\frac{\dot{E}}{\dot{L}} \,,
\end{align}
where the energy flux $\dot{E}$ and angular momentum flux $\dot{L}$ are calculated from the extrapolated ($N=4$) and CoM corrected strain and using all the available modes (up to $l=8$).
\end{enumerate}

\begin{figure}[t]
\includegraphics[width = 0.98 \columnwidth]{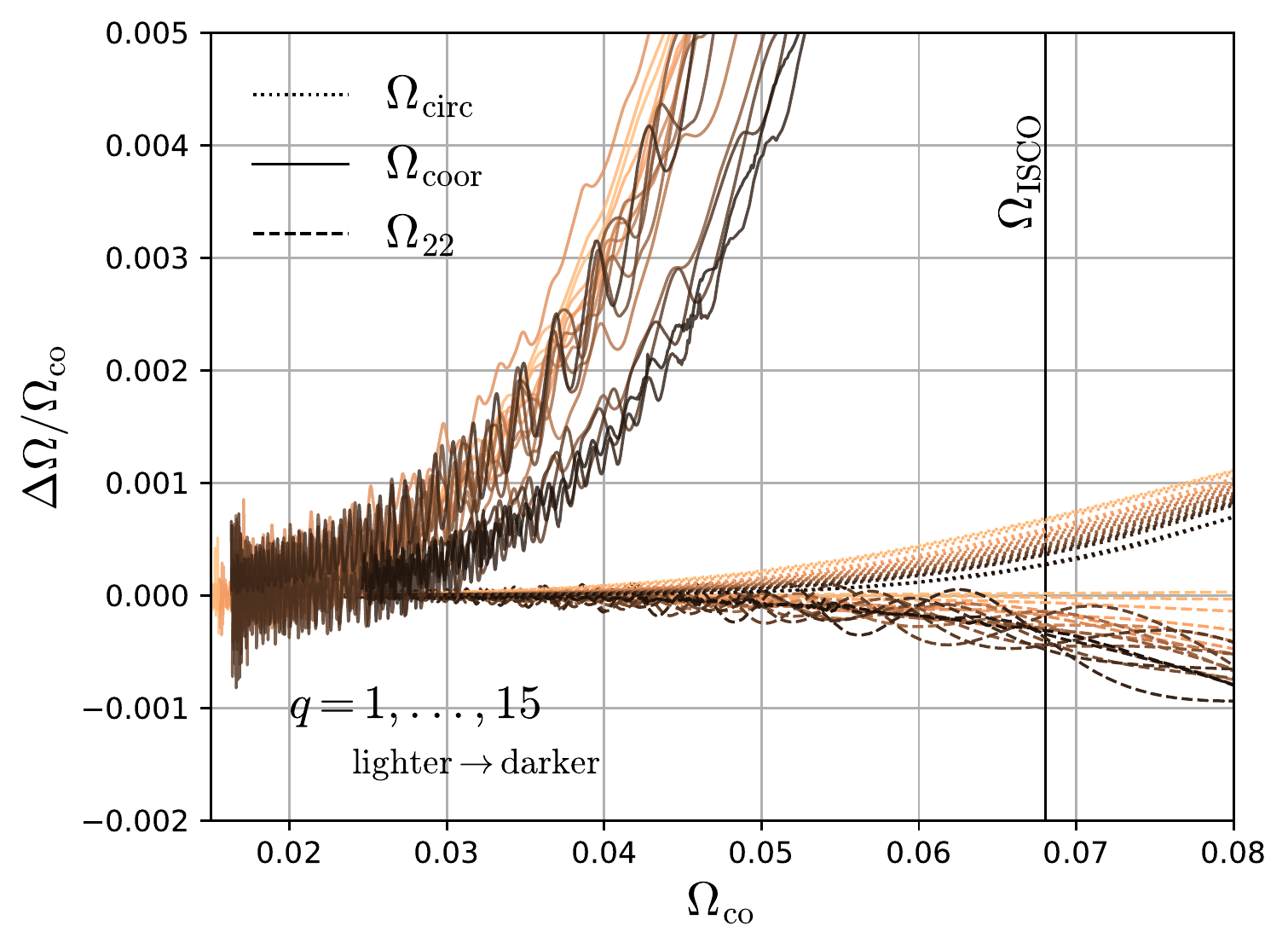}
\caption{\label{omegas} Relative difference between the corotating frequency $\Omega_{\text{co}}$ and the other measures of the orbital frequency: $\Omega_{\text{coor}}$ (dash-dotted), $\Omega_{\text{circ}}$ (dotted), $\Omega_{\text{22}}$ (dashed) . 
Differences are shown for the range of mass ratios covered by our simulations at the highest resolution.}
\end{figure}

In Fig.~\ref{omegas} we compare these using $\Omega_{\rm{co}}$ as a baseline. 
In this and all our figures, the shading of the curves ranges from lighter to darker as we move from lower $q$ to higher $q$.
The four definitions coincide early in the inspiral. 
The largest deviations correspond to $\Delta\Omega_{\text{coor}}$, with relative differences growing much faster than the others. 
This may be expected, since the definition is based on the gauge-dependent quantity $\vec{r}$ defined with respect to the simulation coordinates.
The relative differences between $\Omega_{22}$, $\Omega_{\rm circ}$ and $\Omega_{\rm{co}}$ are below $0.01\%$ for all of the inspiral (before the ISCO frequency $M\Omega_{\rm{ISCO}}=6^{-3/2}$). 
We also see that $\Omega_{22}$ approaches $\Omega_{\text{co}}$ for more equal mass ratios. 
In contrast, $\Omega_{\rm circ}$ approaches $\Omega_{\text{co}}$ for smaller mass ratios (larger $q$).
This is a desired behaviour, since at a fixed frequency we expect smaller departures from circularity at smaller mass ratios (larger $q$). 
We speculate that the reason that $\Omega_{\text{co}}$ better limits to the expected behavior at small mass ratios is that at these mass ratios, where emission from higher angular harmonics is more important, it better captures the overall phase evolution of the binary than the leading $l=2$, $m=2$ mode.
For this reason we use $\Omega_{\text{co}}$ as the orbital frequency in our SMR analysis. 
However we have checked that using $\Omega_{22}$ or $\Omega_{\rm circ}$ doesn't substantially change any of the results presented here. 
Results of the analysis using $\Omega_{\rm{coor}}$ are shown in Appendix~\ref{sec:OmegaCompAppx}, and while this choice of frequency shows larger discrepancies with our preferred choice $\Omega_{\rm co}$, it does not change our main conclusions.

\subsection{Correcting CoM-induced redshift oscillations}
\label{CoMoscillations}

\begin{figure}[tb]
\includegraphics[width = 0.98\columnwidth]{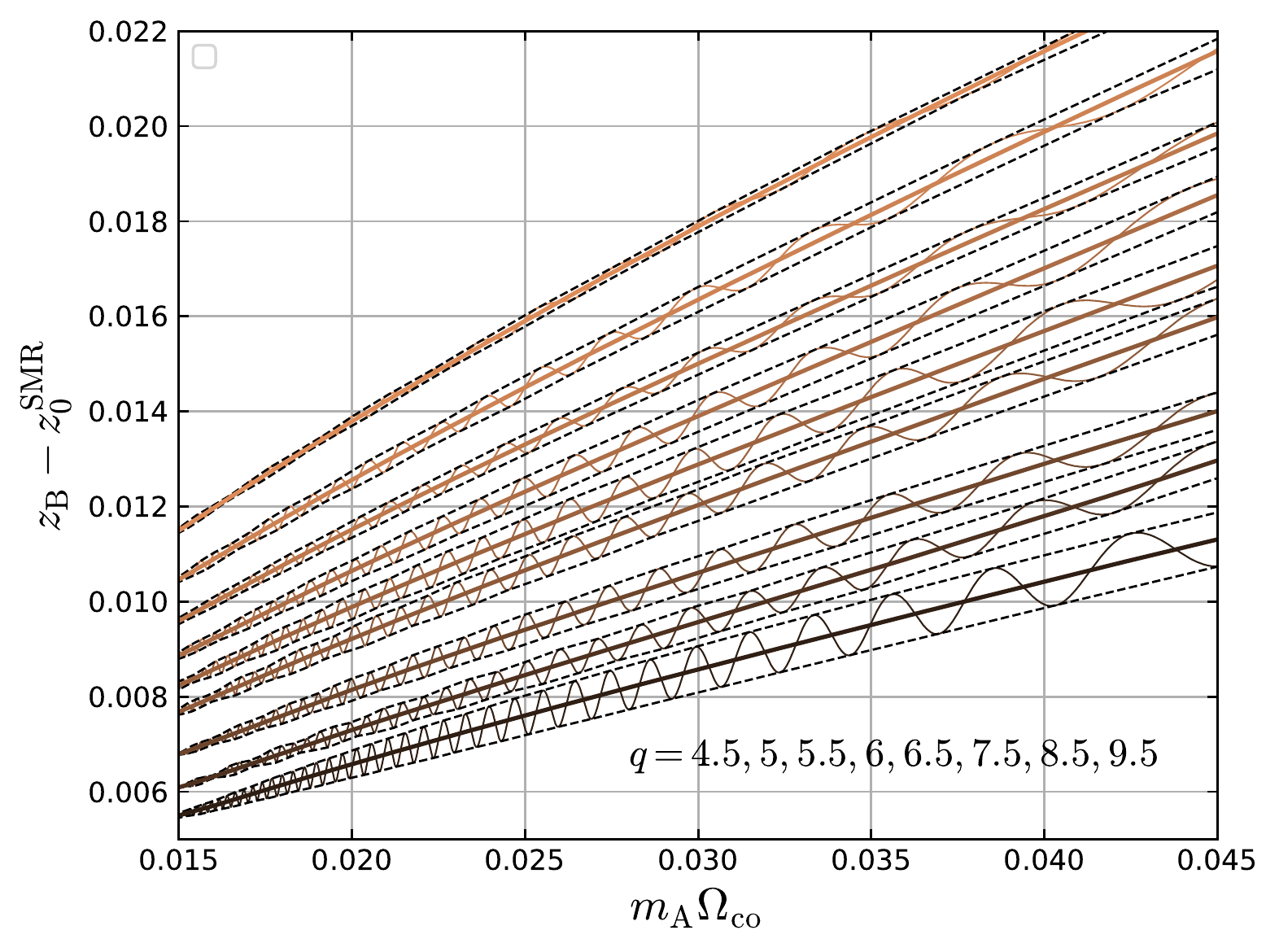}
\caption{Uncorrected redshifts (thin lines), displaying the CoM motion-induced oscillations. Also plotted are the corrected redshifts (thick lines) and upper and lower envelopes (dashed lines), found using the sampling method. 
The envelopes are quadratic splines through the $z(m_A\Omega_{\rm{co}}(t_i))$ points defined by $\Omega_{\text{coor}}(t_i)-\Omega_{B,\text{coor}}(t_i)=0$. 
The amplitude of the oscillations is larger for higher $q$ (smaller mass ratio). The modulations in the envelope are consistent with the residual eccentricity.}
\label{envelopes}
\end{figure}

One challenge encountered by our analysis is that our simulations exhibit center of mass (CoM) motion that induces small oscillations in the extracted redshift. 
These oscillations in $z_B(m_A \Omega_{\text{co}})$ are illustrated in Fig.~\ref{envelopes} for a subset of our simulations which clearly display this affect.
The oscillations are quite small, and to display them we first subtract out the geodesic predictions $z_0^{\text{SMR}}$. Nevertheless they contaminate our SMR analysis, which requires high precision.
They do not appear to be due to orbital eccentricity: they grow during the simulations, although orbital eccentricity is expected to decay; further they are generally larger than $e_0^2$ effects we would expect from our initial eccentricities.
We also plot the upper and lower envelopes of the oscillation in Fig.~\ref{envelopes}, as well as our final corrected values for the redshift, using the procedure described below.

First, we argue that these oscillations are due to the CoM motion.
As discussed in Sec.~\ref{sec:DynamicalRedshift} the redshift factor relies on a normalization of the approximate HKVF at asymptotic infinity, which requires a choice of an asymptotic inertial frame. 
In the PN approximation this frame is centered with respect to to the binary's center of mass in the limit of large separation, and in the SMR approximation it is centered around the larger black hole.
By contrast, in our NR simulations we cannot {\it a priori} precisely select the asymptotic inertial frame, and it is in general different for each simulation.
Different asymptotic inertial frames in general measure a different redshift (provided there is a map of asymptotic quantities onto the horizon, which there is if one assumes a HKVF).

\begin{figure}[tb]
\includegraphics[width = 0.98\columnwidth]{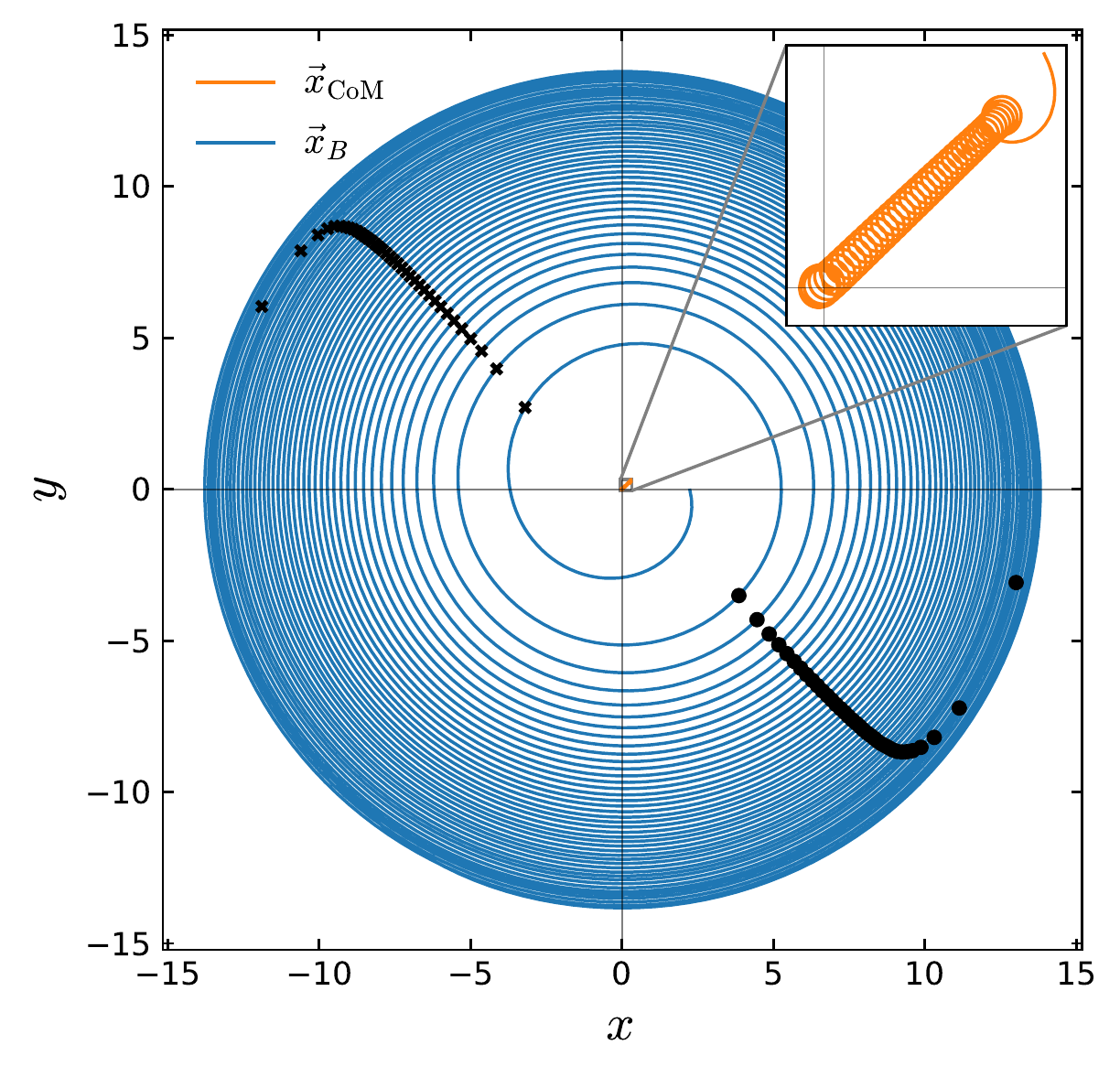}
\caption{Trajectory of the smaller black hole center $\vec{x}_B$ and CoM $\vec{x}_{\rm{CoM}}$ in simulation coordinates. The points marked with dots correspond to the ``minima" of the redshift while the crosses correspond to the ``maxima". One can see from the figure that they roughly match the points where the small black hole's velocity is maximized and minimized by picking up a contribution from the average CoM drift velocity.}
\label{sampling2}
\end{figure}

\begin{figure}[tb]
\includegraphics[width = 0.98\columnwidth]{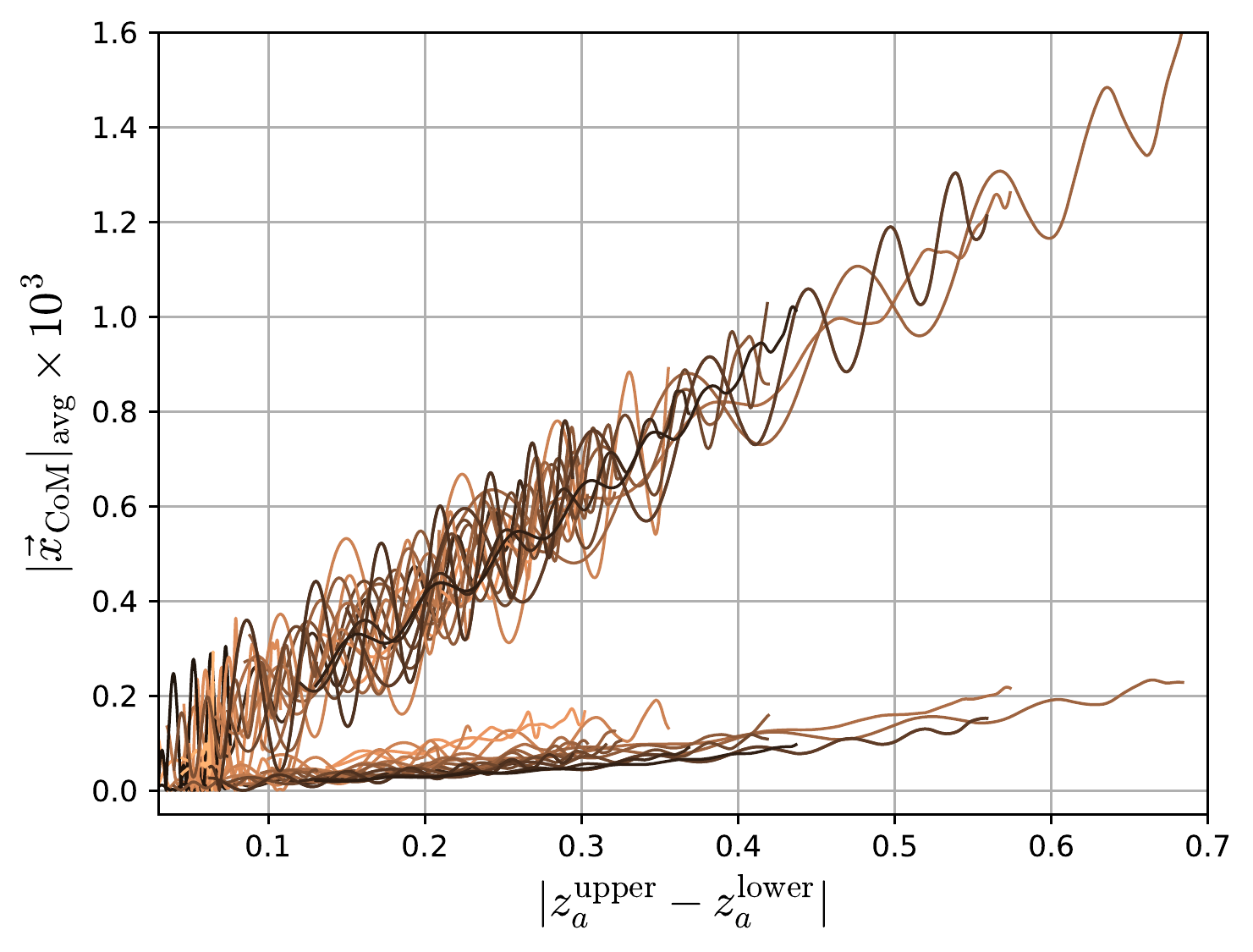}
\caption{Instantaneous average displacement of the CoM plotted against the amplitude of the oscillations in the redshifts for both the larger and smaller black holes.
The larger black hole displays smaller amplitudes in all cases.
The residual oscillations in these curves correspond to modulations in the amplitude which are well correlated with the residual eccentricity (see Table~\ref{table:1}).}
\label{correlation}
\end{figure}

The coordinate motion of the Newtonian center of mass illustrates the CoM motion.
We can see the binary drifting away from the origin while exhibiting epicyclic motion for our $q = 8.5$ simulation in the inset of Fig.~\ref{sampling2}. 
This motion is a clear sign that the Killing field in Eq.~\eqref{KNR} is not centered with respect to the simulation's inertial coordinates. 
Further, Fig.~\ref{correlation} shows correlation between the amplitude of the redshift oscillations and the average displacement from the origin of the simulation coordinates.
The amplitude for both black holes in each simulations is plotted, and can be differentiated by the fact that the larger black holes always display smaller amplitude oscillations.
The average displacement is found by fitting $x_{\rm{CoM}}(t)$ and $y_{\rm{CoM}}(t)$ to an low degree polynomial, which smooths over the epicycles, and taking the norm $|x_{\rm{CoM}}(t)^2+y_{\rm{CoM}}(t)^2|$. 
The amplitude of the redshift oscillations is the difference between the upper and lower envelopes of the oscillations.
For the smaller black hole, the amplitude grows nearly linearly with the displacement of the CoM, almost independent of the mass ratio. 
For the larger black hole, the amplitude also also grows close to linearly, and we can observe a small dependence on the mass ratio. 
The oscillations in Fig.~\ref{correlation} are due to further modulations in the envelopes, and we find that these secondary modulations are consistent with the orbital eccentricity.

Although there are rigorous methods to correct for the CoM effects on the waveform \cite{Mitman:2021xkq}, these do not apply to the redshift data, since the latter is measured on the apparent horizons rather than at asymptotic infinity. 
An equivalent method to correct for the redshift would require an invariant notion of the surface gravity on a dynamical horizon, the definition of which is beyond the scope of this work. 
Instead we apply an empirical method to remove the oscillations. 
To estimate any possible bias introduced by our chosen method, we compared it with a $q=4$ simulation with the same initial parameters but with highly reduced CoM displacement. 
We found minimal differences between these case, below the error due to the choice of $\Omega$.
This comparison is in Appendix~\ref{sec:DetailedCoM}.

\begin{figure}[tb]
\includegraphics[width = 0.98\columnwidth]{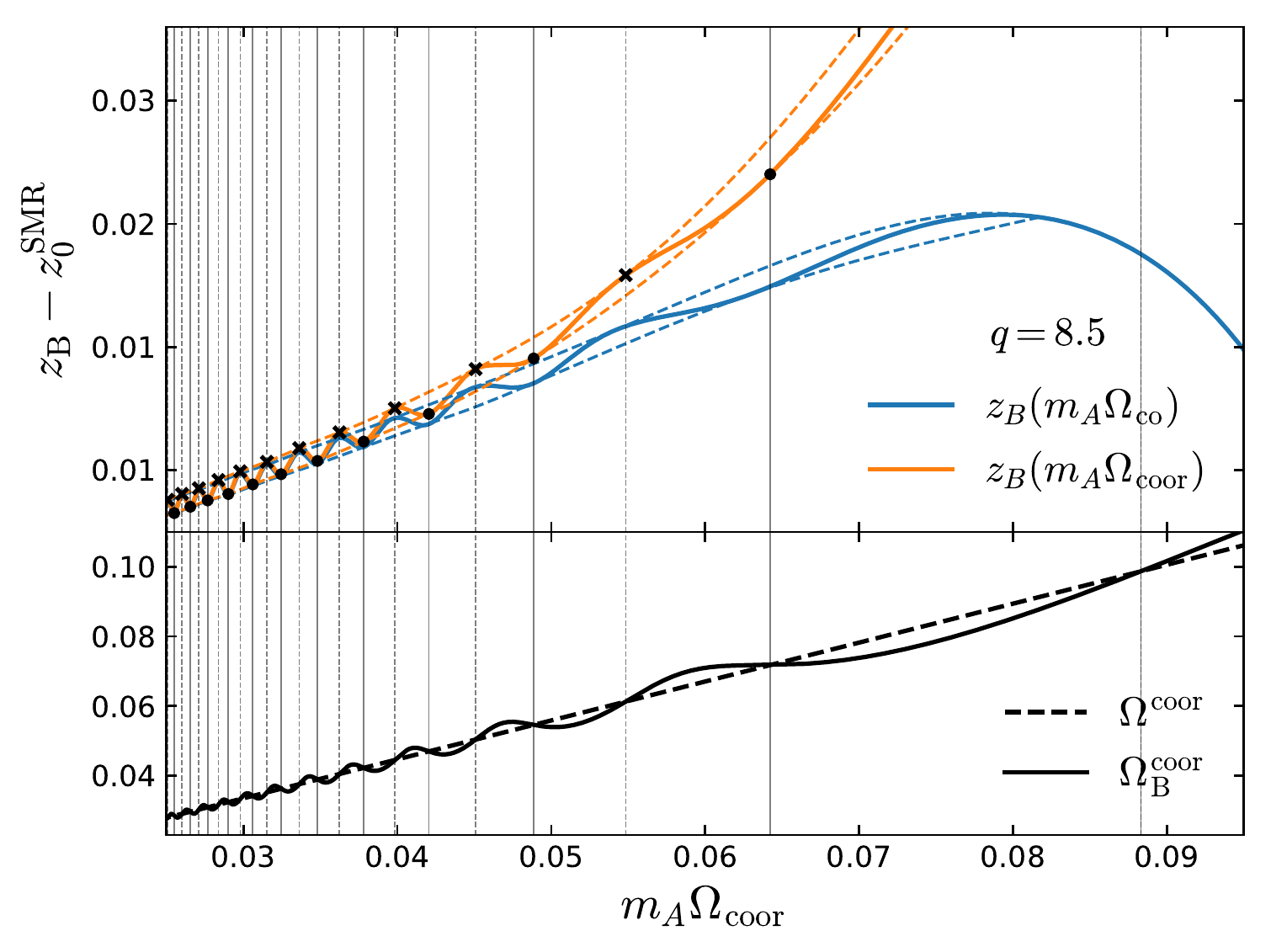}
\caption{Visualization of how the sampling works for the $q=8.5$ SKS simulation (with noticeable oscillations). 
Upper panel: raw redshift data and the corresponding envelopes obtained using the sampling method. 
We show the results for both $z_B(m_A\Omega_{\rm{co}})$ and $z_B(m_A\Omega_{\rm{coor}})$. 
Bottom panel: $\Omega_{B,\mathrm{coor}}$ and $\Omega_{\mathrm{coor}}$ as a function of $m_A \Omega_{\rm{coor}}$. 
The intersection of $\Omega_{B,\mathrm{coor}}$ with the diagonal gives the $t_i$ used to generate the envelopes. 
Quartic splines are constructed from $\Omega_{\rm{co}}(t_i)$ and $\Omega_{\rm{coor}}(t_i)$.}
\label{sampling}
\end{figure}

To remove the oscillations we tried three different methods.
Our preferred method is a sampling method which finds the upper and lower envelopes of the oscillations and takes their mid-line as the corrected version of the redshift. 
For this approach, we find the envelopes by solving for the roots $t_i$ of the function
\begin{align}
\Omega_{\text{coor}}(t_i)-\Omega_{a,\text{coor}}(t_i) \,,
\end{align}
for each black hole, where
\begin{align} 
\label{omega}
\Omega_{a,\text{coor}}=\frac{\rvert \vec{x}_a \times \dot{\vec{x}}_a \rvert}{x_a^2}\,.
\end{align}
Sampling the redshift at $t_i$ gives us points that empirically track the envelopes remarkably well. 
As seen in Fig.~\ref{sampling2}, these roots select out those instants in the orbit when we expect the coordinate velocity to be aligned or anti-aligned with the overall CoM drift. 
However, the magnitude of the redshift oscillations is larger than expected from considering these modulations to be caused by CoM velocity, and so this does not offer a complete explanation for the practical success of this method.
With the points $t_i$ in hand, we use quadratic interpolation to get the envelopes. 
The value of $z$ at any other $t$ is given by the mid-line between the two interpolants. 
This procedure can be visualized in Fig.~\ref{sampling} for the mass ratio $q=8.5$, which has noticeable redshift oscillations, as a function of both $m_A\Omega_{\rm{co}}$ and $m_A\Omega_{\rm{coor}}$. 

In addition to the sampling method, we tried removing the oscillations in $z$ using a rolling average over an orbital cycle and a rolling linear fit over an orbital cycle.
We find that the sampling method performs best, and we use it for our fiducial analysis, but we present the other methods and a comparison between them and our $q=4$ SHK simulation in Appendix~\ref{sec:DetailedCoM}.
For all of the methods we use the local measure of the orbital frequency $\Omega=\Omega_{\rm{coor}}$ to correct the redshift. 
This choice was made because the oscillations were found to correlate with $\Omega_{\rm{coor}}$ better than with $\Omega_{\rm{co}}$.
However, when analyzing the corrected $z(t)$ as a function of $\Omega(t)$, we choose $\Omega=\Omega_{\rm{co}}$. 
The difference between the two choices ($\Omega_{\rm{co}}$ vs $\Omega_{\rm{coor}}$) is shown in the upper panel of Fig.~\ref{sampling} after subtracting $z_0^{\rm{SMR}}$.
Our choice of $z(m_A\Omega_{\rm{co}})$ is further justified by the behaviour at large frequency, where we expect the redshift to decrease as the smaller black hole plunges into the larger, as occurs in the geodesic limit.

Figure~\ref{envelopes} shows the resulting envelopes for a range of mass ratios as a function of $m_A\Omega_{\rm{co}}$.
We observed that SKS simulations show more oscillations than the SHK. 
This is expected since SKS initial data has more junk radiation which can add initial momentum to the binary than seen in SHK simulations~\cite{Varma:2018sqd}.
For the SKS simulations, oscillations also tend to be more prominent for lower mass ratios (higher $q$). 
The small modulations on the envelopes, more noticeable at low frequencies, are well correlated with the presence of small initial eccentricity.

\begin{figure}[tb]
\includegraphics[width = 0.98\columnwidth]{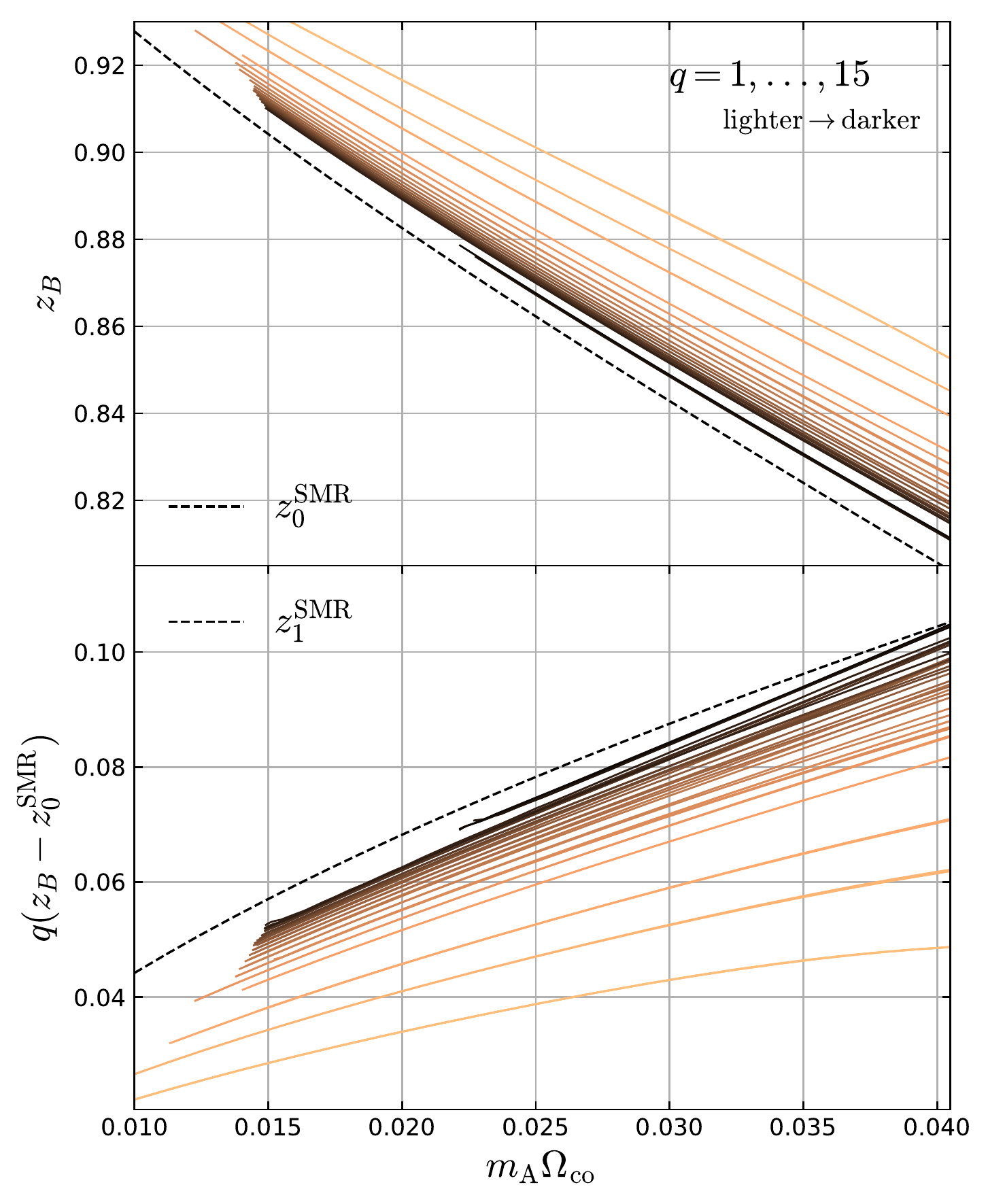}
\caption{\label{correctedzB} Upper panel: Corrected $z_B$ for all mass ratios in Table~\ref{table:1}  and the leading (geodesic) SMR prediction.  The two highest resolutions for each simulation are plotted in the figure. Bottom panel: Corrected $z_B$ after subtracting the leading (geodesic) SMR prediction and multiplying by the expected $q$ scaling. $z_1^{\rm{SMR}}$ (dashed line) is the prediction from self-force calculations. The two highest resolutions for each simulation are plotted in the figure.}
\end{figure}

\begin{figure}[tb]
\includegraphics[width = 0.98\columnwidth]{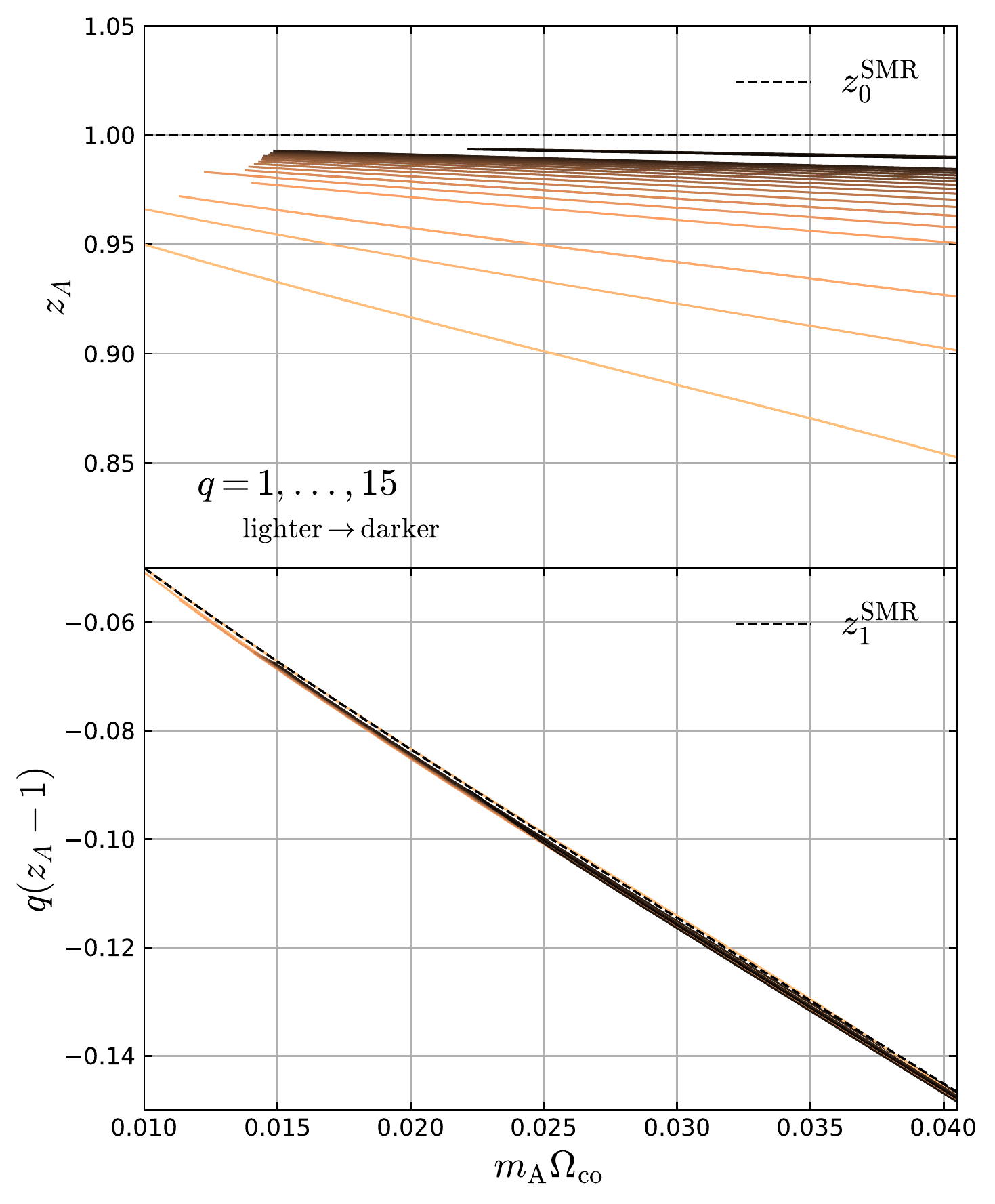}
 \caption{\label{correctedzA} Upper panel: Corrected $z_A$ for all mass ratios in Table~\ref{table:1}  and the leading SMR prediction.  The two highest resolutions for each simulation are plotted in the figure. Bottom panel: Corrected $z_A$ after subtracting the leading (geodesic) SMR prediction and multiplying by the expected $q$ scaling. $z_1^{\rm{SMR}}$ (dashed line) is the prediction from \cite{Gralla:2012dm} (after neglecting the small spin). The two highest resolutions for each simulation are plotted in the figure.}
\end{figure}

Figures~\ref{correctedzB} and \ref{correctedzA} show the corrected redshift curves that are used in the SMR analysis alongside the SMR predictions.
The lower panels in those figures show the resulting curves after subtracting the leading SMR prediction and dividing by the mass ratio.
Note that the GSF predictions give $z_a$ as a function of the dimensionless frequency $m_A \Omega$, not the natural frequency $M \Omega$ of the simulations.
Thus when plotting multiple simulations together, a fixed $m_A \Omega_{\text{co}}$ represents a later portion of the simulation for more equal-mass binaries than for lower mass ratios.
Meanwhile, our methods cannot capture the plunge dynamics near ISCO.
Therefore, the range of frequencies we can treat is limited by when our equal-mass simulations approach the ISCO frequency.
On the other hand, we can provide results to higher frequencies when plotting against $m \Omega_{\text {co}}$.
Similarly, the lowest frequencies we can access are controlled by the lowest frequencies achieved across our simulations, which is limited by the simulations with lowest mass ratio (highest $q$).

In the upper panel of Fig.~\ref{correctedzB} we see the clear clustering of the curves toward the test particle predictions as we move to smaller mass ratios (larger $q$).
In the lower panel, we similarly see the convergence of these curves to the known first SMR correction computed from GSF at low mass ratios, with the difference between each simulation and the dashed curve illustrating as-yet-unknown second-order and higher SMR corrections.
Meanwhile, in Fig.~\ref{correctedzA} we again see the convergence in the upper panel to the leading-order, trivial prediction $z_{A,0}^{\text{SMR}} = 1$ for the larger black hole.
The lower panel shows simultaneously the convergence to the first SMR correction of $z_A$, and the remarkable fact that higher order corrections are numerically very small.
In the next section, we show that these extracted redshift factors have a consistent SMR expansion in powers of the small mass ratio $\epsilon$, compare them to PN and GSF predictions, and measure both non-adiabatic corrections to these predictions and unknown, higher order terms in the SMR expansion.

\section{Results}
\label{sec:Results}

\subsection{SMR limit in NR and PN comparisons}
\label{sec:PN}

As discussed in Sec.~\ref{sec:GSFredshift}, the redshift factor for a point particle on a circular orbit is a well-defined invariant of the conservative dynamics. 
In a PN expansion, one can derive it from the 3.5PN metric \cite{Blanchet:2013haa} after neglecting the radiation reaction terms at 2.5PN and 3.5PN order and using the definition
\begin{align}
    z_a=\frac{d\tau_a}{dt}=\sqrt{-g_{\mu\nu}(x_a) u_a^\mu u_a^\nu} \,,
\end{align}
where $x_a$ is the coordinate location of each point particle.
In the SMR approximation, the conservative redshift is given by the time-symmetric component of the GSF metric perturbation sourced by the circular geodesic of frequency $\Omega$. 
When comparing both conservative approximations, PN and SMR successfully converge to one another in their respective domains of validity. 

\begin{figure}[tb]
\includegraphics[width = 0.98\columnwidth]{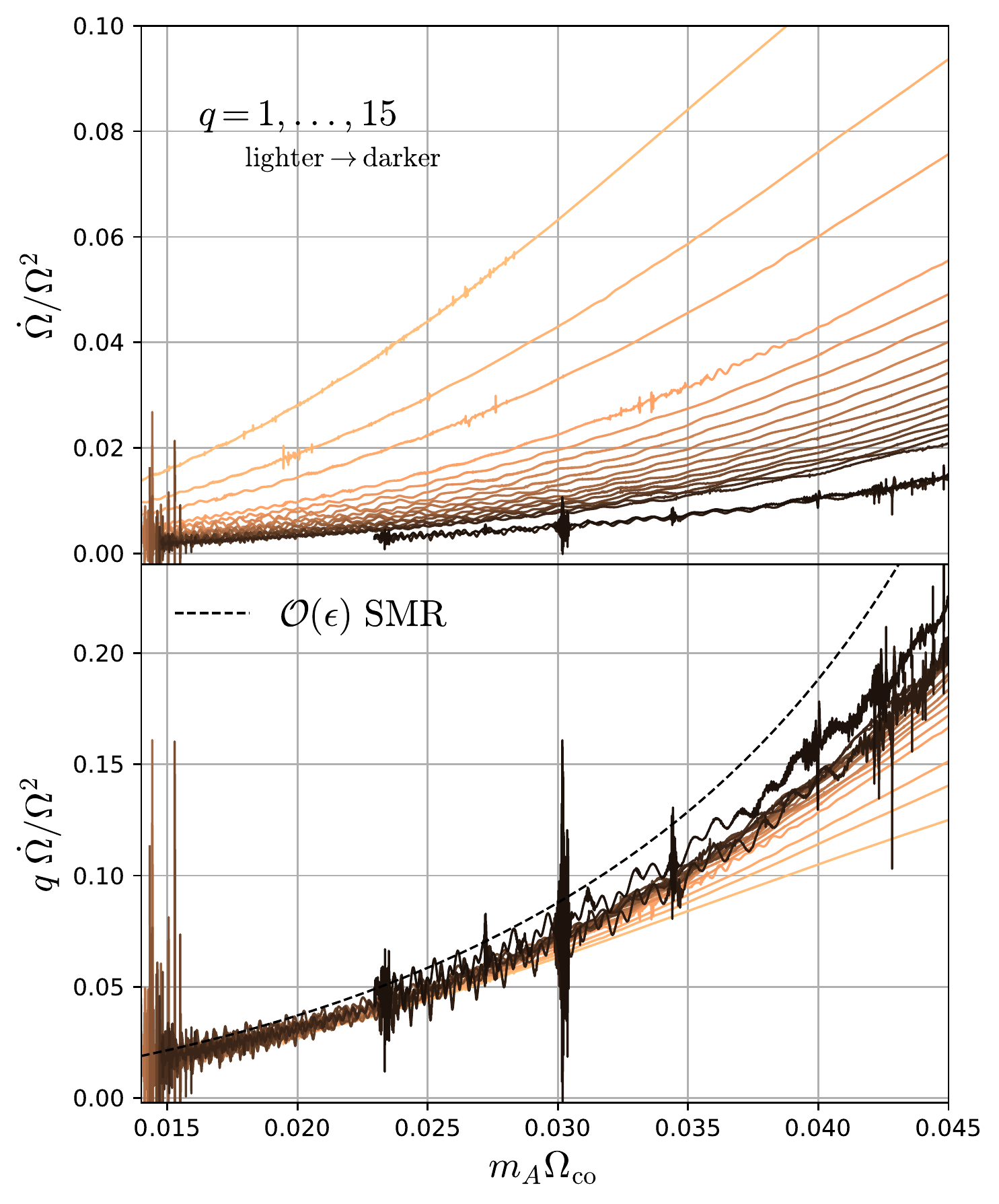}
\caption{Upper panel: Non-adiabaticity as measured by $\dot{\Omega}/\Omega^2$ using $\Omega=\Omega_{\rm{co}}$. 
As expected, it grows with frequency and is smaller for lower mass ratios. 
Bottom panel: Same parameter re-scaling by $q=\epsilon^{-1}$. The overlapping of the curves at low frequency indicates that $\dot{\Omega}/\Omega^2\sim\mathcal{O}(\epsilon)$. 
We also show the $\mathcal{O}(\epsilon) \ \rm{SMR}$ prediction from expanding $\dot{\Omega}=(dE/dt)(\partial E / \partial \Omega)^{-1}$ to first order in $\epsilon$.  We used the energy flux data for circular orbits provided by \cite{BHPToolkit}.}
\label{non-adiabaticity}
\end{figure}

In NR simulations such splitting between conservative and dissipative dynamics is not available, and our redshift definition in Eq.~\eqref{zfromkappa} can only coincide with the conservative redshift in the adiabatic approximation. 
A measure of the non-adiabaticity in our simulations is given by the quantity $\dot{\Omega}/\Omega^2$. 
Figure~\ref{non-adiabaticity} shows the value of this quantity for the range of frequencies and mass ratios used in this analysis. 
Non-adiabatic effects grow with frequency and they vanish in the SMR limit as expected. 
The bottom panel of Fig.~\ref{non-adiabaticity} confirms that they scale as $\mathcal{O}(\epsilon)$.
From this, we see that non-adiabatic corrections to the SMR predictions are expected to arise at $O(\epsilon)$, at a level of several percent.
To confirm this expectation, we compare our numerical redshifts to PN predictions and their SMR limit.

\begin{figure}[tb]
\includegraphics[width = 0.98\columnwidth]{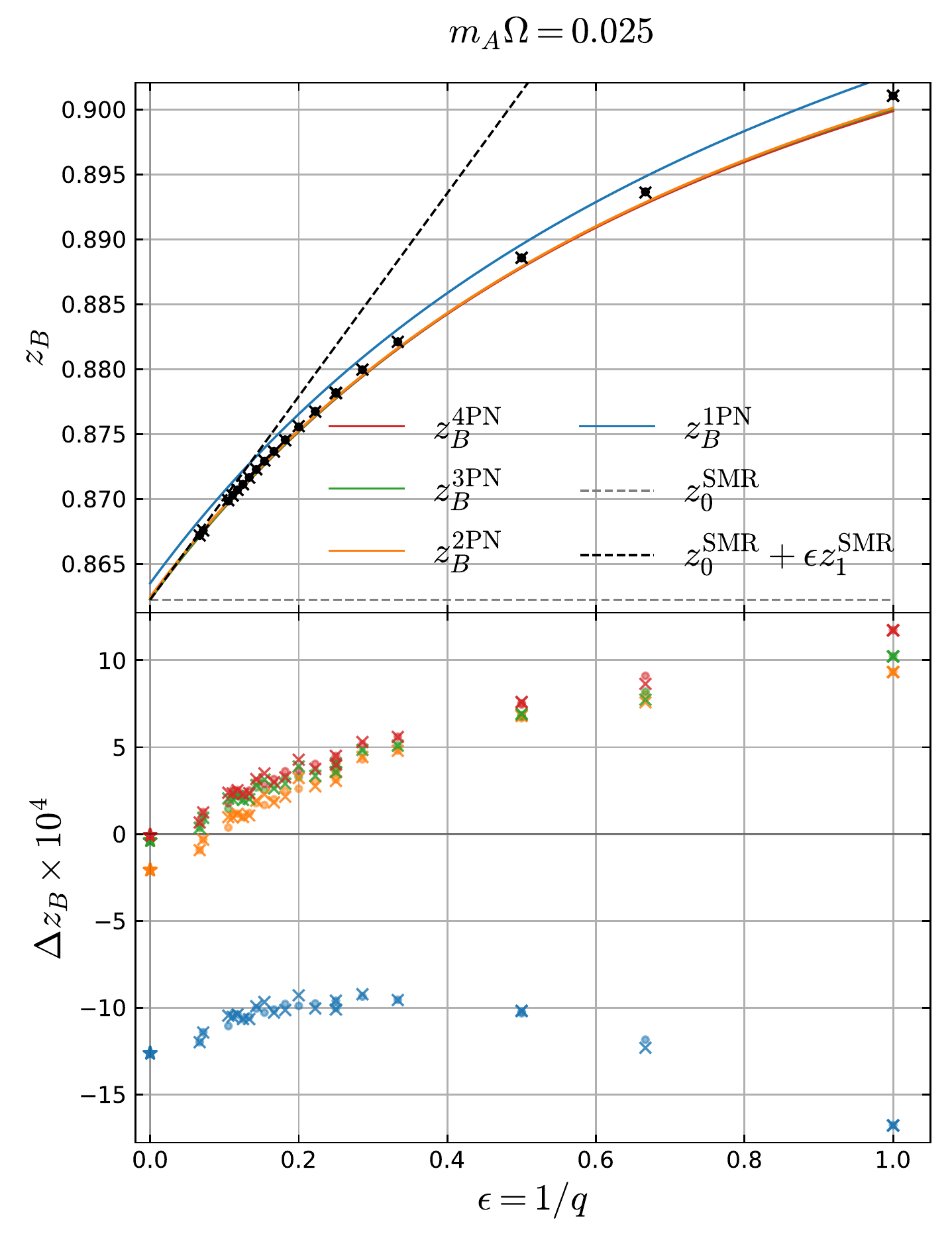}
\caption{\label{qPNresidualsB} Upper panel: Redshift of the smaller black hole across mass ratios, PN predictions (solid lines) and SMR predictions (dashed lines). Bottom panel: PN residuals as a function of the mass ratio for the smaller black hole. The points marked with a star correspond to the expected PN minus SMR value at $\epsilon=0$. Dots correspond to highest resolution data while crosses are the lower available resolution data.}
\end{figure}

\begin{figure}[tb]
\includegraphics[width = 0.98 \columnwidth]{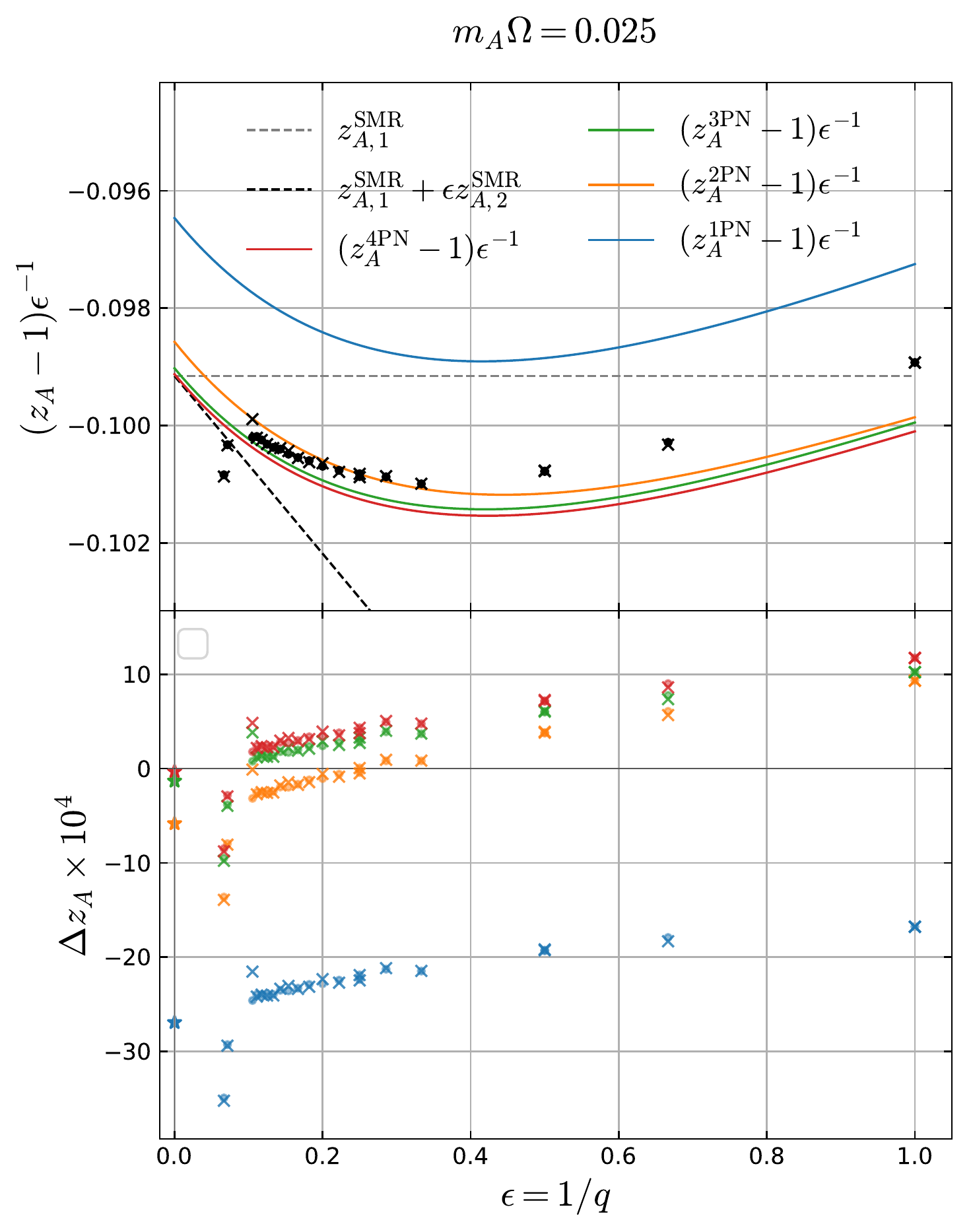}
\caption{\label{qPNresidualsA}Upper panel: Reference quantity $(z_A-1)\epsilon^{-1}$ of the larger black hole across mass ratios (points), PN prediction (solid lines) and SMR predictions (dashed line).
Bottom panel: PN residuals as a function of the mass ratio for the smaller black hole. The points marked with a star correspond to the expected PN minus SMR value at $\epsilon=0$. Dots correspond to highest resolution data while crosses are the lower available resolution data.}
\end{figure}

Figures~\ref{qPNresidualsB} and~\ref{qPNresidualsA} show the NR redshifts as a function of the mass ratio at a reference $m_A\Omega=0.025$ and the corresponding PN predictions from Eq.~(4.2) in \cite{LeTiec:2017ebm} (upper panels).
We plot each PN order separately as well as their residuals $z_a^{\rm{NR}}-z_a^{\rm{PN}}$ (bottom panels). 
We selected this $m_A\Omega$ as the smallest frequency that allows us to also show the $q=14$ and $q=15$ redshifts after their relaxation times, since these simulations start at a higher frequency.

Figure~\ref{qPNresidualsB} shows no improvement in the 3PN and 4PN residuals over the 2PN. 
These PN predictions are conservative, and so this is not surprising.
Dissipative effects, arising first at the 2.5PN order, can begin to contaminate the extracted $z_B$ in our NR simulations, and so we cannot expect our residuals to improve past 2PN at finite mass ratios.
However, a known feature of the 3.5PN equations of motion for circular, non-spinning binaries is that in the limit $\epsilon \rightarrow 0$ the 2.5PN and 3.5PN terms, entering at order $\mathcal{O}(\nu)$, vanish~\cite{Blanchet:2013haa}. 
One is left with the ``even" PN terms which contain $\mathcal{O}(\nu^{0})$ terms. 
Thus, the even PN series alone must converge to the geodesic limit. 
This means that as we approach $\epsilon=0$, radiation-reaction is suppressed and the residuals between $z_B^{\rm{NR}}$ and $z_B^{\rm{PN}}$ should match the residuals between $z^{\rm{SMR}}_{B,0}$ and $z_B^{\rm{PN}}$ from analytic theory. 
These last are marked for each PN order with a star on the vertical axis of the bottom panel of Fig.~\ref{qPNresidualsB}. 
We can see that the trend in the residuals is indeed consistent with their expected value at $\epsilon=0$.
This shows that the NR data approaches the SMR prediction for smaller $\epsilon$.
Further, we expect that NR simulations with an even smaller mass ratio than those presented here would follow the trend in our residuals, so that eventually the 3PN and 4PN predictions would out perform 2PN.
We also note that for the finite mass ratios and for all frequencies of our analysis, the 2PN prediction for $z_B$ always outperforms the SMR prediction from GSF.

For the larger black hole the situation is slightly different. When $\epsilon= 0$ all PN orders other than 0PN vanish and $z_A=1$ (the black hole is at rest). 
Thus we show the residuals for $z_A$ after subtracting the 0PN result and multiplying by $\epsilon^{-1}$ in Fig.~\ref{qPNresidualsA}. 
In the limit $\epsilon\rightarrow 0$, $(z^{\rm{SMR}}_A-1)\epsilon^{-1}$ converges to a finite value for each PN order. 
These are again marked with a star on the vertical axis of the bottom panel of Fig.~\ref{qPNresidualsA}. 
Except for the $q=14$ and $q=15$, the trend in these residuals is again consistent with the first-order SMR prediction in \cite{Gralla:2012dm} as well as the second-order prediction in Eq.~\eqref{zA2}. 
The figure also shows the PN prediction converging towards the SMR prediction in the limit of small mass ratio.

The residuals for the $q=14$ and $q=15$ depart from the general trend. 
These simulations are clear outliers. 
Although they were carried out at lower resolution settings than our other simulations at high $q$, one can argue that the nearly equal residuals $\Delta z_A$ between Levs 2 and 3 indicates this is not due to resolution effects.
However, it may be that much higher resolution near the larger black hole is required to accurately extract $z_A$. 
We also note that these two simulations are our only simulations with SHK initial data at large $q$, and a careful investigation of $z_A$ at early times $t\sim 1000M$ reveals transient behavior that differs from that of our other simulations.
This may indicate some additional effect present in these two simulations that we have not been able to identify or account for.
Due to their clear departure from the trends of the residuals of the other simulations,
we omit them from our SMR fits in Sec.~\ref{sec:fits}.
In the future, higher resolution simulations at high $q$, and using alternative initial data prescriptions, may provide key confirmation of our findings at lower $q$.

We stress that the results of these PN comparisons are consistent with the appearance of dissipative effects at 2.5PN. 
To account for dissipation effects of $\mathcal{O}(\epsilon)$, when comparing to conservative SMR predictions in the next section we must adopt an agnostic strategy. 
Instead of subtracting the successive SMR predictions to our NR data and analysing their residuals, we fit the NR redshift data directly to a series expansion in $\epsilon$, only afterward comparing the resulting coefficients of the fit to the SMR adiabatic prediction.
As we shall see, the leading SMR prediction is recovered to great accuracy, which allows us to repeat the fit after calibrating with the leading order prediction.
This is not the case at the next-to-leading order.

\subsection{Extracting the SMR approximation from NR for the smaller black hole}
\label{sec:fits}

In the SMR approximation the redshift of the small black hole is written as a series expansion in integer powers of the mass ratio $\epsilon$ of the form
\begin{align} 
z_B=\sum_{k=0}^N \epsilon^k z_{k}(m_A\Omega)\,.
\label{fit}
\end{align}
The leading term in this series, $z^{\rm{SMR}}_0(m_A\Omega)$, corresponds to the smaller black hole's ``effective'' center of mass moving on a circular geodesic, and is given by Eq.~\eqref{z0}. 
To find the linear correction $z^{\rm{SMR}}_1(m_A\Omega)$, one typically solves the linearized Einstein equation sourced by the circular geodesic. 
This first-order redshift is given by Eq.~\eqref{z1tilde}. 
Going beyond linear order implies solving successive higher order approximations to the Einstein equation. 

Note that outside of the radius of convergence of Eq.~\eqref{fit}, there is no guarantee that a fit of the data to a power series should recover the SMR approximation. 
In other words, we should be cautious in extrapolating the NR data to $\epsilon =0$ and drawing conclusions about the SMR coefficients from this.
Only if Eq.~\eqref{fit} converges to the exact result for all mass ratios we are guaranteed to recover the ``true'' coefficients from NR fits. 
Our results suggest that this is in fact the case.
To validate our extrapolation method we provide convergence tests in Appendix~\ref{sec:ConvergenceTest}.

\begin{figure}[tb]
\includegraphics[width = 0.98\columnwidth]{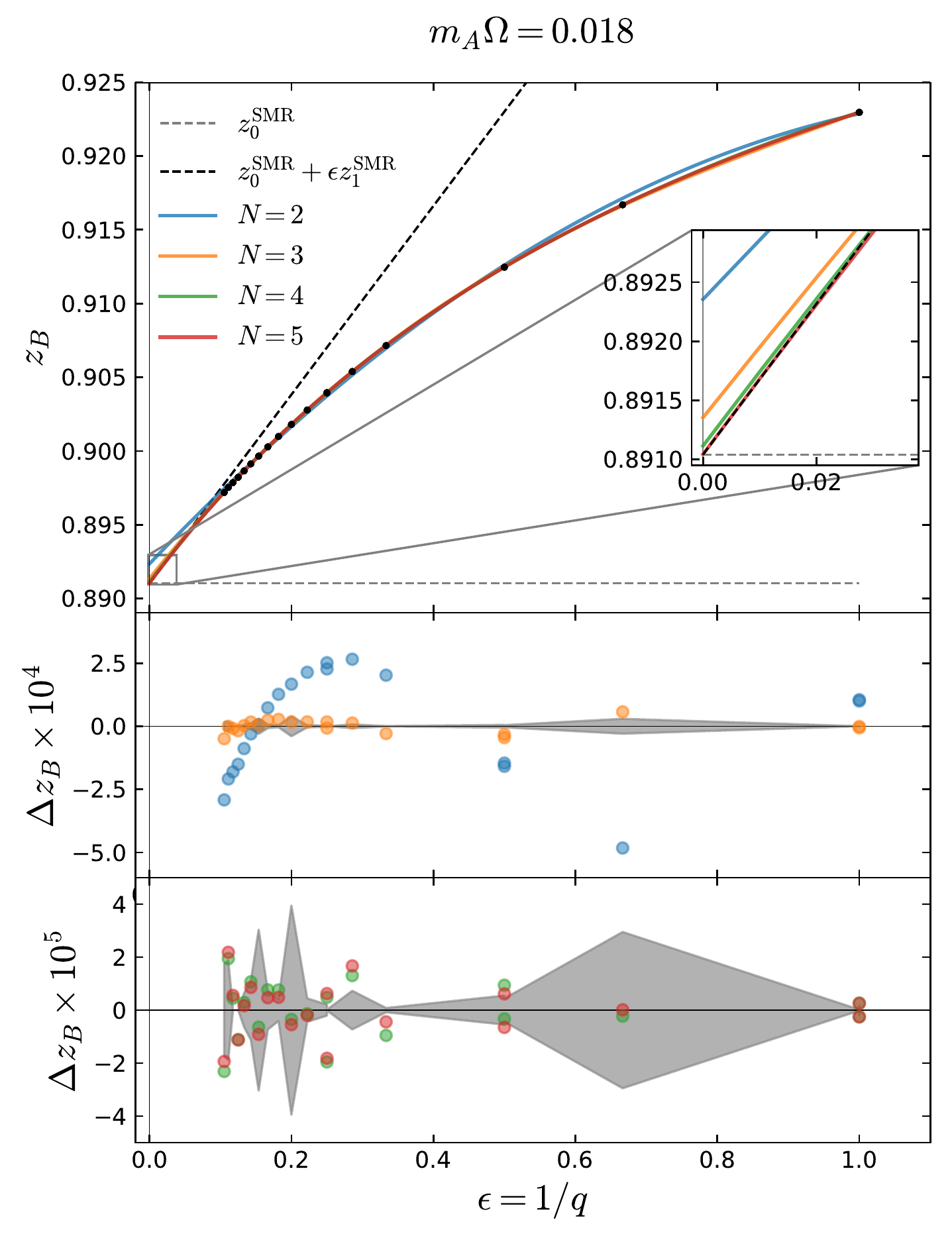}
\caption{\label{N4N5} Upper panel: NR fits for $N=2,3,4,5$ at a fixed $m_A \Omega=0.018$ (solid lines) and the SMR predictions (dashed lines). Middle panel: residuals of the $N=2,3$. Bottom panel: residuals of the $N=4,5$. Shaded area corresponds to the difference between the two highest resolutions available.}
\label{fitplot}
\end{figure}

\begin{figure}[tb]
\includegraphics[width = 0.98\columnwidth]{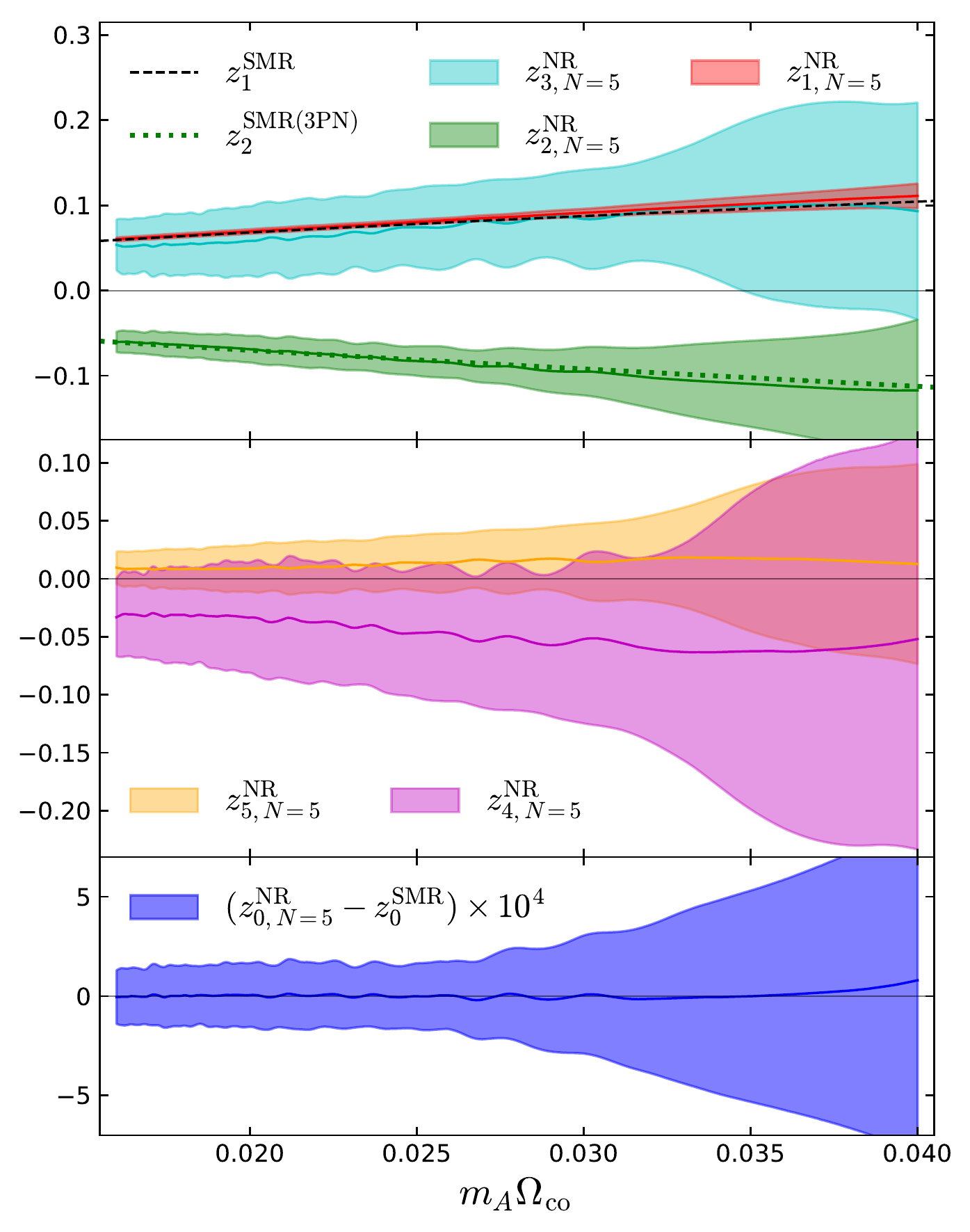}
\caption{\label{qfit_coef} Upper two panels: extracted coefficients from the NR fit (colored bands) and the SMR predictions. Dashed line corresponds to the 21.5PN formula for $z_{B,1}^{\rm{SMR}}$. Dotted lines represent the SMR coefficient generated from the 3PN redshift series. Bottom panel: difference between the leading order SMR prediction and the extracted $z_0^{\rm{NR}}$ coefficient from the NR fit with $N=5$. The color bands correspond to a conservative error estimated by the range of repeated calculations using: a lower resolution, $N=4$ fit and one-sigma deviation from the $N=5$ fit.}
\end{figure}

In order to extract the SMR coefficients from our NR data we do the following: at a fixed $m_A\Omega$, we perform a least squares fit of the redshift to Eq.~\eqref{fit} for different values of $N$. 
Figure~\ref{N4N5} shows the results of these fits for the (lowest available) reference frequency, $m_A\Omega=0.018$. 
The $N=1,2,3$ fits clearly leave behind features in the data, seen as structures in the residuals in the middle panel of Fig.~\ref{N4N5}.
The residuals for the $N=4,5$ fits meanwhile do not seem to favor one over the other. 

It is tempting to select $N=4$ to avoid overfitting the data. 
However, a more careful study of the convergence of the fit coefficients with $N$ shows that they converge exponentially for our highest resolution simulations up until $N=5$.
Beyond this, we do not see convergence with increasing $N$. 
In addition, we employ several metrics of goodness of fit (AIC, BIC, and the adjusted $R$-squared tests). 
These show improvement until $N=4$, with no improvement beyond this.
Our final, decisive criterion is seen in the inset of the top panel of Fig.~\ref{N4N5}: we continue to see convergence of the fitted $z_B$ to the geodesic prediction in the $\epsilon \to 0$ limit until $N=5$.
For these reasons, we conclude that $N=5$ terms are required in our SMR fit.
We present our convergence tests in Appendix~\ref{sec:ConvergenceTest}.

Figure~\ref{qfit_coef} shows the extracted values of the coefficients for $N=5$ fit for $z_B(m_A\Omega)$. 
The error bands correspond to the largest of: the range in variation in these coefficients obtained by repeating the fit using a lower resolution while keeping $N=5$ and in repeating the fit at the highest high resolution but using $N=4$; and the one-sigma deviation obtained from the high resolution ($N=5$) least squares fit.
These fits are one of the primary results of our study.

Remarkably, for the range of frequencies analyzed here, the leading order coefficient $z^{\rm{NR}}_0$ of the fit agrees at the level of a $10^{-5}$ relative difference with $z^{\rm{SMR}}_0$.
The next coefficient, $z^{\rm{NR}}_1$, deviates from the  SMR prediction at the level of non-adiabatic contributions which are $2 \% \lesssim \dot{\Omega}/\Omega^2 \lesssim 10\%$ for the equal mass binary. 
We see that $z^{\rm{NR}}_2$, which is currently unavailable from GSF calculations, agrees at the same level with the SMR prediction generated from the 3PN series.
We also find that higher order coefficients alternate in sign and decrease in magnitude. 

\subsection{Result of calibrating the fits for $z_B$ with $z_0^{\rm{SMR}}$}

The agreement between $z_0^{\rm{NR}}$ and $z_0^{\rm{SMR}}$ suggests that we can use the leading SMR result to calibrate our fit by fitting instead the residuals after subtracting $z_0^{\rm{SMR}}$. 
This is equivalent to forcing the fits through the SMR prediction at $\epsilon=0$. This calibration is further justified by noting that non-adiabatic effects are not expected to have an effect at leading order, $\dot{\Omega}/\Omega^2=\mathcal{O}(\epsilon)$.

\begin{figure}[tb]
\includegraphics[width = 0.98 \columnwidth]{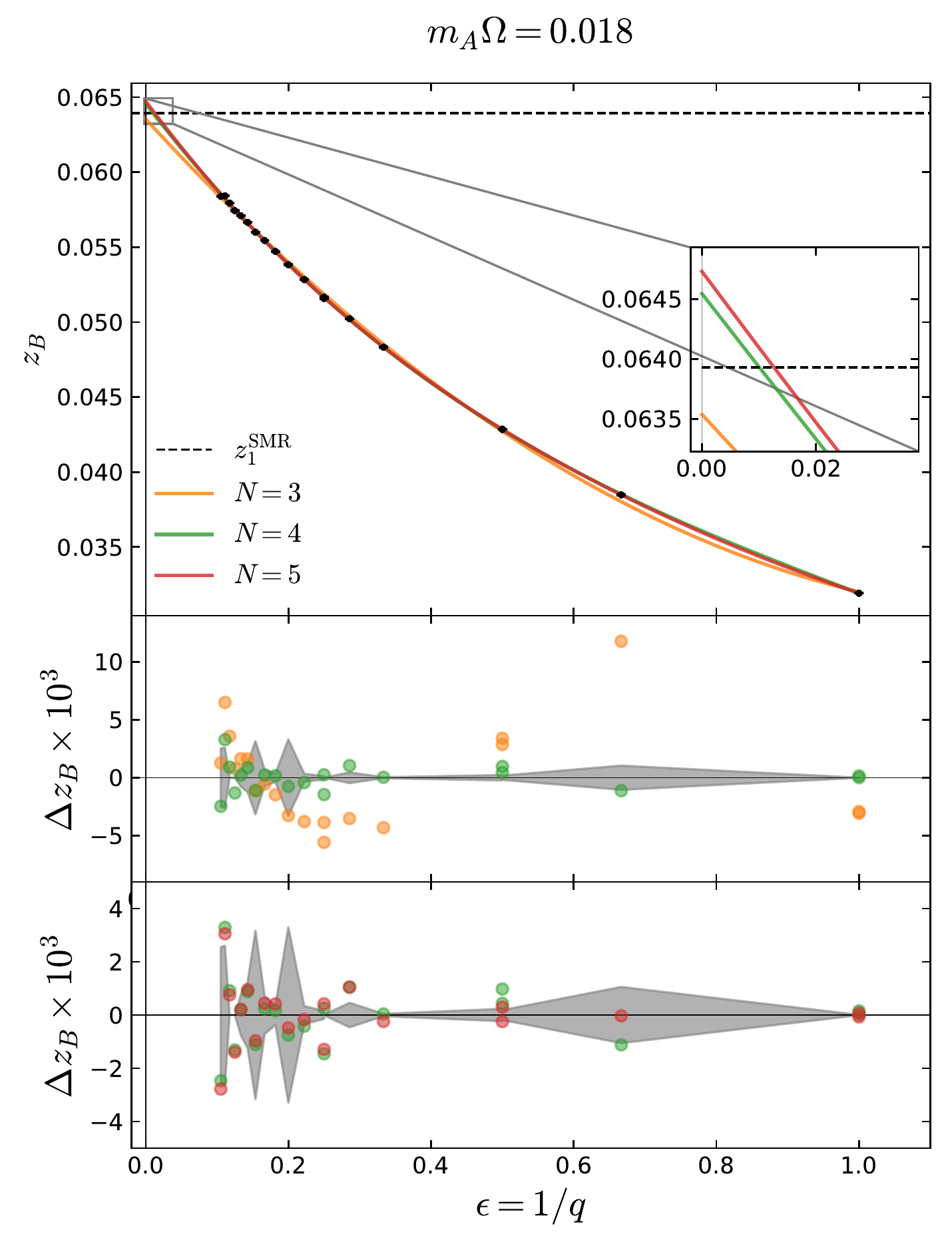}
\caption{ Upper panel: Calibrated NR fits for $N=3,4,5$ at a fixed $m_A \Omega=0.018$ (solid lines) and the SMR prediction (dashed line). Middle panel: residuals of the $N=3,4$ fits. Bottom panel: residuals of the $N=4,5$ fits.  Shaded area corresponds to the difference between the two highest resolutions available.}
\label{qfit_calib}
\end{figure}

\begin{figure}[tb]
\includegraphics[width = 0.98 \columnwidth]{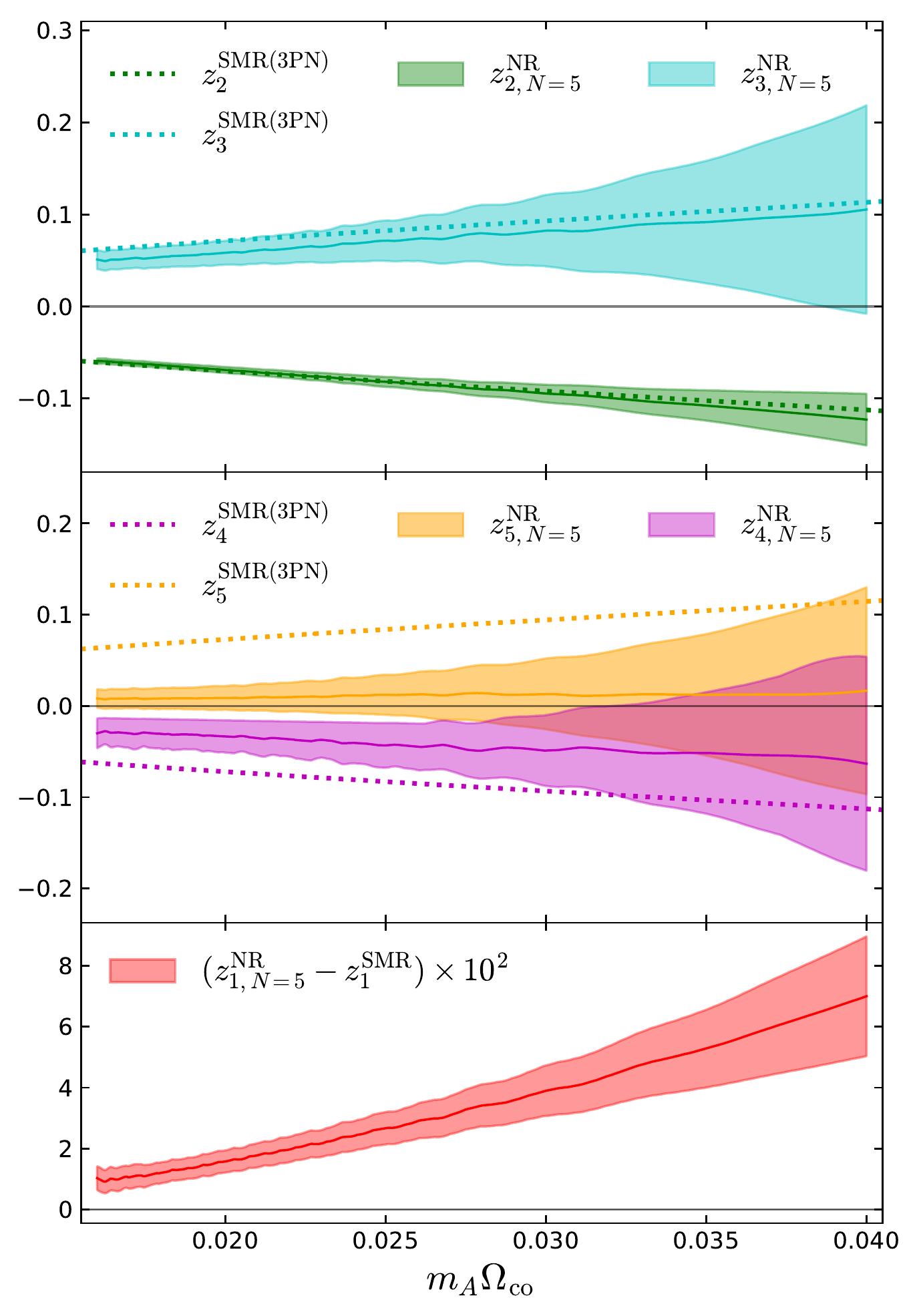}
\caption{ Upper two panels: extracted coefficients from the calibrated NR fit (colored bands) and the SMR predictions (dotted lines) generated from the 3PN series. Bottom panel: relative difference between the $z^{\rm{SMR}}_1$ prediction and the extracted $z^{\rm{NR}}_1$ coefficient. The color bands correspond to a conservative error estimated by the range of repeated calculations using: a lower resolution, $N=4$ fit and one-sigma deviation from the $N=5$ fit.}
\label{qfit_calib_coeff}
\end{figure}

Figure~\ref{qfit_calib} shows the fit to $(z_0^{\rm{NR}}-z_0^{\rm{SMR}})\epsilon^{-1}$. 
The intercept of the new fit corresponds to $z_1^{\rm{NR}}$. 
The extracted coefficients as a function of $m_A\Omega$ after this calibration are given in Fig.~\ref{qfit_calib_coeff}. 
The error bars are estimated in the same way as for the uncalibrated fit.
After calibrating the fit, the predicted coefficients don't change significantly, however the error bands are significantly reduced. 
From the calibrated fit one can more confidently see that there is a deviation from the $z^{\rm{SMR}}_1$ conservative prediction. 
This percent-level deviation is consistent with non-adiabatic effects of the same order. 
This is also the reason why we do not further calibrate our fits using $z^{\rm{SMR}}_1$.
Finally, although they agree within the NR error bars, the difference between $z^{\rm{NR}}_2$ and $z^{\rm{SMR(3PN)}}_2$ and $z^{\rm{NR}}_3$ and $z^{\rm{SMR(3PN)}}_3$ is also consistent with this non-adiabatic effect. 
The higher order terms $z_4^{\rm{NR}}$ and $z_5^{\rm{NR}}$ extracted here show a clear departure from their corresponding 3PN prediction. Although we do not have the same level of confidence in our fits to these higher-order coefficients, we speculate that they are consistent with an alternating, convergent series even for $\epsilon=1$. 
However, more accurate measurements of higher-order coefficients would be required to establish that.
As an aside, we point out that this alternating-in-sign behaviour is reproduced by the SMR expansion of the conservative 3PN redshift series, shown as dashed lines in Fig.~\ref{qfit_calib_coeff}. 
There, as opposed to the NR result, successive higher order coefficients increase in magnitude.

\subsection{Re-expansion in the symmetric mass ratio $\nu$}

Previous comparisons between the SMR approximations and NR have suggested that a re-expansion of the SMR series in terms of the symmetric mass ratio $\nu$ can extend the regime of validity of the SMR series to comparable-mass binaries. 
Especially promising are indications that $\mathcal O(\epsilon)$ predictions provide good agreement with NR results.
A common feature of the quantities for which the $\nu$ re-expansion is effective is the symmetry under the exchange $m_A \leftrightarrow m_B$ (e.g.~binding energy, periastron advance, gravitational wave phase). 
The quantities $z_A(m_A\Omega)$ and $z_B(m_A\Omega)$ separately don't have this property.
However the sum of the two $Z \coloneqq z_A+z_B$ as a function of $m\Omega$ is invariant under this transformation. 
Moreover, Ref.~\cite{LeTiec:2017ebm} shows the first law implies that the redshift factors take the schematic form $z_a=f(\nu,x)\pm g(\nu,x)\sqrt{1-4\nu}$, with $+$ selecting the larger body and $-$ the smaller.
This suggests that the direct sum of the two redshifts cancels part of the $\nu$ dependence. 
These facts motivate us to explore the simple sum $Z$ in our simulations, and expand it in terms of $\nu$ as
\begin{align}
Z(m\Omega)\coloneqq z_A+z_B = \sum_{k=0}^N \nu^{k}Z_k \,.
\end{align}
We verified that an attempt to fit $z_a$ to an integer power series in $\nu$ shows no convergence, which is justified by the expected functional dependence of $z_a$ on $\sqrt{1-4\nu}$, as discussed above and in~\cite{LeTiec:2017ebm}.

The coefficients $Z_0$ and $Z_1$ are known in the SMR approximation and are given by Eqs.~\eqref{z0}, \eqref{z1tilde}, \eqref{zA0} and \eqref{zAletiec} after re-expanding them in terms of $m\Omega$. This requires taking into account an extra $\mathcal{O}(\epsilon)$ term from expanding $m_A=m(1-\epsilon)+\mathcal{O}(\epsilon^2)$ in $z^{\rm{SMR}}_{B,0}(m_A\Omega)$. This is exactly cancelled by the first order term $z^{\rm{SMR}}_{A,1}(m\Omega)$ due to the first law equality
 \begin{align}
    \frac{\partial z_B}{\partial m_A}=\frac{\partial z_A}{\partial m_B} \,,
 \end{align}
which is a convenient feature of $Z$.

\begin{figure}[tb]
\includegraphics[width = 0.98 \columnwidth]{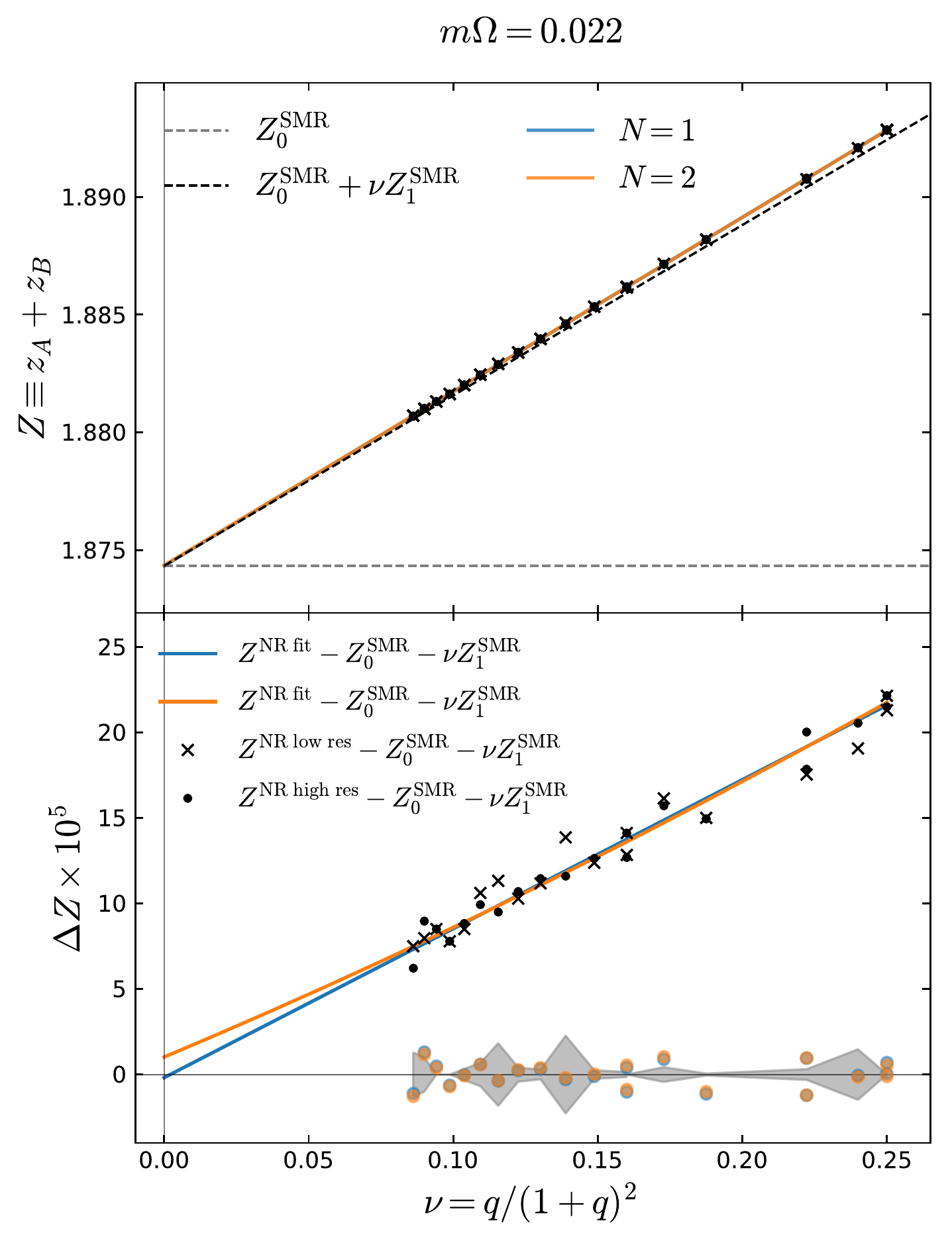}
\caption{\label{Zfit}Upper panel: NR fits for $N=1,2$ at a fixed $m\Omega=0.022$ (solid lines) and the SMR predictions (dashed lines). Bottom panel: residuals of the $N=1,2$ fits (blue and orange) and the residuals with respect to the SMR prediction (black). The SMR residuals follow a linear trend, illustrating the missing $\mathcal{O}(\nu)$ non-adiabatic contribution. Shaded area corresponds to the difference between the two highest resolutions available.
}
\end{figure}

\begin{figure}[tb]
\includegraphics[width = 0.98\columnwidth]{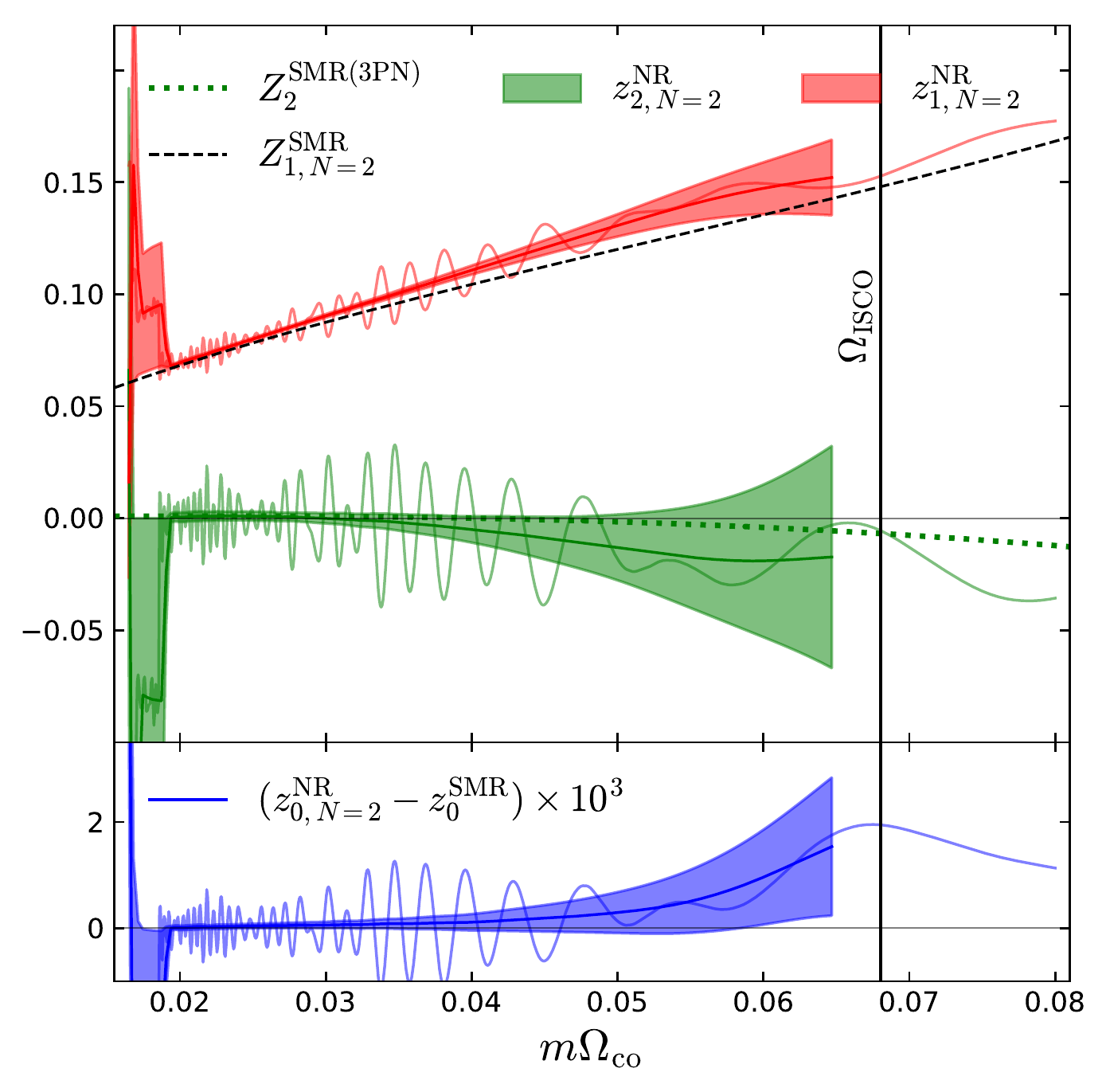}
\caption{\label{symmetric_coeff_fit} Upper panel: extracted coefficients from the NR fit (colored bands) and the SMR predictions (dashed and dotted lines). Bottom panel: difference between the leading order SMR prediction and the extracted $Z_0^{\rm{NR}}$ coefficient from the NR fit with $N=2$.  The color bands correspond to a conservative error estimated by the range of repeated calculations using: a lower resolution, $N=1$ fit and one-sigma deviation from the $N=2$ fit.}
\end{figure}

In Figure~\ref{Zfit} we show $Z$ as a function of the symmetric mass ratio at a reference $m\Omega=0.022$ (the lowest available for all mass ratios to have achieved relaxation time). 
The distinct linear trend in the data is a clear indication that the $\mathcal{O}(\nu)$ term alone captures most of the variation across all mass ratios, including the equal mass case, and that $\mathcal{O}(\nu^2)$ corrections are very small. 
From this and a convergence study of the coefficients $Z_k$ (Appendix~\ref{sec:ConvergenceTest}), we conclude that $N=2$ terms are sufficient for fitting our data.

Figure~\ref{symmetric_coeff_fit} shows the NR-predicted value of the coefficients truncating the series at $N=2$ for a range of frequencies up to the geodesic ISCO frequency, as well as the SMR predictions and our estimated error bands.
This lower-order polynomial fit for $Z$ is more stable to variations in the data, it gives comparable results even if we do not correct for the CoM-induced oscillations discussed in Sec~\ref{CoMoscillations}, and so
we also show the results of the fit if we don't correct for the oscillations (thin lines).
The midline trend of the oscillating, uncorrected coefficients is consistent with our corrected results.

We find good agreement between our fit coefficients and the leading $\mathcal O(\nu^0)$ prediction (bottom panel) up to a few cycles before merger, where we expect the quasi-circular approximation to break down. 
Similarly note the smallness of the $\mathcal{O}(\nu^2)$ coefficient and good agreement with its PN predicted value. 
Finally, note how NR predicted $\mathcal{O}(\nu)$ coefficient approaches the SMR prediction towards the more adiabatic region of the inspiral (lower frequencies) while the disagreement grows towards the less adiabatic region (higher frequencies).
The percent-level deviations from the conservative prediction are again consistent with the non-adiabaticity of the system as measured by $\dot{\Omega}/\Omega^2$ in Fig.~\ref{non-adiabaticity}.

\section{Conclusions}

In this work we give a detailed analysis of the Detweiler redshift factor as extracted from NR simulations using the surface gravity on apparent horizons.
We find that CoM motion imprints small oscillations in the extracted redshifts, and demonstrate a method for removing these effects.
With our corrected redshift factors, we give a detailed analysis showing that the NR results admit a consistent SMR expansion
including good agreement with analytic predictions and a clear measurement of the imprint of non-adiabatic effects on the redshift beginning at $\mathcal O(\epsilon)$.

By fitting the NR redshift to a series expansion in $\epsilon$ we recover with great accuracy the leading (geodesic) coefficient of the SMR approximation, when fitting the data to a $N=5$ polynomial. 
This gives us confidence in our SMR extraction procedure and allowed us to estimate the value of higher order coefficients.
In particular, we provide a prediction for the $\mathcal O(\epsilon^2)$ term in the SMR expansion, which has not been predicted by GSF methods to date.
When considering the symmetric quantity $Z\coloneqq z_A+z_B$, a re-expansion in $\nu$ is very effective, with the $\mathcal{O}(\nu)$ term capturing most variability in the data. 
In all cases, the disagreement that we observe at first order with the conservative SMR predictions is consistent with percent-level non-adiabatic effects at the same order.

Similar analysis of the SMR limit in NR have been done in the past.
Some of the earliest analyses treated quantities measured locally by the trajectories, such as the periastron advance~\cite{LeTiec:2011bk} or other ratios of orbital frequencies~\cite{Lewis:2016lgx}, but which can in principle be measured from gravitational waves at infinity.
Many others considered quantities encoded directly in the gravitational waves, such as the binding energy \cite{LeTiec:2011dp} and the gravitational wave phase \cite{vandeMeent:2020xgc}.
In contrast, the redshift in our simulations is computed from quantities measured on the black hole horizons, providing a direct point of comparison with local self-force calculations.

In the past, most analysis have dealt with direct comparisons between NR and conservative SMR predictions at fixed mass ratios. 
In all cases the SMR approximation worked remarkably well after re-expanding in terms of the symmetric mass ratio, even in the presence of dissipation. 
Our results further illuminate previous studies by showing the extent to which non-adiabatic effects are important when comparing to NR simulations. In particular, this analysis shows that they appear as percent contributions at first order in the SMR expansion of the redshift. 
However, some caution is merited in interpreting these non-adiabatic effects, since there is currently no preferred definition of the redshift factor in the presence of dissipation.
In particular we already made use of the connection between the redshift and the surface gravity which is strict only in the adiabatic approximation when defining our surface gravity in Eq.~\eqref{zfromkappa}. 

Recently, the two-timescale expansion \cite{Miller:2020bft} has provided a framework for the first direct calculation of the second order GSF, which was used to calculate the binding energy and energy flux~\cite{Pound:2019lzj,Warburton:2021kwk} for circular orbits in a Schwarzschild background to post-adiabatic accuracy. Both are in good agreement with NR. 
Using these results in combination with the Bondi-Sachs mass-loss formula, the corresponding (post-adiabatic) waveforms were generated in \cite{Wardell:2021fyy}.
As suggested in \cite{Miller:2020bft}, this formalism can potentially be used to calculate the second order local GSF. 
With it, one could calculate the first order correction to the orbital frequency and the second order redshift. 
Our work can provide a comparison for these results in the future.

There are a number of avenues to extend our results. 
In recent years there has been progress on formalizing and calculating the gauge invariant redshift for eccentric orbits, with comparisons between SMR and PN. 
A similar analysis to the one presented here will be given for eccentric orbits in a forthcoming paper.
Another natural and important extension of this work would be to investigate the redshift factor of spinning black holes in NR simulations, first for circular and then eccentric and precessing orbits.
The redshift factor contains important information about the conservative dynamics of these generic orbits, through its relation to the interaction Hamiltonian between the two bodies~\cite{Fujita:2016igj}.
Especially interesting would be the development of an improved measure of the redshift factor in numerical spacetimes, perhaps one that can account for the non-adiabatic effects we have measured.
A possible direction here is the calculation of the best approximate HKFV in the simulation, as is done in SpEC to measure black hole spins use approximate axial Killing vectors~\cite{Cook:2007wr,Lovelace:2008tw,Boyle:2019kee}.

Finally, a limitation of our analysis is that it neglects departures from non-circularity due to radiation-reaction.
The effect of radiation-reaction on the redshift is not considered in the GSF and PN calculations with a HKVF (exact or averaged in the case of eccentric binaries) used in our comparison, which can partly explain the disagreement we found at first order. A future comparison in the light of two-timescale calculations \cite{Miller:2020bft}, which necessarily take into account the secular change to $\Omega$, can help better understand the limitations of assuming a HKVF. 
The work in \cite{Compere:2021iwh} also showed that for a consistent matching between the adiabatic inspiral and the transition regime one should take into account the secular change to $\Omega$ during the adiabatic inspiral due to radiation-reaction.
We hope to extend the NR-SMR comparison to the transition dynamics in future work.
 
\acknowledgements
We would like to thank Serguei Ossokine for sharing with us the $q=3.5$ to $q=9.5$ SKS simulations used in this work. We also thank Jooheon Yoo for sharing the $q=15$ simulation and Keefe Mitman for sharing the redshift data from the recent SHK simulations of $q=1$ and $q=4$.
For the simulations used in this work, computations were performed on the Wheeler cluster at Caltech, which is supported by the Sherman Fairchild Foundation and by Caltech; and on Frontera at the Texas Advanced Computing Center~\cite{frontera}.
We also thank the developers of \texttt{Scri} \cite{Boyle:2015nqa,Boyle_scri_2020}, which was used to calculate the energy and angular momentum fluxes and angular velocity of the co-rotating frame. 
We thank the participants of a number of Capra conferences for valuable discussions on the topics of this work over several years, especially Takahiro Tanaka, Adam Pound, Leor Barack, Abraham Harte, Soichiro Isoyama, Eric Poisson, and Alexandre Le Tiec.
S.N.A.~and A.Z.~are supported by NSF Grant Number PHY-1912578. 
M.G.~is supported by NSF Grant Number PHY-1912081 at Cornell.
M.A.S.~is supported in part by the Sherman Fairchild Foundation and by National Science Foundation (NSF) Grant Nos. PHY-2011961, PHY-2011968, and OAC-1931266 at Caltech.

\appendix

\section{Results using $m\Omega_{\rm{coor}}$}
\label{sec:OmegaCompAppx}
\begin{figure}[tb]
\includegraphics[width = 0.98\columnwidth]{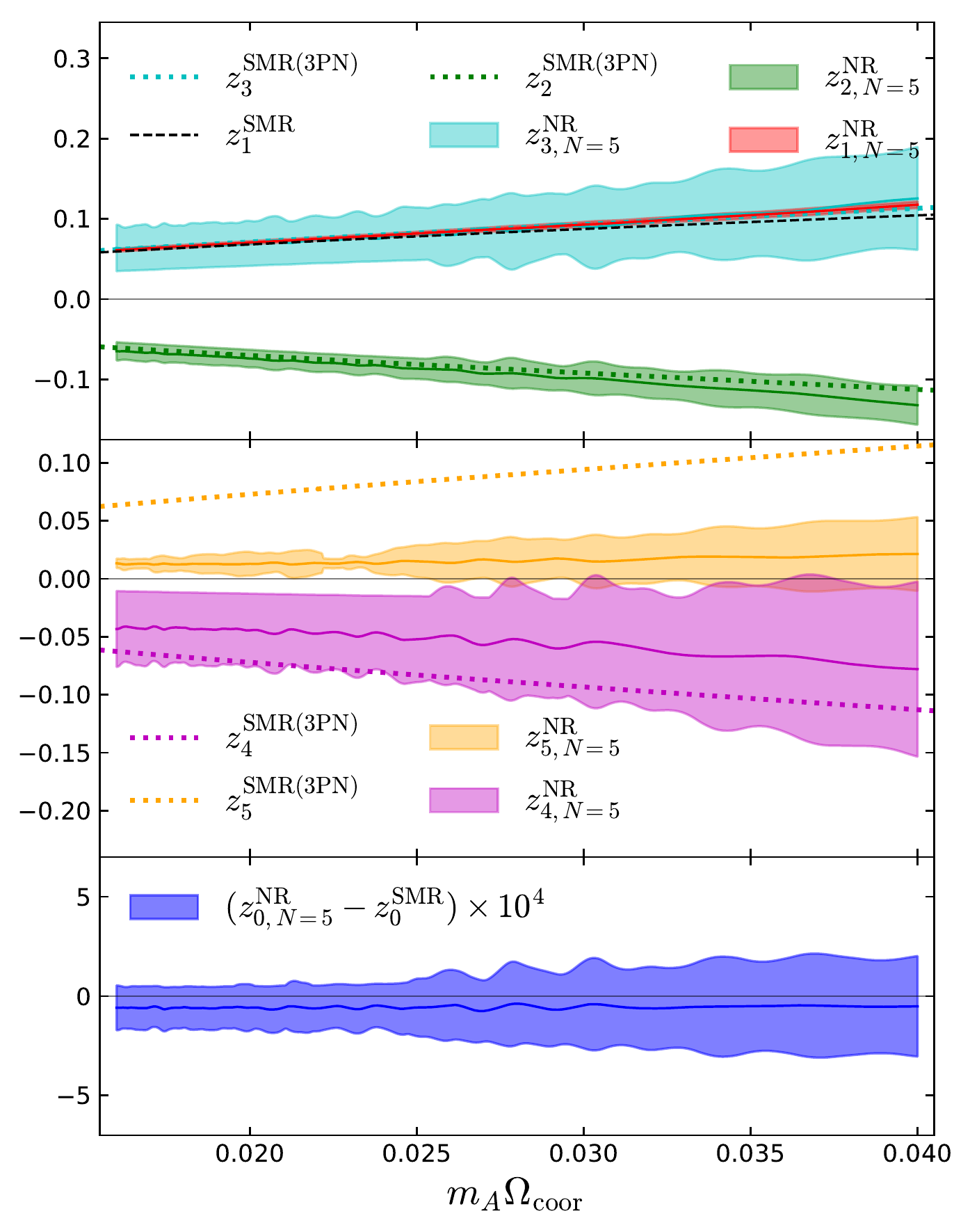}
\caption{Equivalent to Fig.~\ref{qfit_coef} using $z_B(m_A\Omega_{\rm{coor}})$ instead.}
\label{fig:zCoordFreq}
\end{figure}

In Sec.~\ref{sec:omega} we discussed the different choices of $\Omega$ used to compare our results to analytic approximations and extract the SMR expansion for $z_a(m_A \Omega)$. 
Figure~\ref{fig:zCoordFreq} shows the results of our SMR fits when applied using $\Omega_{\rm coor}$, a local measure of the orbital frequency constructed from the coordinate centers of the black holes.
We see that the leading order term $z^{\rm NR}_0(m_A \Omega_{\rm {coor}})$ is in worse agreement with the prediction from geodesic theory than our fiducial analysis, differing by $10^{-4}$ rather than $\lesssim 10^{-5}$ throughout the range of our analysis.
We take this as evidence that gauge-invariant frequency choices based on gravitational waves are preferred for understanding the SMR limit of NR simulations and for comparison to analytic results.
Broadly speaking however, the SMR coefficients extracted with this choice follow the same patterns as our fiducial analysis, and both the first and second order SMR coefficients agree with analytic approximations up to the dissipative effects not captured by conservative predictions.
This analysis shows clear evidence of these non-adiabatic corrections to the first SMR correction to geodesic theory as our fiducial analysis, with the same sign an approximate size.
 
\section{Center of mass correction validation: comparison of SHK and SKS cases}
\label{sec:DetailedCoM}

\begin{figure}[t]
\includegraphics[width = 0.98 \columnwidth]{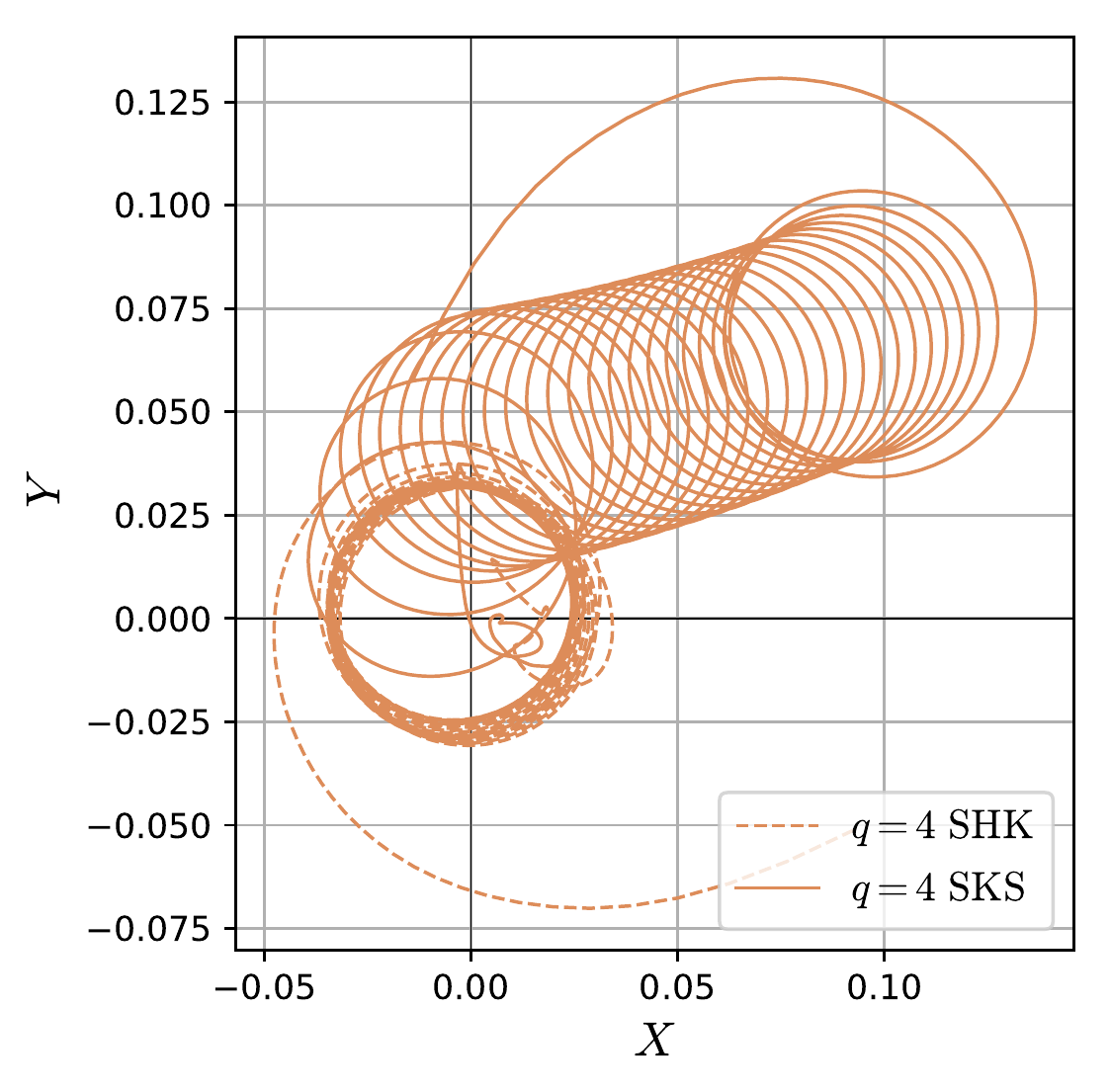}
\caption{\label{q4CoM} Trajectory of the $q=4$ CoM motion from $\vec{r}_{\rm{CoM}}=(m_A \vec{x}_A + m_B \vec{x}_B)/(m_A+m_B)$ for the SHK  initial data (dashed) and the SKS initial data (solid). The CoM drift is an order of magnitude smaller for the SHK data. See Table~\ref{table:1} for an estimate of the CoM coordinate velocities.}
\end{figure}

To estimate the potential bias introduced by the sampling method to correct for the CoM induced oscillations we have compared the $q=4$ SKS with $|v|\approx 1.6\times10^{-5}$ and $|\delta x_{\rm{CoM}}|=0.032$ to the $q=4$ SHK with $|v|\approx 2\times10^{-6}$ and $|\delta x_{\rm{CoM}}|=0.001$.
Figure~\ref{q4CoM} shows the trajectory of the CoM with respect to the simulations coordinates for each of these simulations (using highest resolution data). 
It is clear from here that the SHK initial data does better at keeping the binary centered in these coordinates, and it has negligible redshift oscillations.
We can therefore use the SHK as a ground truth reference for our approaches for correcting the redshift.

In addition to the sampling method, we tried two additional methods for correcting $z_a$:
\begin{enumerate}[label=(\roman*)]
\item {\it Rolling average}. The corrected redshift at any $m_A\Omega(t)$ is given by
\begin{align}
    z(t)=\frac{1}{T}\int_{t-T/2}^{t+T/2} z(t') \, dt' \,.
\end{align}
where $T$ is the oscillation period $T=1/(2\pi\Omega)$.
\item {\it Rolling linear fit}. The corrected redshift at any $m_A\Omega(t)$ is given by
\begin{align}
    z(t)= a(t) + b(t) m_A\Omega(t) \,,
\end{align}
where $a$ and $b$ are the coefficients of a linear fit to the redshift data over a window size of one period $T$ centered at $t$, and thus vary as we scan over $t$.
\end{enumerate}

\begin{figure}[t]
\includegraphics[width = 0.98 \columnwidth]{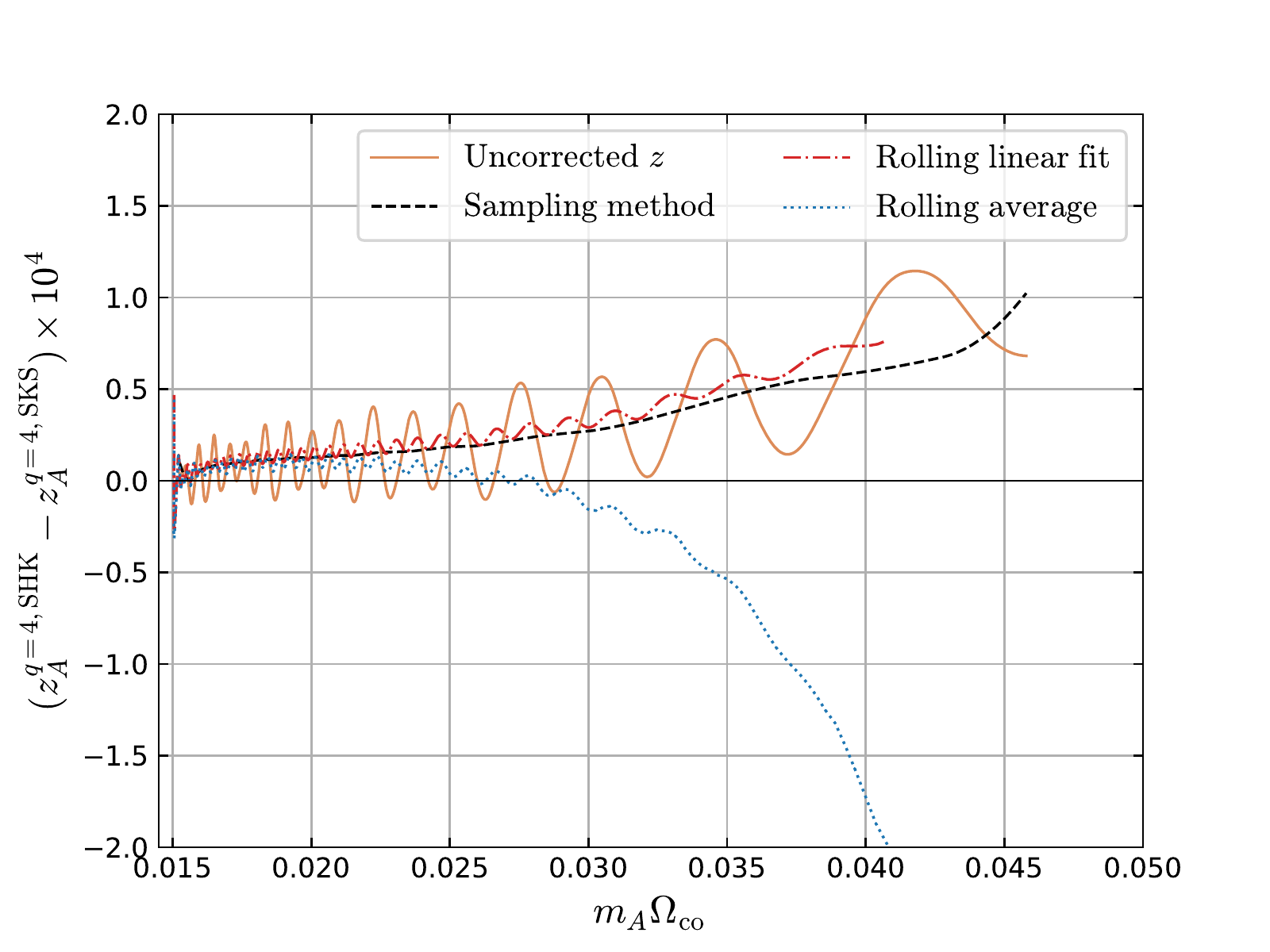}\\
\includegraphics[width = 0.98 \columnwidth]{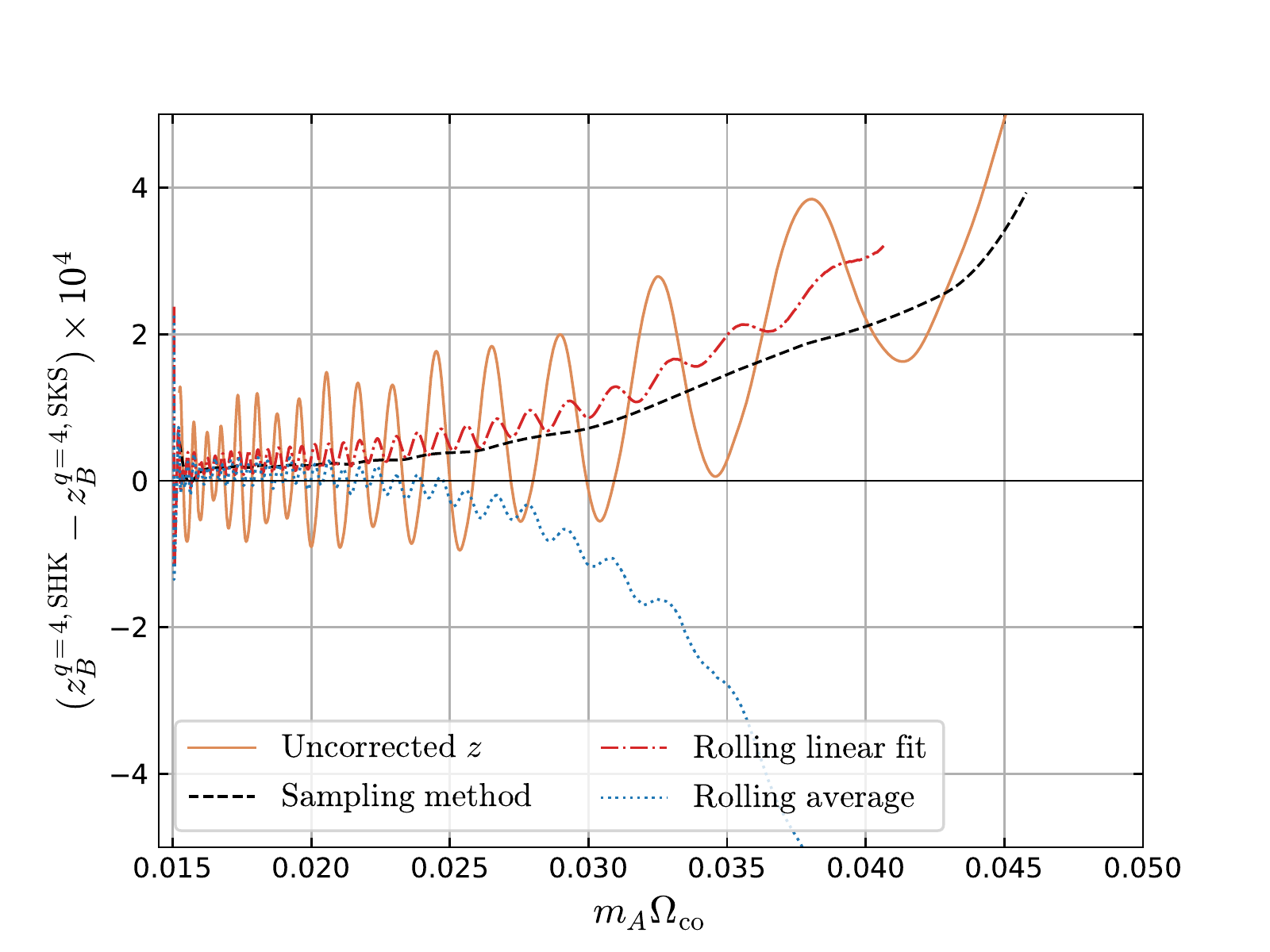}
\caption{Upper panel: Comparison between different methods to correct for the redshift oscillations as a function of $\Omega_{\rm{co}}$ for black hole $A$. The corrections shown have been applied to the SKS $q=4$ simulation and are compared against the SHK $q=4$.
Lower panel: The same comparison as above for black hole $B$.}
\label{q4comparedA_omegaco_nolog}
\end{figure}

Figure~\ref{q4comparedA_omegaco_nolog} shows the how these two methods and our fiducial method for correcting the redshift factor via sampling and averaging the envelopes compare.
We plot the difference between the $z_a(m_A \Omega_{\text{co}})$ extracted from the SKS simulation, with and without corrections, and the SHK simulation.
We see that the sampling method and rolling linear fit both perform similarly, keeping to the midline of the uncorrected result and remaining close to the SHK redshift for all frequencies.
The sampling method is preferred as it further smooths over the small residual modulations seen int he rolling linear fit.
Meanwhile, the rolling average remains very close to the SHK results at lower frequencies (earlier times), but diverges strongly at later times, which is why do not prefer it.

\begin{figure}[t]
\includegraphics[width = 0.98 \columnwidth]{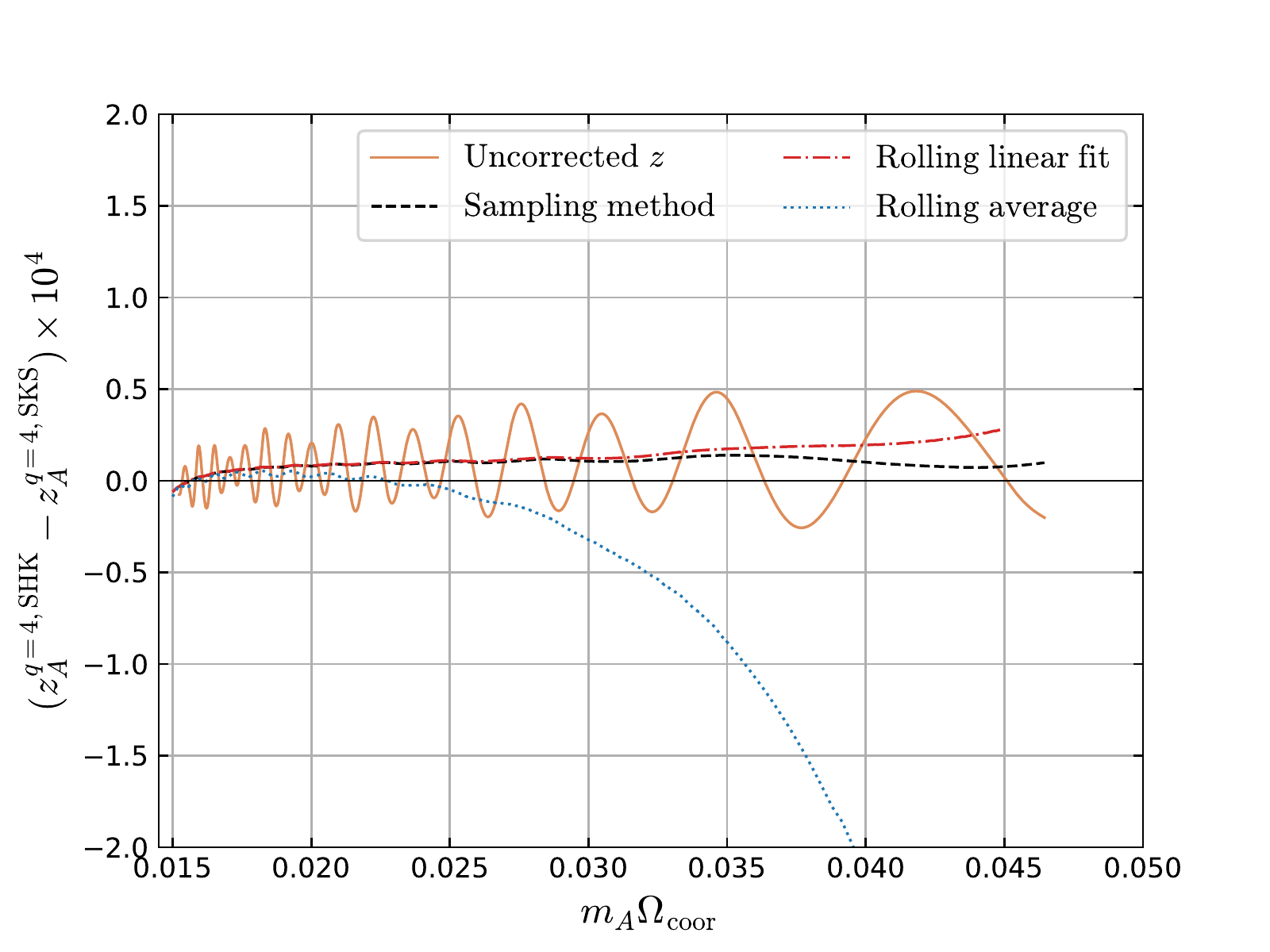}\\
\includegraphics[width = 0.98 \columnwidth]{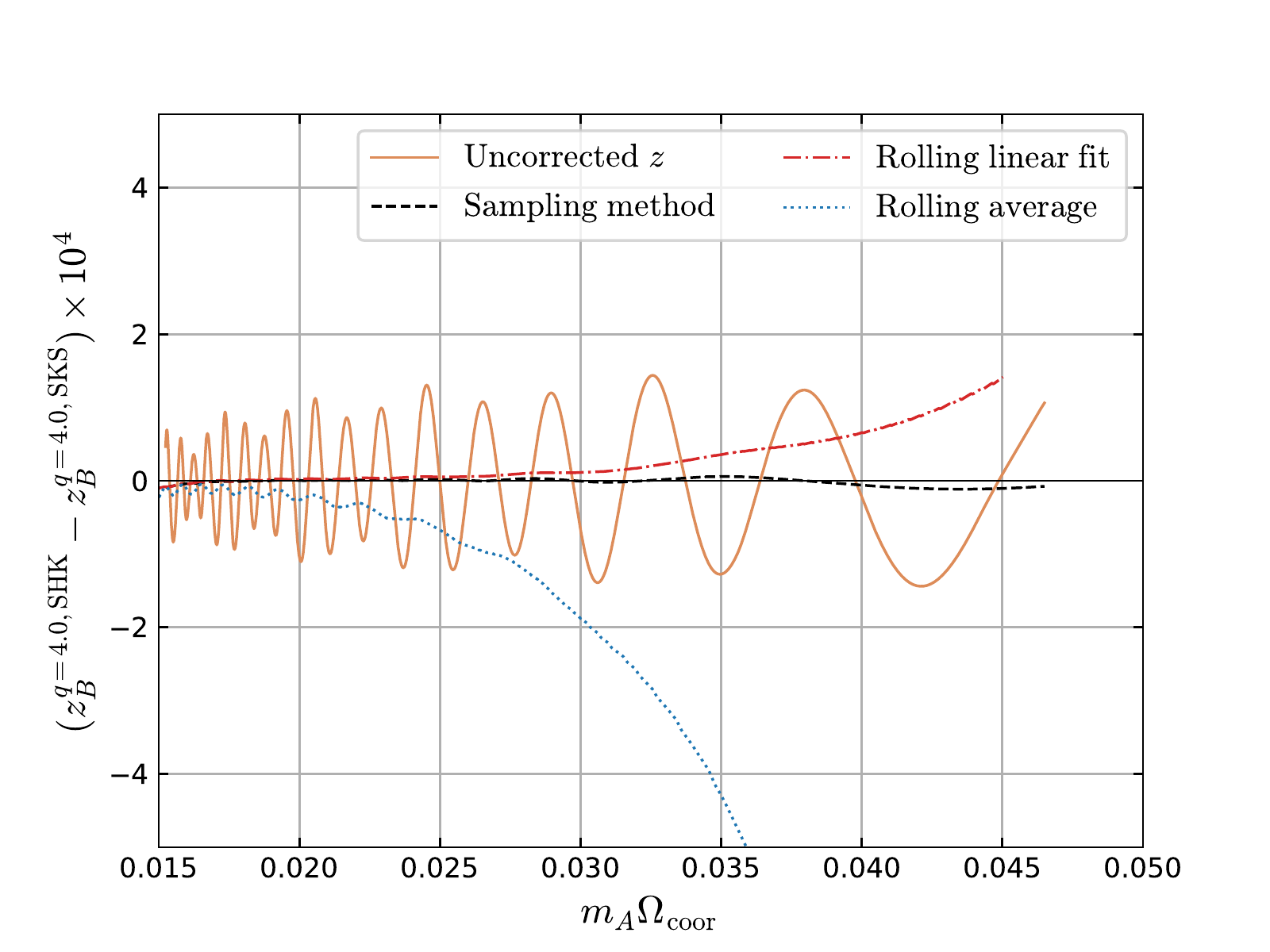}
\caption{Upper panel: Comparison between different methods to correct for the redshift oscillations as a function of $\Omega_{\rm{coor}}$ for black hole $A$. The corrections shown have been applied to the $q=4$ SKS simulation and are compared against the $q=4$ SHK.
Lower panel: The same comparison as above for black hole $B$.
\label{q4comparedA_coor_nolog}}
\end{figure}

We also show the same comparisons for $z_a(m_A\Omega_{\text {coor}})$ in Fig.~\ref{q4comparedA_coor_nolog}, where the agreement between SKS and SHK is even better.
This again illustrates that the sampling method performs better than the other methods we tried.
Interestingly, the difference between the comparisons for each frequency parametrization indicate that the local orbital dynamics of the two cases is very similar, but that their orbital frequencies measured asymptotically from the gravitational waves features a slight offset. 
This in turn may be due to the differences in CoM motion.

\section{Convergence tests to select SMR fit orders $N$}
\label{sec:ConvergenceTest}

When fitting the NR data to the power series in Eq.~\eqref{fit} we need to make a choice of $N$. 
If it is too low the fit won't capture all the ``real" features in the data, as it is clear from the $N=2$ residuals in Fig.~\ref{N4N5}. Conversely, if $N$ is too large we will start to fit the noise in the data and extrapolation of the SMR approximation won't be reliable anymore.
An optimal $N=5$ was selected by looking at the convergence of the coefficients $z^{\rm{NR}}_k$ with $N$. 
For completeness we also calculated the AICc and BIC metrics for model selection and the (unbiased) adjusted $R$-squared.

\begin{figure}[tb]
\includegraphics[width = 0.98\columnwidth]{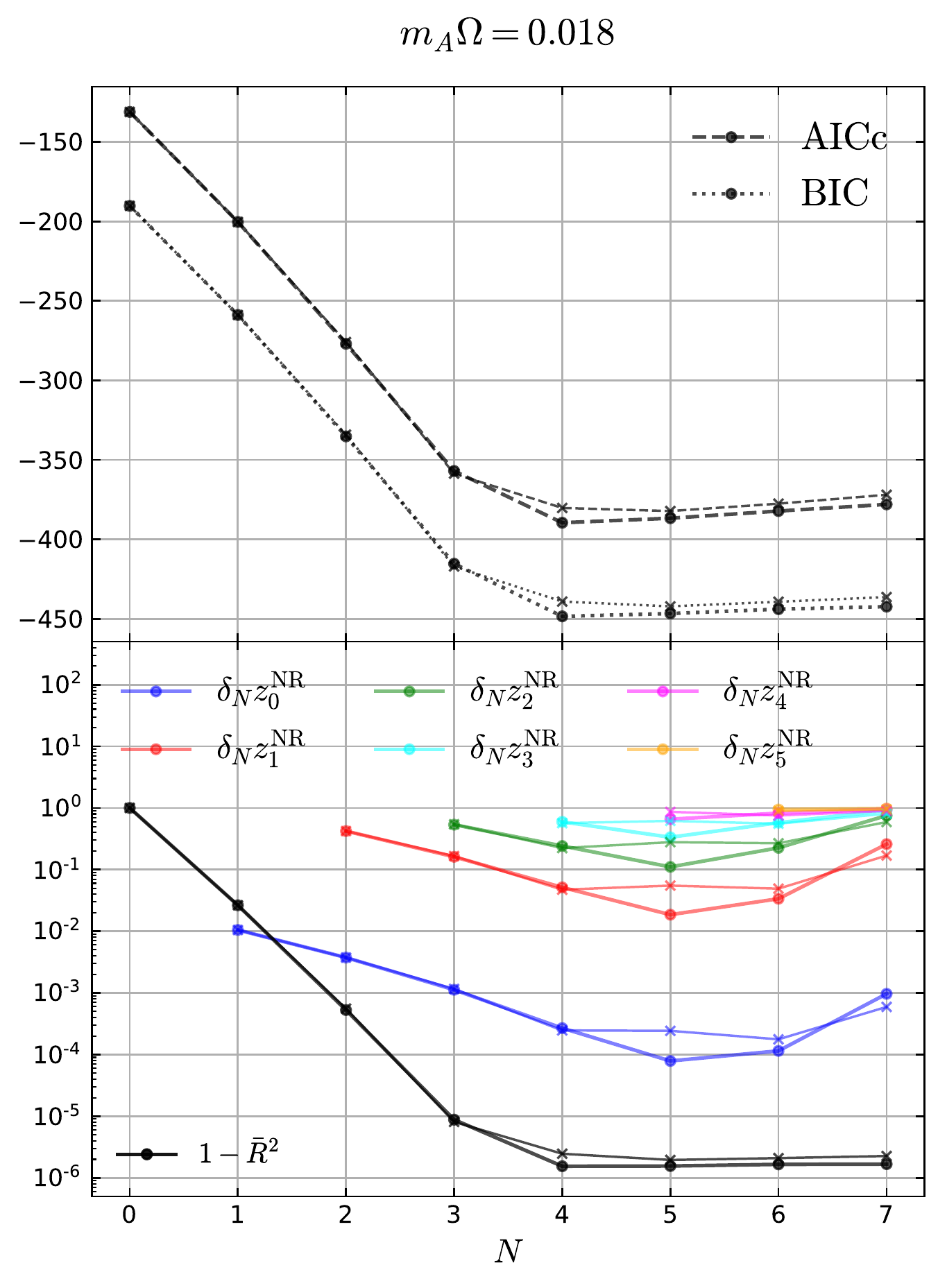}
\caption{Upper panel: $\rm{AICc}$ and $\rm{BIC}$ values. Bottom panel: Value of $1-\bar{R}^2$ and the convergence of the coefficients with $N$. Calculated at a fixed $m_A\Omega=0.018$. Dots indicate results using higher resolution data and crosses correspond to the result using lower resolution data.}
\label{goodnessfit}
\end{figure}

Figure~\ref{goodnessfit} shows the values of the model selection metrics as well as the convergence of the coefficients with $N$, for both high- and low-resolution simulations. 
This is shown at a reference frequency $m_A\Omega=0.018$, but the qualitative behaviour of these metrics is similar for all frequencies analyzed.

All metrics clearly reject fits with $N<4$. 
For $N>4$, the values of AICc, BIC and $\log{(1-\bar{R}^2)}$ are not as informative. 
Confidence intervals for these metrics tend to increase for smaller sample size ($n=20$ in our case) and the number of fit parameters ($N$). We can see how they vary when using the lower resolution data as an estimate of their variance. 
The $N=5$ values for these metrics are too close to the $N=4$ to discard that model.
To establish an upper bound on $N$, we look instead at the convergence of each of the coefficients with $N$. 
Since we are interested in testing whether we can recover the SMR approximation from NR, it is relevant to consider whether $z^{\rm{NR}}_k$ converges with $N$. 
The convergence of $z^{\rm{NR}}_k$ should also serve as a test to prevent over-fitting.
With the high resolution data, the convergence is exponential and the relative differences between resolutions are minimized for $N=5$.
Following this criteria, we have selected $N=5$ in our fits for $z_B$. 
Nevertheless the $N=4$ result is used to establish the error bands in our final results, as discussed in the main text.

\begin{figure}[t!]
\includegraphics[width = 0.98\columnwidth]{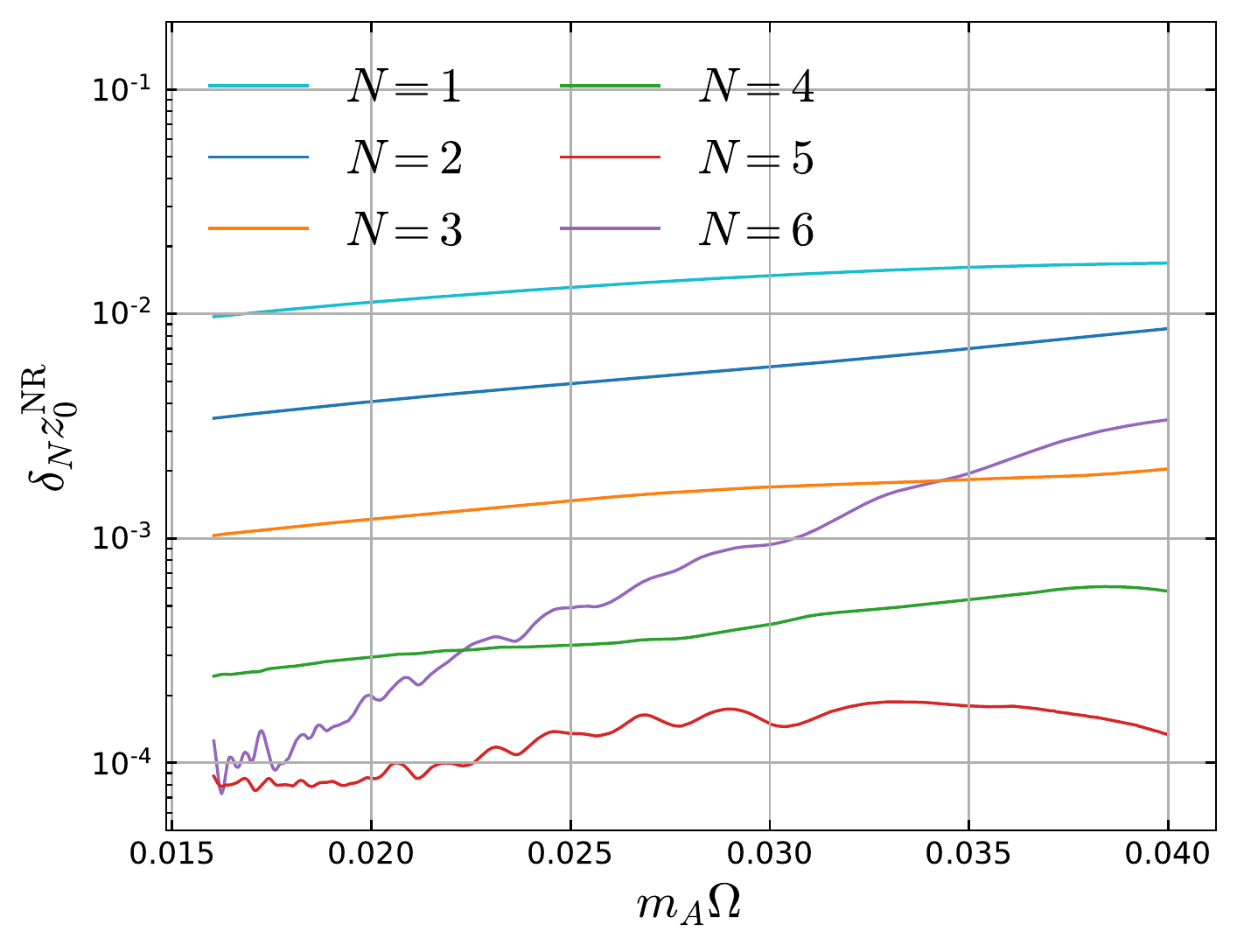}\\
\includegraphics[width = 0.98\columnwidth]{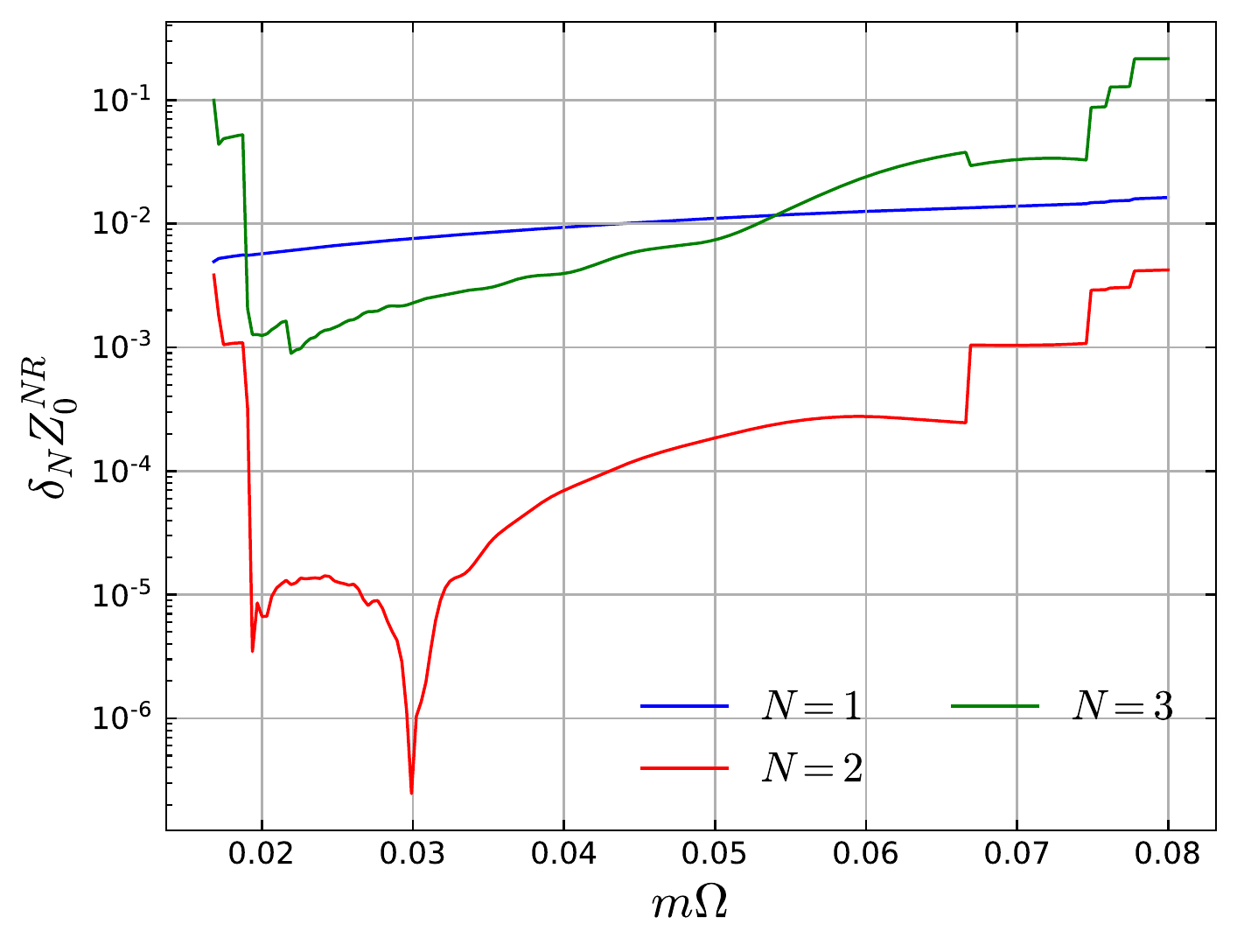}
\caption{Upper panel: Convergence of fit coefficient $z^{\text{NR}}_0$ with $N$ as a function of $m_A\Omega$, illustrated using the relative differences $\delta_N z_0^{\text{NR}}$ between successive $N$ fits. 
The lowest relative differences between the fit with $N$ coefficients and $N-1$ coefficients is achieved for $N=5$.
Lower panel: Convergence of $Z^{\text{NR}}_0$ fit coefficient of $N$ for the fit of $Z(m\Omega_{\text{co}})$ to a series expansion in $\nu$. 
The lowest relative differences in this case are achieved for $N=2$.}
\label{z0convergence} 
\end{figure}

Figure~\ref{z0convergence} shows the value of 
\begin{align}
\delta_N z^{\rm NR}_0 \coloneqq (z^{\rm{NR}}_{0,N}-z^{\rm{NR}}_{0,N-1})/z^{\rm{NR}}_{0,N}
\end{align} 
for $N\leq 6$, as a function of $m_A \Omega$. 
Similar results are obtained for the higher order coefficients. 
They all show convergence for $N \leq 5$. 
At the same time, it is $z^{\rm{NR}}_{0,N=5}$ that comes closest to the $z^{\rm{SMR}}_0$ prediction.
Fig.~\ref{N4N5} shows in detail the shift of the intercept for different choices of $N$ at a reference $m_A\Omega=0.018$, illustrating this result.

Finally, we show the convergence of the coefficients of the expansion of $Z(m\Omega)$ in Fig.~\ref{symmetric_coeff_fit}.
Here the relative differences $\delta_N Z$ show a sharp drop going from $N=1$ to $N=2$, with the latter displaying a noisiness which may be due to numerical truncation.
Going beyond this to $N=3$ results in an increase in the relative differences, so we cannot achieve convergence with our dataset beyond $N=2$.

\bibliography{apssamp.bbl}

\providecommand{\noopsort}[1]{}\providecommand{\singleletter}[1]{#1}%
\begin{thebibliography}{80}%
\makeatletter
\providecommand \@ifxundefined [1]{%
 \@ifx{#1\undefined}
}%
\providecommand \@ifnum [1]{%
 \ifnum #1\expandafter \@firstoftwo
 \else \expandafter \@secondoftwo
 \fi
}%
\providecommand \@ifx [1]{%
 \ifx #1\expandafter \@firstoftwo
 \else \expandafter \@secondoftwo
 \fi
}%
\providecommand \natexlab [1]{#1}%
\providecommand \enquote  [1]{``#1''}%
\providecommand \bibnamefont  [1]{#1}%
\providecommand \bibfnamefont [1]{#1}%
\providecommand \citenamefont [1]{#1}%
\providecommand \href@noop [0]{\@secondoftwo}%
\providecommand \href [0]{\begingroup \@sanitize@url \@href}%
\providecommand \@href[1]{\@@startlink{#1}\@@href}%
\providecommand \@@href[1]{\endgroup#1\@@endlink}%
\providecommand \@sanitize@url [0]{\catcode `\\12\catcode `\$12\catcode
  `\&12\catcode `\#12\catcode `\^12\catcode `\_12\catcode `\%12\relax}%
\providecommand \@@startlink[1]{}%
\providecommand \@@endlink[0]{}%
\providecommand \url  [0]{\begingroup\@sanitize@url \@url }%
\providecommand \@url [1]{\endgroup\@href {#1}{\urlprefix }}%
\providecommand \urlprefix  [0]{URL }%
\providecommand \Eprint [0]{\href }%
\providecommand \doibase [0]{https://doi.org/}%
\providecommand \selectlanguage [0]{\@gobble}%
\providecommand \bibinfo  [0]{\@secondoftwo}%
\providecommand \bibfield  [0]{\@secondoftwo}%
\providecommand \translation [1]{[#1]}%
\providecommand \BibitemOpen [0]{}%
\providecommand \bibitemStop [0]{}%
\providecommand \bibitemNoStop [0]{.\EOS\space}%
\providecommand \EOS [0]{\spacefactor3000\relax}%
\providecommand \BibitemShut  [1]{\csname bibitem#1\endcsname}%
\let\auto@bib@innerbib\@empty
\bibitem [{\citenamefont {Amaro-Seoane}\ \emph {et~al.}(2017)\citenamefont
  {Amaro-Seoane} \emph {et~al.}}]{Audley:2017drz}%
  \BibitemOpen
  \bibfield  {author} {\bibinfo {author} {\bibfnamefont {P.}~\bibnamefont
  {Amaro-Seoane}} \emph {et~al.} (\bibinfo {collaboration} {LISA}),\ }\bibfield
   {title} {\bibinfo {title} {{Laser Interferometer Space Antenna}},\
  }\href@noop {} {\  (\bibinfo {year} {2017})},\ \Eprint
  {https://arxiv.org/abs/1702.00786} {arXiv:1702.00786 [astro-ph.IM]}
  \BibitemShut {NoStop}%
\bibitem [{\citenamefont {Aasi}\ \emph {et~al.}(2015)\citenamefont {Aasi} \emph
  {et~al.}}]{LIGOScientific:2014pky}%
  \BibitemOpen
  \bibfield  {author} {\bibinfo {author} {\bibfnamefont {J.}~\bibnamefont
  {Aasi}} \emph {et~al.} (\bibinfo {collaboration} {LIGO Scientific}),\
  }\bibfield  {title} {\bibinfo {title} {{Advanced LIGO}},\ }\href
  {https://doi.org/10.1088/0264-9381/32/7/074001} {\bibfield  {journal}
  {\bibinfo  {journal} {Class. Quant. Grav.}\ }\textbf {\bibinfo {volume}
  {32}},\ \bibinfo {pages} {074001} (\bibinfo {year} {2015})},\ \Eprint
  {https://arxiv.org/abs/1411.4547} {arXiv:1411.4547 [gr-qc]} \BibitemShut
  {NoStop}%
\bibitem [{\citenamefont {Acernese}\ \emph {et~al.}(2015)\citenamefont
  {Acernese} \emph {et~al.}}]{VIRGO:2014yos}%
  \BibitemOpen
  \bibfield  {author} {\bibinfo {author} {\bibfnamefont {F.}~\bibnamefont
  {Acernese}} \emph {et~al.} (\bibinfo {collaboration} {VIRGO}),\ }\bibfield
  {title} {\bibinfo {title} {{Advanced Virgo: a second-generation
  interferometric gravitational wave detector}},\ }\href
  {https://doi.org/10.1088/0264-9381/32/2/024001} {\bibfield  {journal}
  {\bibinfo  {journal} {Class. Quant. Grav.}\ }\textbf {\bibinfo {volume}
  {32}},\ \bibinfo {pages} {024001} (\bibinfo {year} {2015})},\ \Eprint
  {https://arxiv.org/abs/1408.3978} {arXiv:1408.3978 [gr-qc]} \BibitemShut
  {NoStop}%
\bibitem [{\citenamefont {Akutsu}\ \emph {et~al.}(2021)\citenamefont {Akutsu}
  \emph {et~al.}}]{KAGRA:2020tym}%
  \BibitemOpen
  \bibfield  {author} {\bibinfo {author} {\bibfnamefont {T.}~\bibnamefont
  {Akutsu}} \emph {et~al.} (\bibinfo {collaboration} {KAGRA}),\ }\bibfield
  {title} {\bibinfo {title} {{Overview of KAGRA: Detector design and
  construction history}},\ }\href {https://doi.org/10.1093/ptep/ptaa125}
  {\bibfield  {journal} {\bibinfo  {journal} {PTEP}\ }\textbf {\bibinfo
  {volume} {2021}},\ \bibinfo {pages} {05A101} (\bibinfo {year} {2021})},\
  \Eprint {https://arxiv.org/abs/2005.05574} {arXiv:2005.05574
  [physics.ins-det]} \BibitemShut {NoStop}%
\bibitem [{\citenamefont {Amaro-Seoane}(2018)}]{Amaro-Seoane:2018gbb}%
  \BibitemOpen
  \bibfield  {author} {\bibinfo {author} {\bibfnamefont {P.}~\bibnamefont
  {Amaro-Seoane}},\ }\bibfield  {title} {\bibinfo {title} {{Detecting
  Intermediate-Mass Ratio Inspirals From The Ground And Space}},\ }\href
  {https://doi.org/10.1103/PhysRevD.98.063018} {\bibfield  {journal} {\bibinfo
  {journal} {Phys. Rev. D}\ }\textbf {\bibinfo {volume} {98}},\ \bibinfo
  {pages} {063018} (\bibinfo {year} {2018})},\ \Eprint
  {https://arxiv.org/abs/1807.03824} {arXiv:1807.03824 [astro-ph.HE]}
  \BibitemShut {NoStop}%
\bibitem [{\citenamefont {Dwyer}\ \emph {et~al.}(2015)\citenamefont {Dwyer},
  \citenamefont {Sigg}, \citenamefont {Ballmer}, \citenamefont {Barsotti},
  \citenamefont {Mavalvala},\ and\ \citenamefont {Evans}}]{Dwyer:2014fpa}%
  \BibitemOpen
  \bibfield  {author} {\bibinfo {author} {\bibfnamefont {S.}~\bibnamefont
  {Dwyer}}, \bibinfo {author} {\bibfnamefont {D.}~\bibnamefont {Sigg}},
  \bibinfo {author} {\bibfnamefont {S.~W.}\ \bibnamefont {Ballmer}}, \bibinfo
  {author} {\bibfnamefont {L.}~\bibnamefont {Barsotti}}, \bibinfo {author}
  {\bibfnamefont {N.}~\bibnamefont {Mavalvala}},\ and\ \bibinfo {author}
  {\bibfnamefont {M.}~\bibnamefont {Evans}},\ }\bibfield  {title} {\bibinfo
  {title} {{Gravitational wave detector with cosmological reach}},\ }\href
  {https://doi.org/10.1103/PhysRevD.91.082001} {\bibfield  {journal} {\bibinfo
  {journal} {Phys. Rev. D}\ }\textbf {\bibinfo {volume} {91}},\ \bibinfo
  {pages} {082001} (\bibinfo {year} {2015})},\ \Eprint
  {https://arxiv.org/abs/1410.0612} {arXiv:1410.0612 [astro-ph.IM]}
  \BibitemShut {NoStop}%
\bibitem [{\citenamefont {Evans}\ \emph {et~al.}(2021)\citenamefont {Evans}
  \emph {et~al.}}]{Evans:2021gyd}%
  \BibitemOpen
  \bibfield  {author} {\bibinfo {author} {\bibfnamefont {M.}~\bibnamefont
  {Evans}} \emph {et~al.},\ }\bibfield  {title} {\bibinfo {title} {{A Horizon
  Study for Cosmic Explorer: Science, Observatories, and Community}},\
  }\href@noop {} {\  (\bibinfo {year} {2021})},\ \Eprint
  {https://arxiv.org/abs/2109.09882} {arXiv:2109.09882 [astro-ph.IM]}
  \BibitemShut {NoStop}%
\bibitem [{\citenamefont {Ryan}(1997)}]{Ryan:1997hg}%
  \BibitemOpen
  \bibfield  {author} {\bibinfo {author} {\bibfnamefont {F.~D.}\ \bibnamefont
  {Ryan}},\ }\bibfield  {title} {\bibinfo {title} {{Accuracy of estimating the
  multipole moments of a massive body from the gravitational waves of a binary
  inspiral}},\ }\href {https://doi.org/10.1103/PhysRevD.56.1845} {\bibfield
  {journal} {\bibinfo  {journal} {Phys. Rev. D}\ }\textbf {\bibinfo {volume}
  {56}},\ \bibinfo {pages} {1845} (\bibinfo {year} {1997})}\BibitemShut
  {NoStop}%
\bibitem [{\citenamefont {Barausse}\ \emph {et~al.}(2020)\citenamefont
  {Barausse} \emph {et~al.}}]{Barausse:2020rsu}%
  \BibitemOpen
  \bibfield  {author} {\bibinfo {author} {\bibfnamefont {E.}~\bibnamefont
  {Barausse}} \emph {et~al.},\ }\bibfield  {title} {\bibinfo {title}
  {{Prospects for Fundamental Physics with LISA}},\ }\href
  {https://doi.org/10.1007/s10714-020-02691-1} {\bibfield  {journal} {\bibinfo
  {journal} {Gen. Rel. Grav.}\ }\textbf {\bibinfo {volume} {52}},\ \bibinfo
  {pages} {81} (\bibinfo {year} {2020})},\ \Eprint
  {https://arxiv.org/abs/2001.09793} {arXiv:2001.09793 [gr-qc]} \BibitemShut
  {NoStop}%
\bibitem [{\citenamefont {Fernando}\ \emph {et~al.}(2018)\citenamefont
  {Fernando}, \citenamefont {Neilsen}, \citenamefont {Lim}, \citenamefont
  {Hirschmann},\ and\ \citenamefont {Sundar}}]{Fernando:2018mov}%
  \BibitemOpen
  \bibfield  {author} {\bibinfo {author} {\bibfnamefont {M.}~\bibnamefont
  {Fernando}}, \bibinfo {author} {\bibfnamefont {D.}~\bibnamefont {Neilsen}},
  \bibinfo {author} {\bibfnamefont {H.}~\bibnamefont {Lim}}, \bibinfo {author}
  {\bibfnamefont {E.}~\bibnamefont {Hirschmann}},\ and\ \bibinfo {author}
  {\bibfnamefont {H.}~\bibnamefont {Sundar}},\ }\bibfield  {title} {\bibinfo
  {title} {{Massively Parallel Simulations of Binary Black Hole
  Intermediate-Mass-Ratio Inspirals}}\ }\href
  {https://doi.org/10.1137/18M1196972} {10.1137/18M1196972} (\bibinfo {year}
  {2018}),\ \Eprint {https://arxiv.org/abs/1807.06128} {arXiv:1807.06128
  [gr-qc]} \BibitemShut {NoStop}%
\bibitem [{\citenamefont {Lousto}\ and\ \citenamefont
  {Healy}(2020)}]{2006.04818}%
  \BibitemOpen
  \bibfield  {author} {\bibinfo {author} {\bibfnamefont {C.~O.}\ \bibnamefont
  {Lousto}}\ and\ \bibinfo {author} {\bibfnamefont {J.}~\bibnamefont {Healy}},\
  }\bibfield  {title} {\bibinfo {title} {{Exploring the Small Mass Ratio Binary
  Black Hole Merger via Zeno\textquoteright{}s Dichotomy Approach}},\ }\href
  {https://doi.org/10.1103/PhysRevLett.125.191102} {\bibfield  {journal}
  {\bibinfo  {journal} {Phys. Rev. Lett.}\ }\textbf {\bibinfo {volume} {125}},\
  \bibinfo {pages} {191102} (\bibinfo {year} {2020})},\ \Eprint
  {https://arxiv.org/abs/2006.04818} {arXiv:2006.04818 [gr-qc]} \BibitemShut
  {NoStop}%
\bibitem [{\citenamefont {Dhesi}\ \emph {et~al.}(2021)\citenamefont {Dhesi},
  \citenamefont {R\"uter}, \citenamefont {Pound}, \citenamefont {Barack},\ and\
  \citenamefont {Pfeiffer}}]{Dhesi:2021yje}%
  \BibitemOpen
  \bibfield  {author} {\bibinfo {author} {\bibfnamefont {M.}~\bibnamefont
  {Dhesi}}, \bibinfo {author} {\bibfnamefont {H.~R.}\ \bibnamefont {R\"uter}},
  \bibinfo {author} {\bibfnamefont {A.}~\bibnamefont {Pound}}, \bibinfo
  {author} {\bibfnamefont {L.}~\bibnamefont {Barack}},\ and\ \bibinfo {author}
  {\bibfnamefont {H.~P.}\ \bibnamefont {Pfeiffer}},\ }\bibfield  {title}
  {\bibinfo {title} {{Worldtube excision method for intermediate-mass-ratio
  inspirals: Scalar-field toy model}},\ }\href
  {https://doi.org/10.1103/PhysRevD.104.124002} {\bibfield  {journal} {\bibinfo
   {journal} {Phys. Rev. D}\ }\textbf {\bibinfo {volume} {104}},\ \bibinfo
  {pages} {124002} (\bibinfo {year} {2021})},\ \Eprint
  {https://arxiv.org/abs/2109.03531} {arXiv:2109.03531 [gr-qc]} \BibitemShut
  {NoStop}%
\bibitem [{\citenamefont {Poisson}\ \emph {et~al.}(2011)\citenamefont
  {Poisson}, \citenamefont {Pound},\ and\ \citenamefont
  {Vega}}]{Poisson:2011nh}%
  \BibitemOpen
  \bibfield  {author} {\bibinfo {author} {\bibfnamefont {E.}~\bibnamefont
  {Poisson}}, \bibinfo {author} {\bibfnamefont {A.}~\bibnamefont {Pound}},\
  and\ \bibinfo {author} {\bibfnamefont {I.}~\bibnamefont {Vega}},\ }\bibfield
  {title} {\bibinfo {title} {{The Motion of point particles in curved
  spacetime}},\ }\href {https://doi.org/10.12942/lrr-2011-7} {\bibfield
  {journal} {\bibinfo  {journal} {Living Rev. Rel.}\ }\textbf {\bibinfo
  {volume} {14}},\ \bibinfo {pages} {7} (\bibinfo {year} {2011})},\ \Eprint
  {https://arxiv.org/abs/1102.0529} {arXiv:1102.0529 [gr-qc]} \BibitemShut
  {NoStop}%
\bibitem [{\citenamefont {Barack}\ and\ \citenamefont
  {Pound}(2019)}]{Barack:2018yvs}%
  \BibitemOpen
  \bibfield  {author} {\bibinfo {author} {\bibfnamefont {L.}~\bibnamefont
  {Barack}}\ and\ \bibinfo {author} {\bibfnamefont {A.}~\bibnamefont {Pound}},\
  }\bibfield  {title} {\bibinfo {title} {{Self-force and radiation reaction in
  general relativity}},\ }\href {https://doi.org/10.1088/1361-6633/aae552}
  {\bibfield  {journal} {\bibinfo  {journal} {Rept. Prog. Phys.}\ }\textbf
  {\bibinfo {volume} {82}},\ \bibinfo {pages} {016904} (\bibinfo {year}
  {2019})},\ \Eprint {https://arxiv.org/abs/1805.10385} {arXiv:1805.10385
  [gr-qc]} \BibitemShut {NoStop}%
\bibitem [{\citenamefont {Warburton}\ \emph {et~al.}(2021)\citenamefont
  {Warburton}, \citenamefont {Pound}, \citenamefont {Wardell}, \citenamefont
  {Miller},\ and\ \citenamefont {Durkan}}]{Warburton:2021kwk}%
  \BibitemOpen
  \bibfield  {author} {\bibinfo {author} {\bibfnamefont {N.}~\bibnamefont
  {Warburton}}, \bibinfo {author} {\bibfnamefont {A.}~\bibnamefont {Pound}},
  \bibinfo {author} {\bibfnamefont {B.}~\bibnamefont {Wardell}}, \bibinfo
  {author} {\bibfnamefont {J.}~\bibnamefont {Miller}},\ and\ \bibinfo {author}
  {\bibfnamefont {L.}~\bibnamefont {Durkan}},\ }\bibfield  {title} {\bibinfo
  {title} {{Gravitational-wave energy flux for compact binaries through second
  order in the mass ratio}},\ }\href@noop {} {\  (\bibinfo {year} {2021})},\
  \Eprint {https://arxiv.org/abs/2107.01298} {arXiv:2107.01298 [gr-qc]}
  \BibitemShut {NoStop}%
\bibitem [{\citenamefont {Wardell}\ \emph {et~al.}(2021)\citenamefont
  {Wardell}, \citenamefont {Pound}, \citenamefont {Warburton}, \citenamefont
  {Miller}, \citenamefont {Durkan},\ and\ \citenamefont
  {Tiec}}]{Wardell:2021fyy}%
  \BibitemOpen
  \bibfield  {author} {\bibinfo {author} {\bibfnamefont {B.}~\bibnamefont
  {Wardell}}, \bibinfo {author} {\bibfnamefont {A.}~\bibnamefont {Pound}},
  \bibinfo {author} {\bibfnamefont {N.}~\bibnamefont {Warburton}}, \bibinfo
  {author} {\bibfnamefont {J.}~\bibnamefont {Miller}}, \bibinfo {author}
  {\bibfnamefont {L.}~\bibnamefont {Durkan}},\ and\ \bibinfo {author}
  {\bibfnamefont {A.~L.}\ \bibnamefont {Tiec}},\ }\bibfield  {title} {\bibinfo
  {title} {{Gravitational waveforms for compact binaries from second-order
  self-force theory}},\ }\href@noop {} {\  (\bibinfo {year} {2021})},\ \Eprint
  {https://arxiv.org/abs/2112.12265} {arXiv:2112.12265 [gr-qc]} \BibitemShut
  {NoStop}%
\bibitem [{\citenamefont {Le~Tiec}\ \emph {et~al.}(2011)\citenamefont
  {Le~Tiec}, \citenamefont {Mroue}, \citenamefont {Barack}, \citenamefont
  {Buonanno}, \citenamefont {Pfeiffer}, \citenamefont {Sago},\ and\
  \citenamefont {Taracchini}}]{LeTiec:2011bk}%
  \BibitemOpen
  \bibfield  {author} {\bibinfo {author} {\bibfnamefont {A.}~\bibnamefont
  {Le~Tiec}}, \bibinfo {author} {\bibfnamefont {A.~H.}\ \bibnamefont {Mroue}},
  \bibinfo {author} {\bibfnamefont {L.}~\bibnamefont {Barack}}, \bibinfo
  {author} {\bibfnamefont {A.}~\bibnamefont {Buonanno}}, \bibinfo {author}
  {\bibfnamefont {H.~P.}\ \bibnamefont {Pfeiffer}}, \bibinfo {author}
  {\bibfnamefont {N.}~\bibnamefont {Sago}},\ and\ \bibinfo {author}
  {\bibfnamefont {A.}~\bibnamefont {Taracchini}},\ }\bibfield  {title}
  {\bibinfo {title} {{Periastron Advance in Black Hole Binaries}},\ }\href
  {https://doi.org/10.1103/PhysRevLett.107.141101} {\bibfield  {journal}
  {\bibinfo  {journal} {Phys. Rev. Lett.}\ }\textbf {\bibinfo {volume} {107}},\
  \bibinfo {pages} {141101} (\bibinfo {year} {2011})},\ \Eprint
  {https://arxiv.org/abs/1106.3278} {arXiv:1106.3278 [gr-qc]} \BibitemShut
  {NoStop}%
\bibitem [{\citenamefont {Le~Tiec}(2014)}]{Tiec:2014lba}%
  \BibitemOpen
  \bibfield  {author} {\bibinfo {author} {\bibfnamefont {A.}~\bibnamefont
  {Le~Tiec}},\ }\bibfield  {title} {\bibinfo {title} {{The Overlap of Numerical
  Relativity, Perturbation Theory and Post-Newtonian Theory in the Binary Black
  Hole Problem}},\ }\href {https://doi.org/10.1142/S0218271814300225}
  {\bibfield  {journal} {\bibinfo  {journal} {Int. J. Mod. Phys. D}\ }\textbf
  {\bibinfo {volume} {23}},\ \bibinfo {pages} {1430022} (\bibinfo {year}
  {2014})},\ \Eprint {https://arxiv.org/abs/1408.5505} {arXiv:1408.5505
  [gr-qc]} \BibitemShut {NoStop}%
\bibitem [{\citenamefont {van~de Meent}\ and\ \citenamefont
  {Pfeiffer}(2020)}]{vandeMeent:2020xgc}%
  \BibitemOpen
  \bibfield  {author} {\bibinfo {author} {\bibfnamefont {M.}~\bibnamefont
  {van~de Meent}}\ and\ \bibinfo {author} {\bibfnamefont {H.~P.}\ \bibnamefont
  {Pfeiffer}},\ }\bibfield  {title} {\bibinfo {title} {{Intermediate mass-ratio
  black hole binaries: Applicability of small mass-ratio perturbation
  theory}},\ }\href {https://doi.org/10.1103/PhysRevLett.125.181101} {\bibfield
   {journal} {\bibinfo  {journal} {Phys. Rev. Lett.}\ }\textbf {\bibinfo
  {volume} {125}},\ \bibinfo {pages} {181101} (\bibinfo {year} {2020})},\
  \Eprint {https://arxiv.org/abs/2006.12036} {arXiv:2006.12036 [gr-qc]}
  \BibitemShut {NoStop}%
\bibitem [{\citenamefont {Detweiler}(2008)}]{Detweiler:2008ft}%
  \BibitemOpen
  \bibfield  {author} {\bibinfo {author} {\bibfnamefont {S.~L.}\ \bibnamefont
  {Detweiler}},\ }\bibfield  {title} {\bibinfo {title} {{A Consequence of the
  gravitational self-force for circular orbits of the Schwarzschild
  geometry}},\ }\href {https://doi.org/10.1103/PhysRevD.77.124026} {\bibfield
  {journal} {\bibinfo  {journal} {Phys. Rev. D}\ }\textbf {\bibinfo {volume}
  {77}},\ \bibinfo {pages} {124026} (\bibinfo {year} {2008})},\ \Eprint
  {https://arxiv.org/abs/0804.3529} {arXiv:0804.3529 [gr-qc]} \BibitemShut
  {NoStop}%
\bibitem [{\citenamefont {Barack}\ and\ \citenamefont
  {Sago}(2011)}]{Barack:2011ed}%
  \BibitemOpen
  \bibfield  {author} {\bibinfo {author} {\bibfnamefont {L.}~\bibnamefont
  {Barack}}\ and\ \bibinfo {author} {\bibfnamefont {N.}~\bibnamefont {Sago}},\
  }\bibfield  {title} {\bibinfo {title} {{Beyond the geodesic approximation:
  conservative effects of the gravitational self-force in eccentric orbits
  around a Schwarzschild black hole}},\ }\href
  {https://doi.org/10.1103/PhysRevD.83.084023} {\bibfield  {journal} {\bibinfo
  {journal} {Phys. Rev. D}\ }\textbf {\bibinfo {volume} {83}},\ \bibinfo
  {pages} {084023} (\bibinfo {year} {2011})},\ \Eprint
  {https://arxiv.org/abs/1101.3331} {arXiv:1101.3331 [gr-qc]} \BibitemShut
  {NoStop}%
\bibitem [{\citenamefont {Fujita}\ \emph {et~al.}(2017)\citenamefont {Fujita},
  \citenamefont {Isoyama}, \citenamefont {Le~Tiec}, \citenamefont {Nakano},
  \citenamefont {Sago},\ and\ \citenamefont {Tanaka}}]{Fujita:2016igj}%
  \BibitemOpen
  \bibfield  {author} {\bibinfo {author} {\bibfnamefont {R.}~\bibnamefont
  {Fujita}}, \bibinfo {author} {\bibfnamefont {S.}~\bibnamefont {Isoyama}},
  \bibinfo {author} {\bibfnamefont {A.}~\bibnamefont {Le~Tiec}}, \bibinfo
  {author} {\bibfnamefont {H.}~\bibnamefont {Nakano}}, \bibinfo {author}
  {\bibfnamefont {N.}~\bibnamefont {Sago}},\ and\ \bibinfo {author}
  {\bibfnamefont {T.}~\bibnamefont {Tanaka}},\ }\bibfield  {title} {\bibinfo
  {title} {{Hamiltonian Formulation of the Conservative Self-Force Dynamics in
  the Kerr Geometry}},\ }\href {https://doi.org/10.1088/1361-6382/aa7342}
  {\bibfield  {journal} {\bibinfo  {journal} {Class. Quant. Grav.}\ }\textbf
  {\bibinfo {volume} {34}},\ \bibinfo {pages} {134001} (\bibinfo {year}
  {2017})},\ \Eprint {https://arxiv.org/abs/1612.02504} {arXiv:1612.02504
  [gr-qc]} \BibitemShut {NoStop}%
\bibitem [{\citenamefont {Le~Tiec}\ \emph
  {et~al.}(2012{\natexlab{a}})\citenamefont {Le~Tiec}, \citenamefont
  {Blanchet},\ and\ \citenamefont {Whiting}}]{LeTiec:2011ab}%
  \BibitemOpen
  \bibfield  {author} {\bibinfo {author} {\bibfnamefont {A.}~\bibnamefont
  {Le~Tiec}}, \bibinfo {author} {\bibfnamefont {L.}~\bibnamefont {Blanchet}},\
  and\ \bibinfo {author} {\bibfnamefont {B.~F.}\ \bibnamefont {Whiting}},\
  }\bibfield  {title} {\bibinfo {title} {{The First Law of Binary Black Hole
  Mechanics in General Relativity and Post-Newtonian Theory}},\ }\href
  {https://doi.org/10.1103/PhysRevD.85.064039} {\bibfield  {journal} {\bibinfo
  {journal} {Phys. Rev. D}\ }\textbf {\bibinfo {volume} {85}},\ \bibinfo
  {pages} {064039} (\bibinfo {year} {2012}{\natexlab{a}})},\ \Eprint
  {https://arxiv.org/abs/1111.5378} {arXiv:1111.5378 [gr-qc]} \BibitemShut
  {NoStop}%
\bibitem [{\citenamefont {Le~Tiec}\ \emph
  {et~al.}(2012{\natexlab{b}})\citenamefont {Le~Tiec}, \citenamefont
  {Barausse},\ and\ \citenamefont {Buonanno}}]{LeTiec:2011dp}%
  \BibitemOpen
  \bibfield  {author} {\bibinfo {author} {\bibfnamefont {A.}~\bibnamefont
  {Le~Tiec}}, \bibinfo {author} {\bibfnamefont {E.}~\bibnamefont {Barausse}},\
  and\ \bibinfo {author} {\bibfnamefont {A.}~\bibnamefont {Buonanno}},\
  }\bibfield  {title} {\bibinfo {title} {{Gravitational Self-Force Correction
  to the Binding Energy of Compact Binary Systems}},\ }\href
  {https://doi.org/10.1103/PhysRevLett.108.131103} {\bibfield  {journal}
  {\bibinfo  {journal} {Phys. Rev. Lett.}\ }\textbf {\bibinfo {volume} {108}},\
  \bibinfo {pages} {131103} (\bibinfo {year} {2012}{\natexlab{b}})},\ \Eprint
  {https://arxiv.org/abs/1111.5609} {arXiv:1111.5609 [gr-qc]} \BibitemShut
  {NoStop}%
\bibitem [{\citenamefont {Barack}\ and\ \citenamefont
  {Sago}(2009)}]{Barack:2009ey}%
  \BibitemOpen
  \bibfield  {author} {\bibinfo {author} {\bibfnamefont {L.}~\bibnamefont
  {Barack}}\ and\ \bibinfo {author} {\bibfnamefont {N.}~\bibnamefont {Sago}},\
  }\bibfield  {title} {\bibinfo {title} {{Gravitational self-force correction
  to the innermost stable circular orbit of a Schwarzschild black hole}},\
  }\href {https://doi.org/10.1103/PhysRevLett.102.191101} {\bibfield  {journal}
  {\bibinfo  {journal} {Phys. Rev. Lett.}\ }\textbf {\bibinfo {volume} {102}},\
  \bibinfo {pages} {191101} (\bibinfo {year} {2009})},\ \Eprint
  {https://arxiv.org/abs/0902.0573} {arXiv:0902.0573 [gr-qc]} \BibitemShut
  {NoStop}%
\bibitem [{\citenamefont {Barausse}\ \emph {et~al.}(2012)\citenamefont
  {Barausse}, \citenamefont {Buonanno},\ and\ \citenamefont
  {Le~Tiec}}]{Barausse:2011dq}%
  \BibitemOpen
  \bibfield  {author} {\bibinfo {author} {\bibfnamefont {E.}~\bibnamefont
  {Barausse}}, \bibinfo {author} {\bibfnamefont {A.}~\bibnamefont {Buonanno}},\
  and\ \bibinfo {author} {\bibfnamefont {A.}~\bibnamefont {Le~Tiec}},\
  }\bibfield  {title} {\bibinfo {title} {{The complete non-spinning
  effective-one-body metric at linear order in the mass ratio}},\ }\href
  {https://doi.org/10.1103/PhysRevD.85.064010} {\bibfield  {journal} {\bibinfo
  {journal} {Phys. Rev. D}\ }\textbf {\bibinfo {volume} {85}},\ \bibinfo
  {pages} {064010} (\bibinfo {year} {2012})},\ \Eprint
  {https://arxiv.org/abs/1111.5610} {arXiv:1111.5610 [gr-qc]} \BibitemShut
  {NoStop}%
\bibitem [{\citenamefont {Akcay}\ \emph {et~al.}(2012)\citenamefont {Akcay},
  \citenamefont {Barack}, \citenamefont {Damour},\ and\ \citenamefont
  {Sago}}]{Akcay:2012ea}%
  \BibitemOpen
  \bibfield  {author} {\bibinfo {author} {\bibfnamefont {S.}~\bibnamefont
  {Akcay}}, \bibinfo {author} {\bibfnamefont {L.}~\bibnamefont {Barack}},
  \bibinfo {author} {\bibfnamefont {T.}~\bibnamefont {Damour}},\ and\ \bibinfo
  {author} {\bibfnamefont {N.}~\bibnamefont {Sago}},\ }\bibfield  {title}
  {\bibinfo {title} {{Gravitational self-force and the effective-one-body
  formalism between the innermost stable circular orbit and the light ring}},\
  }\href {https://doi.org/10.1103/PhysRevD.86.104041} {\bibfield  {journal}
  {\bibinfo  {journal} {Phys. Rev. D}\ }\textbf {\bibinfo {volume} {86}},\
  \bibinfo {pages} {104041} (\bibinfo {year} {2012})},\ \Eprint
  {https://arxiv.org/abs/1209.0964} {arXiv:1209.0964 [gr-qc]} \BibitemShut
  {NoStop}%
\bibitem [{\citenamefont {Antonelli}\ \emph {et~al.}(2020)\citenamefont
  {Antonelli}, \citenamefont {van~de Meent}, \citenamefont {Buonanno},
  \citenamefont {Steinhoff},\ and\ \citenamefont {Vines}}]{Antonelli:2019fmq}%
  \BibitemOpen
  \bibfield  {author} {\bibinfo {author} {\bibfnamefont {A.}~\bibnamefont
  {Antonelli}}, \bibinfo {author} {\bibfnamefont {M.}~\bibnamefont {van~de
  Meent}}, \bibinfo {author} {\bibfnamefont {A.}~\bibnamefont {Buonanno}},
  \bibinfo {author} {\bibfnamefont {J.}~\bibnamefont {Steinhoff}},\ and\
  \bibinfo {author} {\bibfnamefont {J.}~\bibnamefont {Vines}},\ }\bibfield
  {title} {\bibinfo {title} {{Quasicircular inspirals and plunges from
  nonspinning effective-one-body Hamiltonians with gravitational self-force
  information}},\ }\href {https://doi.org/10.1103/PhysRevD.101.024024}
  {\bibfield  {journal} {\bibinfo  {journal} {Phys. Rev. D}\ }\textbf {\bibinfo
  {volume} {101}},\ \bibinfo {pages} {024024} (\bibinfo {year} {2020})},\
  \Eprint {https://arxiv.org/abs/1907.11597} {arXiv:1907.11597 [gr-qc]}
  \BibitemShut {NoStop}%
\bibitem [{\citenamefont {Shah}\ \emph {et~al.}(2011)\citenamefont {Shah},
  \citenamefont {Keidl}, \citenamefont {Friedman}, \citenamefont {Kim},\ and\
  \citenamefont {Price}}]{Shah:2010bi}%
  \BibitemOpen
  \bibfield  {author} {\bibinfo {author} {\bibfnamefont {A.~G.}\ \bibnamefont
  {Shah}}, \bibinfo {author} {\bibfnamefont {T.~S.}\ \bibnamefont {Keidl}},
  \bibinfo {author} {\bibfnamefont {J.~L.}\ \bibnamefont {Friedman}}, \bibinfo
  {author} {\bibfnamefont {D.-H.}\ \bibnamefont {Kim}},\ and\ \bibinfo {author}
  {\bibfnamefont {L.~R.}\ \bibnamefont {Price}},\ }\bibfield  {title} {\bibinfo
  {title} {{Conservative, gravitational self-force for a particle in circular
  orbit around a Schwarzschild black hole in a Radiation Gauge}},\ }\href
  {https://doi.org/10.1103/PhysRevD.83.064018} {\bibfield  {journal} {\bibinfo
  {journal} {Phys. Rev. D}\ }\textbf {\bibinfo {volume} {83}},\ \bibinfo
  {pages} {064018} (\bibinfo {year} {2011})},\ \Eprint
  {https://arxiv.org/abs/1009.4876} {arXiv:1009.4876 [gr-qc]} \BibitemShut
  {NoStop}%
\bibitem [{\citenamefont {Thompson}\ \emph {et~al.}(2019)\citenamefont
  {Thompson}, \citenamefont {Wardell},\ and\ \citenamefont
  {Whiting}}]{Thompson:2018lgb}%
  \BibitemOpen
  \bibfield  {author} {\bibinfo {author} {\bibfnamefont {J.~E.}\ \bibnamefont
  {Thompson}}, \bibinfo {author} {\bibfnamefont {B.}~\bibnamefont {Wardell}},\
  and\ \bibinfo {author} {\bibfnamefont {B.~F.}\ \bibnamefont {Whiting}},\
  }\bibfield  {title} {\bibinfo {title} {{Gravitational Self-Force
  Regularization in the Regge-Wheeler and Easy Gauges}},\ }\href
  {https://doi.org/10.1103/PhysRevD.99.124046} {\bibfield  {journal} {\bibinfo
  {journal} {Phys. Rev. D}\ }\textbf {\bibinfo {volume} {99}},\ \bibinfo
  {pages} {124046} (\bibinfo {year} {2019})},\ \Eprint
  {https://arxiv.org/abs/1811.04432} {arXiv:1811.04432 [gr-qc]} \BibitemShut
  {NoStop}%
\bibitem [{\citenamefont {van~de Meent}(2016)}]{vandeMeent:2016pee}%
  \BibitemOpen
  \bibfield  {author} {\bibinfo {author} {\bibfnamefont {M.}~\bibnamefont
  {van~de Meent}},\ }\bibfield  {title} {\bibinfo {title} {{Gravitational
  self-force on eccentric equatorial orbits around a Kerr black hole}},\ }\href
  {https://doi.org/10.1103/PhysRevD.94.044034} {\bibfield  {journal} {\bibinfo
  {journal} {Phys. Rev. D}\ }\textbf {\bibinfo {volume} {94}},\ \bibinfo
  {pages} {044034} (\bibinfo {year} {2016})},\ \Eprint
  {https://arxiv.org/abs/1606.06297} {arXiv:1606.06297 [gr-qc]} \BibitemShut
  {NoStop}%
\bibitem [{\citenamefont {Blanchet}\ \emph {et~al.}(2010)\citenamefont
  {Blanchet}, \citenamefont {Detweiler}, \citenamefont {Le~Tiec},\ and\
  \citenamefont {Whiting}}]{Blanchet:2009sd}%
  \BibitemOpen
  \bibfield  {author} {\bibinfo {author} {\bibfnamefont {L.}~\bibnamefont
  {Blanchet}}, \bibinfo {author} {\bibfnamefont {S.~L.}\ \bibnamefont
  {Detweiler}}, \bibinfo {author} {\bibfnamefont {A.}~\bibnamefont {Le~Tiec}},\
  and\ \bibinfo {author} {\bibfnamefont {B.~F.}\ \bibnamefont {Whiting}},\
  }\bibfield  {title} {\bibinfo {title} {{Post-Newtonian and Numerical
  Calculations of the Gravitational Self-Force for Circular Orbits in the
  Schwarzschild Geometry}},\ }\href
  {https://doi.org/10.1103/PhysRevD.81.064004} {\bibfield  {journal} {\bibinfo
  {journal} {Phys. Rev. D}\ }\textbf {\bibinfo {volume} {81}},\ \bibinfo
  {pages} {064004} (\bibinfo {year} {2010})},\ \Eprint
  {https://arxiv.org/abs/0910.0207} {arXiv:0910.0207 [gr-qc]} \BibitemShut
  {NoStop}%
\bibitem [{\citenamefont {Bini}\ and\ \citenamefont
  {Damour}(2013)}]{Bini:2013zaa}%
  \BibitemOpen
  \bibfield  {author} {\bibinfo {author} {\bibfnamefont {D.}~\bibnamefont
  {Bini}}\ and\ \bibinfo {author} {\bibfnamefont {T.}~\bibnamefont {Damour}},\
  }\bibfield  {title} {\bibinfo {title} {{Analytical determination of the
  two-body gravitational interaction potential at the fourth post-Newtonian
  approximation}},\ }\href {https://doi.org/10.1103/PhysRevD.87.121501}
  {\bibfield  {journal} {\bibinfo  {journal} {Phys. Rev. D}\ }\textbf {\bibinfo
  {volume} {87}},\ \bibinfo {pages} {121501} (\bibinfo {year} {2013})},\
  \Eprint {https://arxiv.org/abs/1305.4884} {arXiv:1305.4884 [gr-qc]}
  \BibitemShut {NoStop}%
\bibitem [{\citenamefont {Le~Tiec}(2015)}]{LeTiec:2015kgg}%
  \BibitemOpen
  \bibfield  {author} {\bibinfo {author} {\bibfnamefont {A.}~\bibnamefont
  {Le~Tiec}},\ }\bibfield  {title} {\bibinfo {title} {{First Law of Mechanics
  for Compact Binaries on Eccentric Orbits}},\ }\href
  {https://doi.org/10.1103/PhysRevD.92.084021} {\bibfield  {journal} {\bibinfo
  {journal} {Phys. Rev. D}\ }\textbf {\bibinfo {volume} {92}},\ \bibinfo
  {pages} {084021} (\bibinfo {year} {2015})},\ \Eprint
  {https://arxiv.org/abs/1506.05648} {arXiv:1506.05648 [gr-qc]} \BibitemShut
  {NoStop}%
\bibitem [{\citenamefont {Blanchet}\ and\ \citenamefont
  {Le~Tiec}(2017)}]{Blanchet:2017rcn}%
  \BibitemOpen
  \bibfield  {author} {\bibinfo {author} {\bibfnamefont {L.}~\bibnamefont
  {Blanchet}}\ and\ \bibinfo {author} {\bibfnamefont {A.}~\bibnamefont
  {Le~Tiec}},\ }\bibfield  {title} {\bibinfo {title} {{First Law of Compact
  Binary Mechanics with Gravitational-Wave Tails}},\ }\href
  {https://doi.org/10.1088/1361-6382/aa79d7} {\bibfield  {journal} {\bibinfo
  {journal} {Class. Quant. Grav.}\ }\textbf {\bibinfo {volume} {34}},\ \bibinfo
  {pages} {164001} (\bibinfo {year} {2017})},\ \Eprint
  {https://arxiv.org/abs/1702.06839} {arXiv:1702.06839 [gr-qc]} \BibitemShut
  {NoStop}%
\bibitem [{\citenamefont {Bini}\ and\ \citenamefont
  {Geralico}(2019)}]{Bini:2019lcd}%
  \BibitemOpen
  \bibfield  {author} {\bibinfo {author} {\bibfnamefont {D.}~\bibnamefont
  {Bini}}\ and\ \bibinfo {author} {\bibfnamefont {A.}~\bibnamefont
  {Geralico}},\ }\bibfield  {title} {\bibinfo {title} {{New gravitational
  self-force analytical results for eccentric equatorial orbits around a Kerr
  black hole: redshift invariant}},\ }\href
  {https://doi.org/10.1103/PhysRevD.100.104002} {\bibfield  {journal} {\bibinfo
   {journal} {Phys. Rev. D}\ }\textbf {\bibinfo {volume} {100}},\ \bibinfo
  {pages} {104002} (\bibinfo {year} {2019})},\ \Eprint
  {https://arxiv.org/abs/1907.11080} {arXiv:1907.11080 [gr-qc]} \BibitemShut
  {NoStop}%
\bibitem [{\citenamefont {Akcay}\ \emph {et~al.}(2015)\citenamefont {Akcay},
  \citenamefont {Le~Tiec}, \citenamefont {Barack}, \citenamefont {Sago},\ and\
  \citenamefont {Warburton}}]{Akcay:2015pza}%
  \BibitemOpen
  \bibfield  {author} {\bibinfo {author} {\bibfnamefont {S.}~\bibnamefont
  {Akcay}}, \bibinfo {author} {\bibfnamefont {A.}~\bibnamefont {Le~Tiec}},
  \bibinfo {author} {\bibfnamefont {L.}~\bibnamefont {Barack}}, \bibinfo
  {author} {\bibfnamefont {N.}~\bibnamefont {Sago}},\ and\ \bibinfo {author}
  {\bibfnamefont {N.}~\bibnamefont {Warburton}},\ }\bibfield  {title} {\bibinfo
  {title} {{Comparison Between Self-Force and Post-Newtonian Dynamics: Beyond
  Circular Orbits}},\ }\href {https://doi.org/10.1103/PhysRevD.91.124014}
  {\bibfield  {journal} {\bibinfo  {journal} {Phys. Rev. D}\ }\textbf {\bibinfo
  {volume} {91}},\ \bibinfo {pages} {124014} (\bibinfo {year} {2015})},\
  \Eprint {https://arxiv.org/abs/1503.01374} {arXiv:1503.01374 [gr-qc]}
  \BibitemShut {NoStop}%
\bibitem [{\citenamefont {Kavanagh}\ \emph {et~al.}(2015)\citenamefont
  {Kavanagh}, \citenamefont {Ottewill},\ and\ \citenamefont
  {Wardell}}]{Kavanagh:2015lva}%
  \BibitemOpen
  \bibfield  {author} {\bibinfo {author} {\bibfnamefont {C.}~\bibnamefont
  {Kavanagh}}, \bibinfo {author} {\bibfnamefont {A.~C.}\ \bibnamefont
  {Ottewill}},\ and\ \bibinfo {author} {\bibfnamefont {B.}~\bibnamefont
  {Wardell}},\ }\bibfield  {title} {\bibinfo {title} {{Analytical high-order
  post-Newtonian expansions for extreme mass ratio binaries}},\ }\href
  {https://doi.org/10.1103/PhysRevD.92.084025} {\bibfield  {journal} {\bibinfo
  {journal} {Phys. Rev. D}\ }\textbf {\bibinfo {volume} {92}},\ \bibinfo
  {pages} {084025} (\bibinfo {year} {2015})},\ \Eprint
  {https://arxiv.org/abs/1503.02334} {arXiv:1503.02334 [gr-qc]} \BibitemShut
  {NoStop}%
\bibitem [{\citenamefont {Bini}\ \emph {et~al.}(2016)\citenamefont {Bini},
  \citenamefont {Damour},\ and\ \citenamefont {Geralico}}]{Bini:2016qtx}%
  \BibitemOpen
  \bibfield  {author} {\bibinfo {author} {\bibfnamefont {D.}~\bibnamefont
  {Bini}}, \bibinfo {author} {\bibfnamefont {T.}~\bibnamefont {Damour}},\ and\
  \bibinfo {author} {\bibfnamefont {a.}~\bibnamefont {Geralico}},\ }\bibfield
  {title} {\bibinfo {title} {{New gravitational self-force analytical results
  for eccentric orbits around a Schwarzschild black hole}},\ }\href
  {https://doi.org/10.1103/PhysRevD.93.104017} {\bibfield  {journal} {\bibinfo
  {journal} {Phys. Rev. D}\ }\textbf {\bibinfo {volume} {93}},\ \bibinfo
  {pages} {104017} (\bibinfo {year} {2016})},\ \Eprint
  {https://arxiv.org/abs/1601.02988} {arXiv:1601.02988 [gr-qc]} \BibitemShut
  {NoStop}%
\bibitem [{\citenamefont {Kavanagh}\ \emph {et~al.}(2016)\citenamefont
  {Kavanagh}, \citenamefont {Ottewill},\ and\ \citenamefont
  {Wardell}}]{Kavanagh:2016idg}%
  \BibitemOpen
  \bibfield  {author} {\bibinfo {author} {\bibfnamefont {C.}~\bibnamefont
  {Kavanagh}}, \bibinfo {author} {\bibfnamefont {A.~C.}\ \bibnamefont
  {Ottewill}},\ and\ \bibinfo {author} {\bibfnamefont {B.}~\bibnamefont
  {Wardell}},\ }\bibfield  {title} {\bibinfo {title} {{Analytical high-order
  post-Newtonian expansions for spinning extreme mass ratio binaries}},\ }\href
  {https://doi.org/10.1103/PhysRevD.93.124038} {\bibfield  {journal} {\bibinfo
  {journal} {Phys. Rev. D}\ }\textbf {\bibinfo {volume} {93}},\ \bibinfo
  {pages} {124038} (\bibinfo {year} {2016})},\ \Eprint
  {https://arxiv.org/abs/1601.03394} {arXiv:1601.03394 [gr-qc]} \BibitemShut
  {NoStop}%
\bibitem [{BHP()}]{BHPToolkit}%
  \BibitemOpen
  \href@noop {} {\bibinfo {title} {{Black Hole Perturbation Toolkit}}},\
  \bibinfo {howpublished}
  {(\href{http://bhptoolkit.org/}{bhptoolkit.org})}\BibitemShut {NoStop}%
\bibitem [{\citenamefont {Zimmerman}\ \emph {et~al.}(2016)\citenamefont
  {Zimmerman}, \citenamefont {Lewis},\ and\ \citenamefont
  {Pfeiffer}}]{Zimmerman:2016ajr}%
  \BibitemOpen
  \bibfield  {author} {\bibinfo {author} {\bibfnamefont {A.}~\bibnamefont
  {Zimmerman}}, \bibinfo {author} {\bibfnamefont {A.~G.~M.}\ \bibnamefont
  {Lewis}},\ and\ \bibinfo {author} {\bibfnamefont {H.~P.}\ \bibnamefont
  {Pfeiffer}},\ }\bibfield  {title} {\bibinfo {title} {{Redshift factor and the
  first law of binary black hole mechanics in numerical simulations}},\ }\href
  {https://doi.org/10.1103/PhysRevLett.117.191101} {\bibfield  {journal}
  {\bibinfo  {journal} {Phys. Rev. Lett.}\ }\textbf {\bibinfo {volume} {117}},\
  \bibinfo {pages} {191101} (\bibinfo {year} {2016})},\ \Eprint
  {https://arxiv.org/abs/1606.08056} {arXiv:1606.08056 [gr-qc]} \BibitemShut
  {NoStop}%
\bibitem [{\citenamefont {Le~Tiec}\ and\ \citenamefont
  {Grandcl\'ement}(2018)}]{LeTiec:2017ebm}%
  \BibitemOpen
  \bibfield  {author} {\bibinfo {author} {\bibfnamefont {A.}~\bibnamefont
  {Le~Tiec}}\ and\ \bibinfo {author} {\bibfnamefont {P.}~\bibnamefont
  {Grandcl\'ement}},\ }\bibfield  {title} {\bibinfo {title} {{Horizon Surface
  Gravity in Corotating Black Hole Binaries}},\ }\href
  {https://doi.org/10.1088/1361-6382/aac58c} {\bibfield  {journal} {\bibinfo
  {journal} {Class. Quant. Grav.}\ }\textbf {\bibinfo {volume} {35}},\ \bibinfo
  {pages} {144002} (\bibinfo {year} {2018})},\ \Eprint
  {https://arxiv.org/abs/1710.03673} {arXiv:1710.03673 [gr-qc]} \BibitemShut
  {NoStop}%
\bibitem [{\citenamefont {Bonazzola}\ \emph {et~al.}(1997)\citenamefont
  {Bonazzola}, \citenamefont {Gourgoulhon},\ and\ \citenamefont
  {Marck}}]{Bonazzola:1997gc}%
  \BibitemOpen
  \bibfield  {author} {\bibinfo {author} {\bibfnamefont {S.}~\bibnamefont
  {Bonazzola}}, \bibinfo {author} {\bibfnamefont {E.}~\bibnamefont
  {Gourgoulhon}},\ and\ \bibinfo {author} {\bibfnamefont {J.-A.}\ \bibnamefont
  {Marck}},\ }\bibfield  {title} {\bibinfo {title} {{A Relativistic formalism
  to compute quasiequilibrium configurations of nonsynchronized neutron star
  binaries}},\ }\href {https://doi.org/10.1103/PhysRevD.56.7740} {\bibfield
  {journal} {\bibinfo  {journal} {Phys. Rev. D}\ }\textbf {\bibinfo {volume}
  {56}},\ \bibinfo {pages} {7740} (\bibinfo {year} {1997})},\ \Eprint
  {https://arxiv.org/abs/gr-qc/9710031} {arXiv:gr-qc/9710031} \BibitemShut
  {NoStop}%
\bibitem [{\citenamefont {Detweiler}\ and\ \citenamefont
  {Whiting}(2003)}]{Detweiler:2002mi}%
  \BibitemOpen
  \bibfield  {author} {\bibinfo {author} {\bibfnamefont {S.~L.}\ \bibnamefont
  {Detweiler}}\ and\ \bibinfo {author} {\bibfnamefont {B.~F.}\ \bibnamefont
  {Whiting}},\ }\bibfield  {title} {\bibinfo {title} {{Selfforce via a Green's
  function decomposition}},\ }\href
  {https://doi.org/10.1103/PhysRevD.67.024025} {\bibfield  {journal} {\bibinfo
  {journal} {Phys. Rev. D}\ }\textbf {\bibinfo {volume} {67}},\ \bibinfo
  {pages} {024025} (\bibinfo {year} {2003})},\ \Eprint
  {https://arxiv.org/abs/gr-qc/0202086} {arXiv:gr-qc/0202086} \BibitemShut
  {NoStop}%
\bibitem [{\citenamefont {Gibbons}\ and\ \citenamefont
  {Stewart}(1984)}]{Gibbons}%
  \BibitemOpen
  \bibfield  {author} {\bibinfo {author} {\bibfnamefont {G.}~\bibnamefont
  {Gibbons}}\ and\ \bibinfo {author} {\bibfnamefont {J.~M.}\ \bibnamefont
  {Stewart}},\ }\bibfield  {title} {\bibinfo {title} {{Absence of
  asymptotically flat solutions of Einstein's equations which are periodic and
  empty near infinity}},\ }\href@noop {} {\bibfield  {journal} {\bibinfo
  {journal} {Classical general relativity}\ } (\bibinfo {year}
  {1984})}\BibitemShut {NoStop}%
\bibitem [{\citenamefont {Bardeen}\ \emph {et~al.}(1973)\citenamefont
  {Bardeen}, \citenamefont {Carter},\ and\ \citenamefont
  {Hawking}}]{cmp/1103858973}%
  \BibitemOpen
  \bibfield  {author} {\bibinfo {author} {\bibfnamefont {J.~M.}\ \bibnamefont
  {Bardeen}}, \bibinfo {author} {\bibfnamefont {B.}~\bibnamefont {Carter}},\
  and\ \bibinfo {author} {\bibfnamefont {S.~W.}\ \bibnamefont {Hawking}},\
  }\bibfield  {title} {\bibinfo {title} {{The four laws of black hole
  mechanics}},\ }\href {https://doi.org/cmp/1103858973} {\bibfield  {journal}
  {\bibinfo  {journal} {Communications in Mathematical Physics}\ }\textbf
  {\bibinfo {volume} {31}},\ \bibinfo {pages} {161 } (\bibinfo {year}
  {1973})}\BibitemShut {NoStop}%
\bibitem [{\citenamefont {Friedman}\ \emph {et~al.}(2002)\citenamefont
  {Friedman}, \citenamefont {Uryu},\ and\ \citenamefont
  {Shibata}}]{Friedman:2001pf}%
  \BibitemOpen
  \bibfield  {author} {\bibinfo {author} {\bibfnamefont {J.~L.}\ \bibnamefont
  {Friedman}}, \bibinfo {author} {\bibfnamefont {K.}~\bibnamefont {Uryu}},\
  and\ \bibinfo {author} {\bibfnamefont {M.}~\bibnamefont {Shibata}},\
  }\bibfield  {title} {\bibinfo {title} {{Thermodynamics of binary black holes
  and neutron stars}},\ }\href {https://doi.org/10.1103/PhysRevD.70.129904}
  {\bibfield  {journal} {\bibinfo  {journal} {Phys. Rev. D}\ }\textbf {\bibinfo
  {volume} {65}},\ \bibinfo {pages} {064035} (\bibinfo {year} {2002})},\
  \bibinfo {note} {[Erratum: Phys.Rev.D 70, 129904 (2004)]},\ \Eprint
  {https://arxiv.org/abs/gr-qc/0108070} {arXiv:gr-qc/0108070} \BibitemShut
  {NoStop}%
\bibitem [{\citenamefont {Pound}(2015)}]{PoundRedshift:2015}%
  \BibitemOpen
  \bibfield  {author} {\bibinfo {author} {\bibfnamefont {A.}~\bibnamefont
  {Pound}},\ }\bibfield  {title} {\bibinfo {title} {Self-force effects on the
  horizon geometry of a small black hole}} (\bibinfo {year} {2015}),\ \bibinfo
  {note} {unpublished}\BibitemShut {NoStop}%
\bibitem [{\citenamefont {Poisson}(2005)}]{Poisson:2005pi}%
  \BibitemOpen
  \bibfield  {author} {\bibinfo {author} {\bibfnamefont {E.}~\bibnamefont
  {Poisson}},\ }\bibfield  {title} {\bibinfo {title} {{Metric of a tidally
  distorted, nonrotating black hole}},\ }\href
  {https://doi.org/10.1103/PhysRevLett.94.161103} {\bibfield  {journal}
  {\bibinfo  {journal} {Phys. Rev. Lett.}\ }\textbf {\bibinfo {volume} {94}},\
  \bibinfo {pages} {161103} (\bibinfo {year} {2005})},\ \Eprint
  {https://arxiv.org/abs/gr-qc/0501032} {arXiv:gr-qc/0501032} \BibitemShut
  {NoStop}%
\bibitem [{\citenamefont {Gralla}\ and\ \citenamefont
  {Le~Tiec}(2013)}]{Gralla:2012dm}%
  \BibitemOpen
  \bibfield  {author} {\bibinfo {author} {\bibfnamefont {S.~E.}\ \bibnamefont
  {Gralla}}\ and\ \bibinfo {author} {\bibfnamefont {A.}~\bibnamefont
  {Le~Tiec}},\ }\bibfield  {title} {\bibinfo {title} {{Thermodynamics of a
  Black Hole with Moon}},\ }\href {https://doi.org/10.1103/PhysRevD.88.044021}
  {\bibfield  {journal} {\bibinfo  {journal} {Phys. Rev. D}\ }\textbf {\bibinfo
  {volume} {88}},\ \bibinfo {pages} {044021} (\bibinfo {year} {2013})},\
  \Eprint {https://arxiv.org/abs/1210.8444} {arXiv:1210.8444 [gr-qc]}
  \BibitemShut {NoStop}%
\bibitem [{\citenamefont {Ashtekar}\ and\ \citenamefont
  {Krishnan}(2003)}]{Ashtekar:2003hk}%
  \BibitemOpen
  \bibfield  {author} {\bibinfo {author} {\bibfnamefont {A.}~\bibnamefont
  {Ashtekar}}\ and\ \bibinfo {author} {\bibfnamefont {B.}~\bibnamefont
  {Krishnan}},\ }\bibfield  {title} {\bibinfo {title} {{Dynamical horizons and
  their properties}},\ }\href {https://doi.org/10.1103/PhysRevD.68.104030}
  {\bibfield  {journal} {\bibinfo  {journal} {Phys. Rev. D}\ }\textbf {\bibinfo
  {volume} {68}},\ \bibinfo {pages} {104030} (\bibinfo {year} {2003})},\
  \Eprint {https://arxiv.org/abs/gr-qc/0308033} {arXiv:gr-qc/0308033}
  \BibitemShut {NoStop}%
\bibitem [{\citenamefont {Gourgoulhon}\ and\ \citenamefont
  {Jaramillo}(2006)}]{Gourgoulhon:2005ng}%
  \BibitemOpen
  \bibfield  {author} {\bibinfo {author} {\bibfnamefont {E.}~\bibnamefont
  {Gourgoulhon}}\ and\ \bibinfo {author} {\bibfnamefont {J.~L.}\ \bibnamefont
  {Jaramillo}},\ }\bibfield  {title} {\bibinfo {title} {{A 3+1 perspective on
  null hypersurfaces and isolated horizons}},\ }\href
  {https://doi.org/10.1016/j.physrep.2005.10.005} {\bibfield  {journal}
  {\bibinfo  {journal} {Phys. Rept.}\ }\textbf {\bibinfo {volume} {423}},\
  \bibinfo {pages} {159} (\bibinfo {year} {2006})},\ \Eprint
  {https://arxiv.org/abs/gr-qc/0503113} {arXiv:gr-qc/0503113} \BibitemShut
  {NoStop}%
\bibitem [{\citenamefont {Boyle}(2013)}]{Boyle:2013nka}%
  \BibitemOpen
  \bibfield  {author} {\bibinfo {author} {\bibfnamefont {M.}~\bibnamefont
  {Boyle}},\ }\bibfield  {title} {\bibinfo {title} {{Angular velocity of
  gravitational radiation from precessing binaries and the corotating frame}},\
  }\href {https://doi.org/10.1103/PhysRevD.87.104006} {\bibfield  {journal}
  {\bibinfo  {journal} {Phys. Rev. D}\ }\textbf {\bibinfo {volume} {87}},\
  \bibinfo {pages} {104006} (\bibinfo {year} {2013})},\ \Eprint
  {https://arxiv.org/abs/1302.2919} {arXiv:1302.2919 [gr-qc]} \BibitemShut
  {NoStop}%
\bibitem [{\citenamefont {Boyle}(2016)}]{Boyle:2015nqa}%
  \BibitemOpen
  \bibfield  {author} {\bibinfo {author} {\bibfnamefont {M.}~\bibnamefont
  {Boyle}},\ }\bibfield  {title} {\bibinfo {title} {{Transformations of
  asymptotic gravitational-wave data}},\ }\href
  {https://doi.org/10.1103/PhysRevD.93.084031} {\bibfield  {journal} {\bibinfo
  {journal} {Phys. Rev. D}\ }\textbf {\bibinfo {volume} {93}},\ \bibinfo
  {pages} {084031} (\bibinfo {year} {2016})},\ \Eprint
  {https://arxiv.org/abs/1509.00862} {arXiv:1509.00862 [gr-qc]} \BibitemShut
  {NoStop}%
\bibitem [{\citenamefont {Boyle}\ \emph {et~al.}(2020)\citenamefont {Boyle},
  \citenamefont {Iozzo}, \citenamefont {Stein}, \citenamefont {Khairnar},\ and\
  \citenamefont {Rüter}}]{Boyle_scri_2020}%
  \BibitemOpen
  \bibfield  {author} {\bibinfo {author} {\bibfnamefont {M.}~\bibnamefont
  {Boyle}}, \bibinfo {author} {\bibfnamefont {D.}~\bibnamefont {Iozzo}},
  \bibinfo {author} {\bibfnamefont {L.}~\bibnamefont {Stein}}, \bibinfo
  {author} {\bibfnamefont {A.}~\bibnamefont {Khairnar}},\ and\ \bibinfo
  {author} {\bibfnamefont {H.}~\bibnamefont {Rüter}},\ }\href
  {https://doi.org/10.5281/zenodo.4041972} {\bibinfo {title} {{scri}}}
  (\bibinfo {year} {2020})\BibitemShut {NoStop}%
\bibitem [{\citenamefont {Boyle}\ \emph {et~al.}(2019)\citenamefont {Boyle}
  \emph {et~al.}}]{Boyle:2019kee}%
  \BibitemOpen
  \bibfield  {author} {\bibinfo {author} {\bibfnamefont {M.}~\bibnamefont
  {Boyle}} \emph {et~al.},\ }\bibfield  {title} {\bibinfo {title} {{The SXS
  Collaboration catalog of binary black hole simulations}},\ }\href
  {https://doi.org/10.1088/1361-6382/ab34e2} {\bibfield  {journal} {\bibinfo
  {journal} {Class. Quant. Grav.}\ }\textbf {\bibinfo {volume} {36}},\ \bibinfo
  {pages} {195006} (\bibinfo {year} {2019})},\ \Eprint
  {https://arxiv.org/abs/1904.04831} {arXiv:1904.04831 [gr-qc]} \BibitemShut
  {NoStop}%
\bibitem [{SpE()}]{SpECwebsite}%
  \BibitemOpen
  \href@noop {} {}\bibinfo {howpublished}
  {\url{http://www.black-holes.org/SpEC.html}}\BibitemShut {NoStop}%
\bibitem [{\citenamefont {York}(1999)}]{York:1998hy}%
  \BibitemOpen
  \bibfield  {author} {\bibinfo {author} {\bibfnamefont {J.~W.}\ \bibnamefont
  {York}, \bibfnamefont {Jr.}},\ }\bibfield  {title} {\bibinfo {title}
  {{Conformal 'thin sandwich' data for the initial-value problem}},\ }\href
  {https://doi.org/10.1103/PhysRevLett.82.1350} {\bibfield  {journal} {\bibinfo
   {journal} {Phys. Rev. Lett.}\ }\textbf {\bibinfo {volume} {82}},\ \bibinfo
  {pages} {1350} (\bibinfo {year} {1999})},\ \Eprint
  {https://arxiv.org/abs/gr-qc/9810051} {arXiv:gr-qc/9810051} \BibitemShut
  {NoStop}%
\bibitem [{\citenamefont {Pfeiffer}\ and\ \citenamefont
  {York}(2003)}]{Pfeiffer:2002iy}%
  \BibitemOpen
  \bibfield  {author} {\bibinfo {author} {\bibfnamefont {H.~P.}\ \bibnamefont
  {Pfeiffer}}\ and\ \bibinfo {author} {\bibfnamefont {J.~W.}\ \bibnamefont
  {York}, \bibfnamefont {Jr.}},\ }\bibfield  {title} {\bibinfo {title}
  {{Extrinsic curvature and the Einstein constraints}},\ }\href
  {https://doi.org/10.1103/PhysRevD.67.044022} {\bibfield  {journal} {\bibinfo
  {journal} {Phys. Rev. D}\ }\textbf {\bibinfo {volume} {67}},\ \bibinfo
  {pages} {044022} (\bibinfo {year} {2003})},\ \Eprint
  {https://arxiv.org/abs/gr-qc/0207095} {arXiv:gr-qc/0207095} \BibitemShut
  {NoStop}%
\bibitem [{\citenamefont {Pfeiffer}\ \emph {et~al.}(2003)\citenamefont
  {Pfeiffer}, \citenamefont {Kidder}, \citenamefont {Scheel},\ and\
  \citenamefont {Teukolsky}}]{Pfeiffer:2002wt}%
  \BibitemOpen
  \bibfield  {author} {\bibinfo {author} {\bibfnamefont {H.~P.}\ \bibnamefont
  {Pfeiffer}}, \bibinfo {author} {\bibfnamefont {L.~E.}\ \bibnamefont
  {Kidder}}, \bibinfo {author} {\bibfnamefont {M.~A.}\ \bibnamefont {Scheel}},\
  and\ \bibinfo {author} {\bibfnamefont {S.~A.}\ \bibnamefont {Teukolsky}},\
  }\bibfield  {title} {\bibinfo {title} {{A Multidomain spectral method for
  solving elliptic equations}},\ }\href
  {https://doi.org/10.1016/S0010-4655(02)00847-0} {\bibfield  {journal}
  {\bibinfo  {journal} {Comput. Phys. Commun.}\ }\textbf {\bibinfo {volume}
  {152}},\ \bibinfo {pages} {253} (\bibinfo {year} {2003})},\ \Eprint
  {https://arxiv.org/abs/gr-qc/0202096} {arXiv:gr-qc/0202096} \BibitemShut
  {NoStop}%
\bibitem [{\citenamefont {Lindblom}\ and\ \citenamefont
  {Szilagyi}(2009)}]{Lindblom:2009tu}%
  \BibitemOpen
  \bibfield  {author} {\bibinfo {author} {\bibfnamefont {L.}~\bibnamefont
  {Lindblom}}\ and\ \bibinfo {author} {\bibfnamefont {B.}~\bibnamefont
  {Szilagyi}},\ }\bibfield  {title} {\bibinfo {title} {{An Improved Gauge
  Driver for the GH Einstein System}},\ }\href
  {https://doi.org/10.1103/PhysRevD.80.084019} {\bibfield  {journal} {\bibinfo
  {journal} {Phys. Rev. D}\ }\textbf {\bibinfo {volume} {80}},\ \bibinfo
  {pages} {084019} (\bibinfo {year} {2009})},\ \Eprint
  {https://arxiv.org/abs/0904.4873} {arXiv:0904.4873 [gr-qc]} \BibitemShut
  {NoStop}%
\bibitem [{\citenamefont {Choptuik}\ and\ \citenamefont
  {Pretorius}(2010)}]{Choptuik:2009ww}%
  \BibitemOpen
  \bibfield  {author} {\bibinfo {author} {\bibfnamefont {M.~W.}\ \bibnamefont
  {Choptuik}}\ and\ \bibinfo {author} {\bibfnamefont {F.}~\bibnamefont
  {Pretorius}},\ }\bibfield  {title} {\bibinfo {title} {{Ultra Relativistic
  Particle Collisions}},\ }\href
  {https://doi.org/10.1103/PhysRevLett.104.111101} {\bibfield  {journal}
  {\bibinfo  {journal} {Phys. Rev. Lett.}\ }\textbf {\bibinfo {volume} {104}},\
  \bibinfo {pages} {111101} (\bibinfo {year} {2010})},\ \Eprint
  {https://arxiv.org/abs/0908.1780} {arXiv:0908.1780 [gr-qc]} \BibitemShut
  {NoStop}%
\bibitem [{\citenamefont {Szilagyi}\ \emph {et~al.}(2009)\citenamefont
  {Szilagyi}, \citenamefont {Lindblom},\ and\ \citenamefont
  {Scheel}}]{Szilagyi:2009qz}%
  \BibitemOpen
  \bibfield  {author} {\bibinfo {author} {\bibfnamefont {B.}~\bibnamefont
  {Szilagyi}}, \bibinfo {author} {\bibfnamefont {L.}~\bibnamefont {Lindblom}},\
  and\ \bibinfo {author} {\bibfnamefont {M.~A.}\ \bibnamefont {Scheel}},\
  }\bibfield  {title} {\bibinfo {title} {{Simulations of Binary Black Hole
  Mergers Using Spectral Methods}},\ }\href
  {https://doi.org/10.1103/PhysRevD.80.124010} {\bibfield  {journal} {\bibinfo
  {journal} {Phys. Rev. D}\ }\textbf {\bibinfo {volume} {80}},\ \bibinfo
  {pages} {124010} (\bibinfo {year} {2009})},\ \Eprint
  {https://arxiv.org/abs/0909.3557} {arXiv:0909.3557 [gr-qc]} \BibitemShut
  {NoStop}%
\bibitem [{\citenamefont {Hemberger}\ \emph {et~al.}(2013)\citenamefont
  {Hemberger}, \citenamefont {Scheel}, \citenamefont {Kidder}, \citenamefont
  {Szil\'agyi}, \citenamefont {Lovelace}, \citenamefont {Taylor},\ and\
  \citenamefont {Teukolsky}}]{Hemberger:2012jz}%
  \BibitemOpen
  \bibfield  {author} {\bibinfo {author} {\bibfnamefont {D.~A.}\ \bibnamefont
  {Hemberger}}, \bibinfo {author} {\bibfnamefont {M.~A.}\ \bibnamefont
  {Scheel}}, \bibinfo {author} {\bibfnamefont {L.~E.}\ \bibnamefont {Kidder}},
  \bibinfo {author} {\bibfnamefont {B.}~\bibnamefont {Szil\'agyi}}, \bibinfo
  {author} {\bibfnamefont {G.}~\bibnamefont {Lovelace}}, \bibinfo {author}
  {\bibfnamefont {N.~W.}\ \bibnamefont {Taylor}},\ and\ \bibinfo {author}
  {\bibfnamefont {S.~A.}\ \bibnamefont {Teukolsky}},\ }\bibfield  {title}
  {\bibinfo {title} {{Dynamical Excision Boundaries in Spectral Evolutions of
  Binary Black Hole Spacetimes}},\ }\href
  {https://doi.org/10.1088/0264-9381/30/11/115001} {\bibfield  {journal}
  {\bibinfo  {journal} {Class. Quant. Grav.}\ }\textbf {\bibinfo {volume}
  {30}},\ \bibinfo {pages} {115001} (\bibinfo {year} {2013})},\ \Eprint
  {https://arxiv.org/abs/1211.6079} {arXiv:1211.6079 [gr-qc]} \BibitemShut
  {NoStop}%
\bibitem [{\citenamefont {Lovelace}\ \emph {et~al.}(2008)\citenamefont
  {Lovelace}, \citenamefont {Owen}, \citenamefont {Pfeiffer},\ and\
  \citenamefont {Chu}}]{Lovelace:2008tw}%
  \BibitemOpen
  \bibfield  {author} {\bibinfo {author} {\bibfnamefont {G.}~\bibnamefont
  {Lovelace}}, \bibinfo {author} {\bibfnamefont {R.}~\bibnamefont {Owen}},
  \bibinfo {author} {\bibfnamefont {H.~P.}\ \bibnamefont {Pfeiffer}},\ and\
  \bibinfo {author} {\bibfnamefont {T.}~\bibnamefont {Chu}},\ }\bibfield
  {title} {\bibinfo {title} {{Binary-black-hole initial data with
  nearly-extremal spins}},\ }\href {https://doi.org/10.1103/PhysRevD.78.084017}
  {\bibfield  {journal} {\bibinfo  {journal} {Phys. Rev. D}\ }\textbf {\bibinfo
  {volume} {78}},\ \bibinfo {pages} {084017} (\bibinfo {year} {2008})},\
  \Eprint {https://arxiv.org/abs/0805.4192} {arXiv:0805.4192 [gr-qc]}
  \BibitemShut {NoStop}%
\bibitem [{\citenamefont {Varma}\ \emph {et~al.}(2018)\citenamefont {Varma},
  \citenamefont {Scheel},\ and\ \citenamefont {Pfeiffer}}]{Varma:2018sqd}%
  \BibitemOpen
  \bibfield  {author} {\bibinfo {author} {\bibfnamefont {V.}~\bibnamefont
  {Varma}}, \bibinfo {author} {\bibfnamefont {M.~A.}\ \bibnamefont {Scheel}},\
  and\ \bibinfo {author} {\bibfnamefont {H.~P.}\ \bibnamefont {Pfeiffer}},\
  }\bibfield  {title} {\bibinfo {title} {{Comparison of binary black hole
  initial data sets}},\ }\href {https://doi.org/10.1103/PhysRevD.98.104011}
  {\bibfield  {journal} {\bibinfo  {journal} {Phys. Rev. D}\ }\textbf {\bibinfo
  {volume} {98}},\ \bibinfo {pages} {104011} (\bibinfo {year} {2018})},\
  \Eprint {https://arxiv.org/abs/1808.08228} {arXiv:1808.08228 [gr-qc]}
  \BibitemShut {NoStop}%
\bibitem [{\citenamefont {Ossokine}\ \emph {et~al.}(2015)\citenamefont
  {Ossokine}, \citenamefont {Foucart}, \citenamefont {Pfeiffer}, \citenamefont
  {Boyle},\ and\ \citenamefont {Szil\'agyi}}]{Ossokine:2015yla}%
  \BibitemOpen
  \bibfield  {author} {\bibinfo {author} {\bibfnamefont {S.}~\bibnamefont
  {Ossokine}}, \bibinfo {author} {\bibfnamefont {F.}~\bibnamefont {Foucart}},
  \bibinfo {author} {\bibfnamefont {H.~P.}\ \bibnamefont {Pfeiffer}}, \bibinfo
  {author} {\bibfnamefont {M.}~\bibnamefont {Boyle}},\ and\ \bibinfo {author}
  {\bibfnamefont {B.}~\bibnamefont {Szil\'agyi}},\ }\bibfield  {title}
  {\bibinfo {title} {{Improvements to the construction of binary black hole
  initial data}},\ }\href {https://doi.org/10.1088/0264-9381/32/24/245010}
  {\bibfield  {journal} {\bibinfo  {journal} {Class. Quant. Grav.}\ }\textbf
  {\bibinfo {volume} {32}},\ \bibinfo {pages} {245010} (\bibinfo {year}
  {2015})},\ \Eprint {https://arxiv.org/abs/1506.01689} {arXiv:1506.01689
  [gr-qc]} \BibitemShut {NoStop}%
\bibitem [{\citenamefont {Pfeiffer}\ \emph {et~al.}(2007)\citenamefont
  {Pfeiffer}, \citenamefont {Brown}, \citenamefont {Kidder}, \citenamefont
  {Lindblom}, \citenamefont {Lovelace},\ and\ \citenamefont
  {Scheel}}]{Pfeiffer:2007yz}%
  \BibitemOpen
  \bibfield  {author} {\bibinfo {author} {\bibfnamefont {H.~P.}\ \bibnamefont
  {Pfeiffer}}, \bibinfo {author} {\bibfnamefont {D.~A.}\ \bibnamefont {Brown}},
  \bibinfo {author} {\bibfnamefont {L.~E.}\ \bibnamefont {Kidder}}, \bibinfo
  {author} {\bibfnamefont {L.}~\bibnamefont {Lindblom}}, \bibinfo {author}
  {\bibfnamefont {G.}~\bibnamefont {Lovelace}},\ and\ \bibinfo {author}
  {\bibfnamefont {M.~A.}\ \bibnamefont {Scheel}},\ }\bibfield  {title}
  {\bibinfo {title} {{Reducing orbital eccentricity in binary black hole
  simulations}},\ }\href {https://doi.org/10.1088/0264-9381/24/12/S06}
  {\bibfield  {journal} {\bibinfo  {journal} {Class. Quant. Grav.}\ }\textbf
  {\bibinfo {volume} {24}},\ \bibinfo {pages} {S59} (\bibinfo {year} {2007})},\
  \Eprint {https://arxiv.org/abs/gr-qc/0702106} {arXiv:gr-qc/0702106}
  \BibitemShut {NoStop}%
\bibitem [{\citenamefont {Buonanno}\ \emph {et~al.}(2011)\citenamefont
  {Buonanno}, \citenamefont {Kidder}, \citenamefont {Mroue}, \citenamefont
  {Pfeiffer},\ and\ \citenamefont {Taracchini}}]{Buonanno:2010yk}%
  \BibitemOpen
  \bibfield  {author} {\bibinfo {author} {\bibfnamefont {A.}~\bibnamefont
  {Buonanno}}, \bibinfo {author} {\bibfnamefont {L.~E.}\ \bibnamefont
  {Kidder}}, \bibinfo {author} {\bibfnamefont {A.~H.}\ \bibnamefont {Mroue}},
  \bibinfo {author} {\bibfnamefont {H.~P.}\ \bibnamefont {Pfeiffer}},\ and\
  \bibinfo {author} {\bibfnamefont {A.}~\bibnamefont {Taracchini}},\ }\bibfield
   {title} {\bibinfo {title} {{Reducing orbital eccentricity of precessing
  black-hole binaries}},\ }\href {https://doi.org/10.1103/PhysRevD.83.104034}
  {\bibfield  {journal} {\bibinfo  {journal} {Phys. Rev. D}\ }\textbf {\bibinfo
  {volume} {83}},\ \bibinfo {pages} {104034} (\bibinfo {year} {2011})},\
  \Eprint {https://arxiv.org/abs/1012.1549} {arXiv:1012.1549 [gr-qc]}
  \BibitemShut {NoStop}%
\bibitem [{\citenamefont {Mroue}\ and\ \citenamefont
  {Pfeiffer}(2012)}]{Mroue:2012kv}%
  \BibitemOpen
  \bibfield  {author} {\bibinfo {author} {\bibfnamefont {A.~H.}\ \bibnamefont
  {Mroue}}\ and\ \bibinfo {author} {\bibfnamefont {H.~P.}\ \bibnamefont
  {Pfeiffer}},\ }\bibfield  {title} {\bibinfo {title} {{Precessing Binary Black
  Holes Simulations: Quasicircular Initial Data}},\ }\href@noop {} {\
  (\bibinfo {year} {2012})},\ \Eprint {https://arxiv.org/abs/1210.2958}
  {arXiv:1210.2958 [gr-qc]} \BibitemShut {NoStop}%
\bibitem [{\citenamefont {Boyle}\ and\ \citenamefont
  {Mroue}(2009)}]{Boyle:2009vi}%
  \BibitemOpen
  \bibfield  {author} {\bibinfo {author} {\bibfnamefont {M.}~\bibnamefont
  {Boyle}}\ and\ \bibinfo {author} {\bibfnamefont {A.~H.}\ \bibnamefont
  {Mroue}},\ }\bibfield  {title} {\bibinfo {title} {{Extrapolating
  gravitational-wave data from numerical simulations}},\ }\href
  {https://doi.org/10.1103/PhysRevD.80.124045} {\bibfield  {journal} {\bibinfo
  {journal} {Phys. Rev. D}\ }\textbf {\bibinfo {volume} {80}},\ \bibinfo
  {pages} {124045} (\bibinfo {year} {2009})},\ \Eprint
  {https://arxiv.org/abs/0905.3177} {arXiv:0905.3177 [gr-qc]} \BibitemShut
  {NoStop}%
\bibitem [{\citenamefont {Mitman}\ \emph {et~al.}(2021)\citenamefont {Mitman}
  \emph {et~al.}}]{Mitman:2021xkq}%
  \BibitemOpen
  \bibfield  {author} {\bibinfo {author} {\bibfnamefont {K.}~\bibnamefont
  {Mitman}} \emph {et~al.},\ }\bibfield  {title} {\bibinfo {title} {{Fixing the
  BMS Frame of Numerical Relativity Waveforms}},\ }\href@noop {} {\  (\bibinfo
  {year} {2021})},\ \Eprint {https://arxiv.org/abs/2105.02300}
  {arXiv:2105.02300 [gr-qc]} \BibitemShut {NoStop}%
\bibitem [{\citenamefont {Blanchet}(2014)}]{Blanchet:2013haa}%
  \BibitemOpen
  \bibfield  {author} {\bibinfo {author} {\bibfnamefont {L.}~\bibnamefont
  {Blanchet}},\ }\bibfield  {title} {\bibinfo {title} {{Gravitational Radiation
  from Post-Newtonian Sources and Inspiralling Compact Binaries}},\ }\href
  {https://doi.org/10.12942/lrr-2014-2} {\bibfield  {journal} {\bibinfo
  {journal} {Living Rev. Rel.}\ }\textbf {\bibinfo {volume} {17}},\ \bibinfo
  {pages} {2} (\bibinfo {year} {2014})},\ \Eprint
  {https://arxiv.org/abs/1310.1528} {arXiv:1310.1528 [gr-qc]} \BibitemShut
  {NoStop}%
\bibitem [{\citenamefont {Lewis}\ \emph {et~al.}(2017)\citenamefont {Lewis},
  \citenamefont {Zimmerman},\ and\ \citenamefont {Pfeiffer}}]{Lewis:2016lgx}%
  \BibitemOpen
  \bibfield  {author} {\bibinfo {author} {\bibfnamefont {A.~G.~M.}\
  \bibnamefont {Lewis}}, \bibinfo {author} {\bibfnamefont {A.}~\bibnamefont
  {Zimmerman}},\ and\ \bibinfo {author} {\bibfnamefont {H.~P.}\ \bibnamefont
  {Pfeiffer}},\ }\bibfield  {title} {\bibinfo {title} {{Fundamental frequencies
  and resonances from eccentric and precessing binary black hole inspirals}},\
  }\href {https://doi.org/10.1088/1361-6382/aa66f4} {\bibfield  {journal}
  {\bibinfo  {journal} {Class. Quant. Grav.}\ }\textbf {\bibinfo {volume}
  {34}},\ \bibinfo {pages} {124001} (\bibinfo {year} {2017})},\ \Eprint
  {https://arxiv.org/abs/1611.03418} {arXiv:1611.03418 [gr-qc]} \BibitemShut
  {NoStop}%
\bibitem [{\citenamefont {Miller}\ and\ \citenamefont
  {Pound}(2021)}]{Miller:2020bft}%
  \BibitemOpen
  \bibfield  {author} {\bibinfo {author} {\bibfnamefont {J.}~\bibnamefont
  {Miller}}\ and\ \bibinfo {author} {\bibfnamefont {A.}~\bibnamefont {Pound}},\
  }\bibfield  {title} {\bibinfo {title} {{Two-timescale evolution of
  extreme-mass-ratio inspirals: waveform generation scheme for quasicircular
  orbits in Schwarzschild spacetime}},\ }\href
  {https://doi.org/10.1103/PhysRevD.103.064048} {\bibfield  {journal} {\bibinfo
   {journal} {Phys. Rev. D}\ }\textbf {\bibinfo {volume} {103}},\ \bibinfo
  {pages} {064048} (\bibinfo {year} {2021})},\ \Eprint
  {https://arxiv.org/abs/2006.11263} {arXiv:2006.11263 [gr-qc]} \BibitemShut
  {NoStop}%
\bibitem [{\citenamefont {Pound}\ \emph {et~al.}(2020)\citenamefont {Pound},
  \citenamefont {Wardell}, \citenamefont {Warburton},\ and\ \citenamefont
  {Miller}}]{Pound:2019lzj}%
  \BibitemOpen
  \bibfield  {author} {\bibinfo {author} {\bibfnamefont {A.}~\bibnamefont
  {Pound}}, \bibinfo {author} {\bibfnamefont {B.}~\bibnamefont {Wardell}},
  \bibinfo {author} {\bibfnamefont {N.}~\bibnamefont {Warburton}},\ and\
  \bibinfo {author} {\bibfnamefont {J.}~\bibnamefont {Miller}},\ }\bibfield
  {title} {\bibinfo {title} {{Second-Order Self-Force Calculation of
  Gravitational Binding Energy in Compact Binaries}},\ }\href
  {https://doi.org/10.1103/PhysRevLett.124.021101} {\bibfield  {journal}
  {\bibinfo  {journal} {Phys. Rev. Lett.}\ }\textbf {\bibinfo {volume} {124}},\
  \bibinfo {pages} {021101} (\bibinfo {year} {2020})},\ \Eprint
  {https://arxiv.org/abs/1908.07419} {arXiv:1908.07419 [gr-qc]} \BibitemShut
  {NoStop}%
\bibitem [{\citenamefont {Cook}\ and\ \citenamefont
  {Whiting}(2007)}]{Cook:2007wr}%
  \BibitemOpen
  \bibfield  {author} {\bibinfo {author} {\bibfnamefont {G.~B.}\ \bibnamefont
  {Cook}}\ and\ \bibinfo {author} {\bibfnamefont {B.~F.}\ \bibnamefont
  {Whiting}},\ }\bibfield  {title} {\bibinfo {title} {{Approximate Killing
  Vectors on S**2}},\ }\href {https://doi.org/10.1103/PhysRevD.76.041501}
  {\bibfield  {journal} {\bibinfo  {journal} {Phys. Rev. D}\ }\textbf {\bibinfo
  {volume} {76}},\ \bibinfo {pages} {041501} (\bibinfo {year} {2007})},\
  \Eprint {https://arxiv.org/abs/0706.0199} {arXiv:0706.0199 [gr-qc]}
  \BibitemShut {NoStop}%
\bibitem [{\citenamefont {Comp\`ere}\ and\ \citenamefont
  {K\"uchler}(2021)}]{Compere:2021iwh}%
  \BibitemOpen
  \bibfield  {author} {\bibinfo {author} {\bibfnamefont {G.}~\bibnamefont
  {Comp\`ere}}\ and\ \bibinfo {author} {\bibfnamefont {L.}~\bibnamefont
  {K\"uchler}},\ }\bibfield  {title} {\bibinfo {title} {{Self-consistent
  adiabatic inspiral and transition motion}},\ }\href
  {https://doi.org/10.1103/PhysRevLett.126.241106} {\bibfield  {journal}
  {\bibinfo  {journal} {Phys. Rev. Lett.}\ }\textbf {\bibinfo {volume} {126}},\
  \bibinfo {pages} {241106} (\bibinfo {year} {2021})},\ \Eprint
  {https://arxiv.org/abs/2102.12747} {arXiv:2102.12747 [gr-qc]} \BibitemShut
  {NoStop}%
\bibitem [{\citenamefont {Stanzione}\ \emph {et~al.}(2020)\citenamefont
  {Stanzione}, \citenamefont {West}, \citenamefont {Evans}, \citenamefont
  {Minyard}, \citenamefont {Ghattas},\ and\ \citenamefont {Panda}}]{frontera}%
  \BibitemOpen
  \bibfield  {author} {\bibinfo {author} {\bibfnamefont {D.}~\bibnamefont
  {Stanzione}}, \bibinfo {author} {\bibfnamefont {J.}~\bibnamefont {West}},
  \bibinfo {author} {\bibfnamefont {R.~T.}\ \bibnamefont {Evans}}, \bibinfo
  {author} {\bibfnamefont {T.}~\bibnamefont {Minyard}}, \bibinfo {author}
  {\bibfnamefont {O.}~\bibnamefont {Ghattas}},\ and\ \bibinfo {author}
  {\bibfnamefont {D.~K.}\ \bibnamefont {Panda}},\ }\bibinfo {title} {Frontera:
  The evolution of leadership computing at the national science foundation},\
  in\ \href {https://doi.org/10.1145/3311790.3396656} {\emph {\bibinfo
  {booktitle} {Practice and Experience in Advanced Research Computing}}}\
  (\bibinfo  {publisher} {Association for Computing Machinery},\ \bibinfo
  {address} {New York, NY, USA},\ \bibinfo {year} {2020})\ p.\ \bibinfo {pages}
  {106–111}\BibitemShut {NoStop}%
\end{thebibliography}%


\providecommand{\noopsort}[1]{}\providecommand{\singleletter}[1]{#1}%
%

\end{document}